\documentclass[10pt,a4paper,openany]{book}
\usepackage[utf8]{inputenc}
\usepackage[english]{babel}

\usepackage{amsmath}
\allowdisplaybreaks 
\usepackage{amsfonts}
\usepackage{amssymb}
\usepackage{amsopn}
\usepackage{tensor} 
\usepackage{bbm} 
\usepackage{slashed} 
\usepackage{braket} 
\usepackage{pdfpages}

\usepackage[left=3cm,right=3cm,top=3cm,bottom=3cm]{geometry}

\usepackage[small]{caption}

\DeclareMathOperator{\tr}{tr}

\newcommand{\DD}[1]{\mathcal{D}#1\ }
\newcommand{\dd}[1]{d#1\ }
\DeclareMathOperator{\str}{\tilde{tr}} 
\DeclareMathOperator{\Tr}{Tr}
\newcommand{\ie}{i.e.\
}
\newcommand{\eg}{e.g.\
}
\newcommand{\GroupName}[1]{ { \text{#1} } }
\newcommand{\cl}{ { \text{cl} } }
\newcommand{\ad}{ { \text{ad} } }
\newcommand{\tot}{ { \text{tot} } }
\newcommand{\WZ}{ { \text{WZ} } }
\newcommand{\Lagr}{ { \mathcal{L} } }
\newcommand{\YM}{ { \text{YM} } }
\newcommand{\HD}{ { \text{hd} } }
\newcommand{\ym}{Yang-Mills}

\newcommand{\SYM}{ { \text{SYM} } }
\newcommand{\CH}{ { \text{chiral} } }
\newcommand{\Ssrc}{ { S_\text{sources} } }
\newcommand{\sym}{super-Yang-Mills}
\newcommand{\covD}{ { \mathcal{\nabla} } }
\newcommand{\covDs}{ { \mathcal{D} } }
\newcommand{\1}{ { \mathbbm{1} } }
\newcommand{\HR}{  \mathbb{H}\hspace{-0.385em}\mathbb{R}  }

\numberwithin{equation}{section} 

\usepackage{fancyhdr}

\usepackage{pgfplots}

\usepackage{color} 
\usepackage{hyperref} 
\hypersetup{
    colorlinks=true, 
    linkcolor=black, 
    urlcolor=black, 
    citecolor = black,
    linktoc=all 
}

\title{\bf On higher derivative gauge theories}
\author{Lorenzo Casarin}

\usepackage[nouppercase,swapnames]{frontespizio}

\begin{document}

	\pagestyle{empty}
	
	\begin{center}
	\includegraphics[height=5cm]{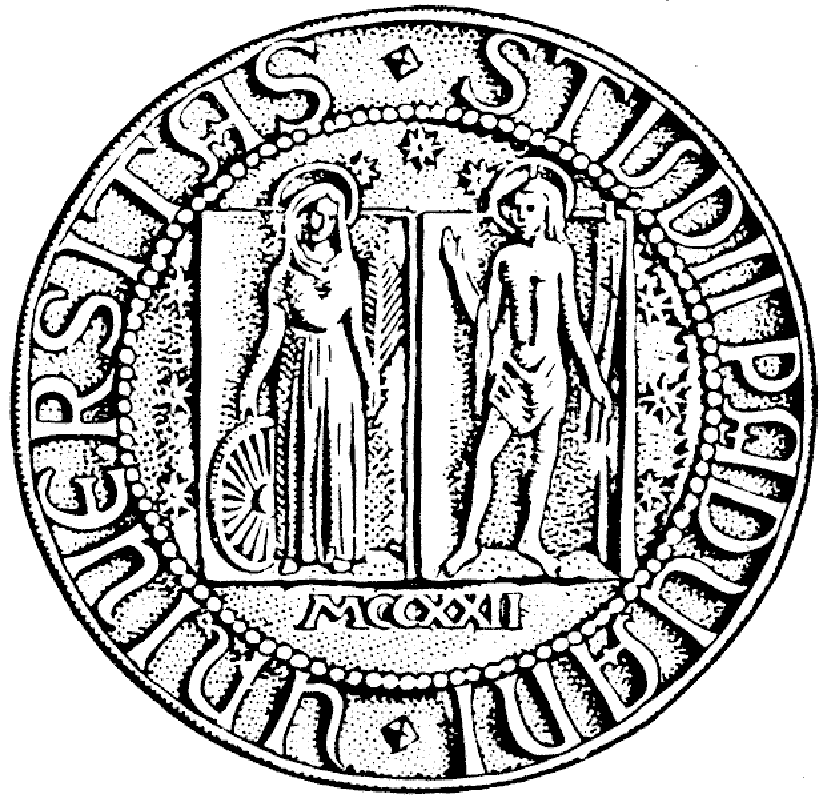}
	\vspace{2em}
	
	\textbf{
	\Large	Universit\`{a} degli Studi di Padova
	}

	\rule{\textwidth}{0.4pt}
	
	{
	\large	
	
	Dipartimento di Fisica e Astronomia ``Galileo Galilei''
	
	Corso di Laurea Magistrale in Fisica
	}	
	
	\end{center}

	\vspace{8em}

	\begin{center}
	\scshape
	\huge
	On Higher-Derivative Gauge Theories
	\end{center}

	\vspace{6em}

	\begin{flushleft}

	{Supervisor:\hfill{}Candidate:}\\[0.5em]	
	\textbf{\Large Prof.\ Arkady Tseytlin\hfill{}Lorenzo Casarin}\\[0.15em]
	{\large Imperial College London}
	
	\vspace{2em}	
	
	{Internal Supervisor:}\\[0.5em]
	\textbf{\Large Prof. Stefano Giusto}\\[0.15em]
	{\large Universit\`{a} degli Studi di Padova}

	\end{flushleft}
	
	\vfill

	\noindent\rule{\textwidth}{0.4pt}

	\begin{center}
	\bfseries 
	\large
	Academic Year 2016/2017
	
	\end{center}

	\newpage
	\pagestyle{empty}

\fancypagestyle{plain}{
	\fancyhead[LE]{ }
	\fancyhead[RO]{ }
	\fancyhead[RE]{ }
	\fancyhead[LO]{ }
	\fancyfoot[CE,CO]{\thepage}
	\fancyfoot[LO,RO]{ }
	\fancyfoot[LE,RE]{ }
	\renewcommand{\headrulewidth}{0pt}
	\renewcommand{\footrulewidth}{0.4pt}
}

	\fancyhead{} 
	\fancyhead[LE]{\scshape \leftmark}
	\fancyhead[RO]{\scshape \rightmark}
	\fancyfoot[LE,RO]{\thepage}
	\fancyfoot[LO,CE]{ }
	\fancyfoot[CO,RE]{ }
	\renewcommand{\headrulewidth}{0.4pt}
	\renewcommand{\footrulewidth}{0.4pt}

	\begin{center}
		\Large
		\textbf{Abstract}
	\end{center}
	
In this work we  study the main properties and the one-loop renormalization of a Yang-Mills theory in which the kinetic term contains also a fourth-order differential operator; in particular, we consider the Euclidean Lagrangian density
\begin{equation*}
\Lagr
	=
- \frac{1}{2 g^2} \tr  F_{\mu\nu} F_{\mu\nu} 
	- \frac{1}{ m^2 g^2 }\tr
		\left[
			 \left( \covD_\mu F_{\mu\nu} \right)^2
			+ \gamma   \: F_{\mu\nu}
				\left[ F_{\mu\lambda},
				F_{\nu\lambda} \right] 
			\right],
\end{equation*}
where
we add to the Yang-Mills term the most general contribution of mass dimension six, weighted with a dimensionful parameter $m$. 
This model is renormalizable; in the literature two values for the $\beta$ function for the gauge coupling $g$ have been reported: one in \cite{Fradkin:1981iu}, using the heat kernel approach, and the other in  \cite{Grinstein:2008qq,Schuster}, obtained via Feynman diagrams.
In this work we repeat the computation using heat kernel techniques confirming the latter result. We also considered coupling with matter.

We then study the supersymmetric extension of the model; this is a nontrivial task because of the complicate structure of the higher-derivative term.
Some partial results were known, but a computation of the $\beta$ functions for the full supersymmetric non-Abelian higher-derivative gauge theory was missing.
We make use of the (unextended) supersymmetric higher-derivative Lagrangian density for the \ym{} field in six spacetime dimensions derived in 
\cite{Ivanov:2005qf}; by dimensional reduction we obtain the $N=1$ and $N=2$ supersymmetric higher-derivative super-Yang-Mills  Lagrangian in four spacetime dimensions, whose $\beta$ function we evaluate using heat kernels. We also deduce the $\beta$ function  for $N=4$ supersymmetry.

\tableofcontents\markboth{}{} 

\pagestyle{plain}

\chapter*{Introduction}
		\addcontentsline{toc}{chapter}{Introduction}\markboth{Introduction}{}

		\fancyhead{} 
			\fancyhead[LE]{\scshape \leftmark}
			\fancyhead[RO]{\scshape \rightmark}
	\fancyfoot[CE,CO]{\thepage}
	\fancyfoot[LO,RO]{ }
	\fancyfoot[LE,RE]{ }
		 		\renewcommand{\headrulewidth}{0.4pt}
			\renewcommand{\footrulewidth}{0.4pt}
		\pagestyle{fancy}
		\renewcommand{\sectionmark}[1]{\markright{\thesection.\ #1}}
		\renewcommand{\chaptermark}[1]{\markboth{\chaptername\ \thechapter.\ #1}{}}
	\fancyfoot[CE,CO]{\thepage}
	\fancyfoot[LO,RO]{ }
	\fancyfoot[LE,RE]{ }

It can be shown that, under quite general hypotheses, a theory is renormalizable if all the couplings have non-negative mass dimension, and this constraints the form of the Lagrangian densities for renormalizable theories. Indeed, since the fields have positive canonical dimension and the derivative operator has mass dimension $1$, only a small number of independent terms can be constructed. Therefore, the presence of derivative interactions is often linked to non-renormalizable theories.

A loophole in the previous result is the fact that it assumes that the propagator scales, for high energies, as the inverse of the squared momentum, but with a carefully chosen derivative term in the equations of motion one can improve the ultraviolet properties of the theory. Indeed, one way to deal with divergent computations in QED, introduced by Pauli and Villars, is the substitution of the photon propagator with
\begin{equation*}
 \frac{i \eta_{\mu\nu}}{p^2} \longrightarrow 
 \frac{i \eta_{\mu\nu}}{p^2}  -  \frac{i \eta_{\mu\nu}}{p^2 - m^2}
 =
- \frac{i \eta_{\mu\nu} m^2}{p^2(p^2-m^2)}
\sim \frac{i \eta_{\mu\nu}}{\left( p^2 \right)^2 },
\end{equation*}
recovering the original theory taking the limit in $m^2 \rightarrow \infty$ at the end of the computation. Lee and Wick, in \cite{Lee:1969fy,Lee:1970iw}, investigated the properties of promoting  the new term arising from such substitution to a fundamental degree of freedom. In the Abelian gauge theory one can obtain such result by adding to the Lagrangian a term with the structure
\(
\sim  F_{\mu\nu} \partial^2  F^{\mu\nu}/ m^2
\). Notice that this kind of modification is peculiar for it is quadratic in the fields, that is the reason why it modifies the propagator.

These motivations suggested Grinstein, O'Connell and Wise  to  discuss in \cite{Grinstein:2007mp,Grinstein:2007iz,Grinstein:2008qq,Wise:2009mi} a whole extension of the Standard Model  inserting a quadratic differential operator in the conventional kinetic term. This was done for all the fields in the theory.
This model enjoys many interesting properties thanks to the improved ultraviolet behaviour; for instance, the Higgs boson mass is free of quadratic divergences in the cut-off, so that the hierarchy problem is solved, and this theory is also compatible with other known possible extensions of the Standard Model such as the see-saw mechanism to account for neutrino masses, as discussed in \cite{Espinosa:2007ny}. Despite the desirable properties, results from LHC seem not to be compatible with an extension of this form (\cite{Rizzo:2007ae}).
However, such models are interesting from a theoretical perspective for their properties and because they arise in various contexts. They also serve as toy models or low energy approximations of other theories, as we are going to discuss.

Quite remarkably, it was discovered (see \cite{Stelle:1976gc,Fradkin:1981iu,Avram}) that there exists an extension (actually, a two-parameter family of extensions) of General Relativity in which the Lagrangian contains quadratic terms in the Ricci tensor that turns out to be renormalizable. This is once again a reflection of the fact that these kind of operators improve the ultraviolet properties of the theory. Also in modern developments of theoretical physics, higher derivative couplings arise as low energy expansions of string theory models, such as the low energy expansion of conformal theories (see for instance \cite{Stelle:1978, Buchbinder:1999jn}).

All these nice properties of the higher derivative theories that we are considering come at some cost. 
We can understand this already at the classical level. The equations of motion  are fourth-order differential equations, so that they admit not only oscillatory solutions, but also solutions that grow exponentially with time (both in the past and in the future). This instability is manifested in the quantum mechanical formulation for the presence of ghost states: Indeed, as suggested in the expansion of the propagator considered above, the four-derivative theory can be expanded into a couple of ordinary-derivative theories, in which one of the two fields has an extra a negative sign in front of the kinetic term, that implies, upon canonical quantization, the presence of negative norm states.
The presence of these extra propagating degrees of freedom is represented also in the fact that the propagator has more poles than in the usual-derivative case; these actually make the Wick rotation problematic, since a proper definition of the contour of integration is a difficult task.


These problems can be partially cured in different ways. One is to impose future and past boundary conditions on the solutions, requiring that in the initial and final states no ghost particle are present. This can be done with a careful definition of the contour of integration that defines the Feynman propagator; this violates the usual causal relations, but ghost partner particles would then contribute only to virtual processes without spoiling the unitarity of the theory. Other approaches, like in \cite{Hawking:2001yt,Ghilencea:2007ex}, allow for a violation of unitarity in spite of maintaining the causal structure of the theory. However such problems might not be an issue if the theory is an effective theory in which the violation of causality or unitarity arises in a region of the parameter space that is out of the regime of applicability of the theory.

\vspace{1.5em}

Another popular candidate (\cite{Tommaso, WB, Sohnius:1985qm}) to describe physics beyond the Standard Model is supersymmetry. It is a symmetry relating the bosonic and fermionic fields in a given model, so that the two kind of degrees of freedom, in a quite  broad meaning, balance each other. 
Supersymmetry in this way provides candidates for dark matter; since bosons and fermions tend to give an opposite contribution to a given process, it also improves the ultraviolet properties of the theory, solving in particular the hierarchy problem discussed above. Supersymmetry also allows for a systematic cancellation of the vacuum energy contribution between the fields, thus hinting to a possible explanation to the smallness of the observed cosmological constant. Supersymmetry is not a manifest symmetry of nature, and therefore it should be broken at our energies. It also ties very well with string theory, that is another widely believed framework in which to unify the Standard Model and General Relativity, and many aspects of modern fundamental physics are nowadays driven by supersymmetric principles. 
It is then natural to wonder how supersymmetry relates with higher derivative theories, both in considering fundamental models and in dealing with low energy effective theories.

A further motivation to work in this direction rises when considering renormalizable theories in higher dimensional spacetime.
It is nowadays widely accepted the idea that our four-dimensional Universe is actually a submanifold embedded in some higher dimensional structure;  it is then of interest to consider the question of finding renormalizable and supersymmetric theories in such spacetime. As discussed in \cite{Smilga:2005pr,Ivanov:2005qf}, besides supersymmetry and renormalizability, another desirable requirement for the fundamental theory, that would ensure its ultraviolet completeness, is the invariance under conformal symmetry. This symmetry extends the Poincar\'e group essentially to include also circular inversions and dilatations. Superconformal algebras have been studied and classified (\cite{VanProeyen:1999ni} is a review), and it turns out that the highest possible dimension in which superconformal symmetry can be realised is six.
Renormalizability is in general a property that is not preserved if one considers a theory in an extended dimensionality; for example, the pure \ym{} theory is non-renormalizable in six spacetime dimensions. On the other hand, adding a higher derivative term of the form discussed above improves the ultraviolet properties also in this case, and exploiting  this fact one can formulate renormalizable theories also in higher dimensions. Indeed, a theory with the structure $F_{\mu\nu} \partial^2 F^{\mu\nu}/f^2$ is renormalizable in six dimensions keeping the canonical dimension $1$ for the gauge field, with $f$ dimensionless.
Therefore, a theory in four dimensions with an extra quadratic differential operator in the free sector, is also of interest as a low-energy manifestation of such more fundamental theories satisfying the aforementioned requirements.

Another aspect to keep into consideration is that, adding a quadratic differential operator, the auxiliary fields, necessary for the supersymmetry algebra to close off-shell, in general become dynamical and appear as ghosts.
However, the link between supersymmetry and higher derivative theories deserves to be studied more deeply. In simple models (some partial result is discussed in \cite{Ferrara:1977mv}) one can show that some kind of supermultiplet structure arises in considering higher derivative supersymmetric theories. However, one can construct supersymmetric theories with higher derivative terms without a dynamical auxiliary field; this is relevant because they are thought to be non-dynamical, and the Euclidean functional integral over the auxiliary field is divergent, while that on the physical fields is formally convergent. Also, in simple cases, it has been shown that the elimination of the dynamical ghost-like auxiliary field can provide mechanisms for supersymmetry breaking, as discussed in \cite{Fujimori:2016udq,Fujimori:2017kyi}.

\vspace{1.5em}

In this work we will study the main properties and the one-loop renormalization of a \ym{} theory in which the kinetic term contains also an extra quadratic differential operator of the type described above, as well as the usual contribution. Then, we will consider its supersymmetric extension. 

In detail, we consider an Euclidean Lagrangian density of the form
\begin{equation*}
\Lagr
	=
- \frac{1}{2 g^2} \tr  F_{\mu\nu} F_{\mu\nu} 
	- \frac{1}{ m^2 g^2 }\tr
		\left[
			 \left( \covD_\mu F_{\mu\nu} \right)^2
			+ \gamma   \: F_{\mu\nu}
				\left[ F_{\mu\lambda},
				F_{\nu\lambda} \right] 
			\right],
\end{equation*}
in which we allowed for adding the most general contribution of mass dimension six. The $\beta$ function for this theory was computed for the first time  using heat kernels in \cite{Fradkin:1981iu}, that studied this Lagrangian as a toy model for higher derivative gravity. More recently, \cite{Grinstein:2008qq,Schuster} considered the theory as a viable extension of the Standard Model and performed the computation with a diagrammatic approach finding a different result; up to now there is no consensus on the correct value of $\beta$.
We repeat in this thesis the computation using heat kernels, and we will confirm the latter result. We also consider some models of coupling with matter.

We then study the supersymmetric extension of the Lagrangian given above; formulating a supersymmetric higher derivative theory is actually a nontrivial task because of the complicate structure of the higher derivative term.  As we will motivate, supersymmetry requires restricting to the case $\gamma = 0$.
The supersymmetric case was somehow discussed in \cite{Buchbinder:1999jn} in terms of superfields, but some contributions have been ignored; \cite{Gama:2011ws} considered the higher-derivative extension of QED. However, an explicit and systematic formulation of the $N=1,2,4$ supersymmetric non-Abelian theory is still missing. Here we will make use of the (unextended) supersymmetric higher-derivative Lagrangian density for the \ym{} field in six dimensions obtained in \cite{Ivanov:2005qf}. In that case, the only  higher derivative term is $\sim  (\covD F)^2 $, weighted with a dimensionless constant,  as we mentioned above; by dimensional reduction  we will then obtain the $N=1$ and $N=2$ supersymmetric higher-derivative \sym{}  Lagrangian in four spacetime dimensions, and we will evaluate the one-loop $\beta$ function for such systems. Remarkably, we will be able to deduce the $\beta$ function also for the case $N=4$.

\vspace{1.5em}

The work is organised in three main Chapters and two Appendices, as follows.

In Chapter 1 we introduce all the technical tools needed for the computation. The general setting of the path integral is introduced, with particular attention to the background field technique. The main technical tool that we will employing to perform the computations, namely the heat kernel method for the computation of functional determinants, is introduced.

In Chapter 2 the higher-derivative extension of \ym{} theory is considered. Some of its properties are described, and in particular the one-loop renormalization is performed and the $\beta$ function of the \ym{} coupling is computed. Simple models of coupling with matter are discussed too, in particular we show that the mass of the scalar field in a class of higher-derivative theory is free of quadratic divergences.

Chapter 3 is devoted to the supersymmetric extension of the higher-derivative \ym{} theory studied in the previous Chapter. The nontrivial aspects of this generalization are introduced, and in order to overcome them the theory is then formulated in six spacetime dimensions. The relevant Lagrangian with $N=1$ and $2$ supersymmetries in four dimension is then obtained by dimensional reduction. The $\beta$ function of the $N=1$, $2$ and even $4$ supersymmetric extension is then computed.

In Appendix~A the notation is established.  In Appendix~B some technical computations are reported, given here favouring the readability of the text.


\chapter{Prolegomena}

This Chapter is devoted to introducing the necessary technology to perform computations in quantum fields theory and to fix the relevant notation. We will mainly quote results from the literature and show how to use them to tackle problems of interest for this thesis; the reader interested in a more detailed and systematic derivation and mathematical theory is invited to check the references.

The literature covering the basic aspects of Quantum Field theory is wide, and we will mainly refer to \cite{Peskin, Ram, WeinbergI, WeinbergII}.

In this Chapter we consider Quantum Field Theories and the path integral in the Euclidean spacetime. This is done to slightly simplify the notation and to deal with a formally convergent functional integral. We will ignore the subtleties in the definition of the Wick rotations, or the insights that Euclidean QFTs might give in other contexts.

\section{Path integral in Quantum Field Theory}

Let us consider a relativistic theory describing several local bosonic and fermionic fields. Bosonic scalar fields are described by real scalar functions
\(
\phi_i(x)
\), being $i$ an index labelling the independent components; in this notation complex fields are represented through their real and imaginary part. Fermionic  fields are represented with complex Grassmann functions
\(
\psi_j(x)
\) and their complex conjugate \(
\bar \psi_j(x)
\), again $j$ labelling independent fields. The fields might as well transform in some representation of a symmetry group, and therefore the functions $\phi_i$ or $\psi_j$ may carry other internal indices. We will consider vector fields as they arise when dealing with gauge theories.

The dynamics of the system is determined assigning a Lagrangian density
\begin{equation}
\Lagr = \Lagr[\phi_i, \psi_j, \bar \psi_j],
\end{equation}
where we made explicit the functional dependence on the fields. The Lagrangian  depends also on derivatives of the fields themselves; usual theories contain, when suitably integrated by parts, up to second derivative of bosonic fields and first derivatives of fermionic ones. We will consider only local theories, namely theories for which $\Lagr$ depends  on one spacetime point only.

\subsection{Generating functionals and the effective action} \label{basic-definitions}

The main tool that we will be employing in order to study the quantum properties of field theories is the path integral. All properties of the system can be encapsulated in the generating functional, that for a Euclidean theory reads
\begin{equation}\label{pi-generating-functional}
Z[J, \eta, \bar \eta ]
	=
\int  \DD{\phi}  \DD{\psi}  \DD{\bar\psi} \exp\left[ { -\frac{1}{\hbar}\left\lbrace S[\phi^i, \psi_j, \bar \psi_j] - \Ssrc \right\rbrace } \right],
\end{equation}
where 
\begin{equation}
S[\phi_i, \psi_j, \bar \psi_j] := \int \dd{^4 x} \Lagr [\phi_i, \psi_j, \bar \psi_j](x).
\end{equation}
is the action of the system seen as a functional of the fields,
\begin{align}
\DD{\phi}  \DD{\psi}  \DD{\bar\psi} &:= \prod_i \DD{\phi_i}
\prod_j \DD{\psi_j}
\prod_{j'} \DD{\bar \psi_{j'}} ,
\end{align}
is the formal integration measure, and the source term reads
\begin{align}
\Ssrc
	:=
 \int  \dd{^4 x} J^i(x) \: \phi_i(x)
 + \int  \dd{^4 x} \eta^i(x) \: \psi_i(x)
 + \int  \dd{^4 x} \bar \psi_i(x) \: \bar \eta^i(x),
\end{align}
being $J^i$, $\eta^i$ and $\bar \eta^i$ generic sources.
We assume here that the measure of the integral is normalized in such a way that the condition
\(
Z[0] = 1
\) 
is satisfied. We might also use the notation
\begin{equation}
J \cdot \phi = \int \dd{^4 x} J^i(x) \: \phi_i(x).
\end{equation}

In the rest of this Chapter we will restrict to the case of bosonic fields only; this simplifies the notation without significant loss of generality -- the main differences for fermions stem from the fact that it is necessary to distinguish between left and right differentiation because of anticommutativity. We will give the relevant results also when fermions are present.

The path integral formalism provides, at least formally, a straightforward way to compute Green's functions. Indeed, they can  be obtained differentiating the functional $Z$ with respect to the sources; for instance, the $n$-point Green function can be expressed as
\begin{equation}\label{pi-green-function}
\braket{ 0 | T\left[ \phi_{i_1}(x_1)\cdot \ldots\cdot \phi_{i_n} (x_n) \right]  | 0 }
	=
\left.
\frac{\delta^n Z[J] }{\delta J^{i_1}(x_1) \ldots J^{i_n}(x_n)}
\right|_{ J = 0 }.
\end{equation}
Such Green's functions are known to be disconnected objects: Though describing physical processes, they do not turn out to be fundamental, at least from a  formal perspective. Their building blocks are the unfactorisable, \ie connected, components; they are generated by the functional $E[J,\eta,\bar \eta]$, given in terms of $Z$ by the relation
\begin{equation}
Z[J] = \exp\left(-\frac{1}{\hbar}E[J]\right).
\end{equation}
The $n$-point connected Green's functions are then
\begin{equation}
G^n_{i_1 \ldots i_n}(x_1,\ldots x_n) 
	=
	-
\left. \frac{\delta^n E[J,\eta,\bar\eta] }{\delta J^{i_1}(x_1) \cdots \delta J^{i_n}(x_n)} \right|_{ J  = 0 }.
\end{equation}

An even more restrictive category of Green's functions is that consisting of one-particle irreducible functions. By this expression we mean those Green's function that remain connected after the elimination of one internal propagator.
To start with, consider the position
\begin{equation}\label{bfq-classical-fields}
\Phi_i(x)
	:=
- \frac{\delta E[J,\eta,\bar\eta]}{\delta J^i(x)}
	=
\frac{1}{Z[J]} \frac{\delta  Z[J] }{ \delta J^i(x) }  , 
	\hspace{1.25em}
\end{equation}
that is the average value of the field in the presence of the current $J$, 
often called the `classical' field. The relation \eqref{bfq-classical-fields} is assumed to be invertible, so that it can be used to define $J[\Phi]$ as a functional of some given configuration $\Phi$. $J[\Phi]$ is then the source term for which \eqref{bfq-classical-fields} holds. Such implicit definition is employed to define the quantum effective action functional \( \Gamma[\Phi] \) through functional Legendre transform
\begin{equation}\label{pi-quantum-effective-action-definition}
\Gamma[\Phi ]
	=
E[J ] + J  \cdot \Phi .
\end{equation}
An interesting property of \(\Gamma\) is that it generates the vertex functions, \ie the one-particle irreducible correlation functions, by differentiation; considering only the bosonic fields,
\begin{equation}
\Gamma_n^{i_1 \ldots i_n} (x_1, \ldots, x_n)
	=
\left.
\frac{\delta^n \Gamma[\Phi ]}{\delta \Phi_{i_1}(x_1)
	 \cdots \delta
	 \Phi_{i_n}(x_n)}
\right|_{\Phi  =0}
\end{equation}
is the $n$-point one-particle irreducible Green's function.

Let us now discuss some remarkable properties of the effective action. As a starting point, consider the relation \eqref{pi-quantum-effective-action-definition} and differentiate with respect to the classical field $\Phi$; simple algebra shows that 
\begin{equation}\label{eff-act-quantum-eom}
\frac{\delta \Gamma}{\delta \Phi_i} = J^i[\Phi] (x).
\end{equation}
Considering free systems, \ie setting the current to zero, the previous relation shows that the the external fields $\Phi$ make $\Gamma$ stationary. This situation is analogous to the classical picture in which solutions of the classical equations of motion are stationary points of the action $S$ itself and if a driving force is present, the variation of the action is indeed proportional to it.

A useful expression for $\Gamma$ is
\begin{equation}\label{eff-act-scalar}
\exp\left[{- \frac{1}{\hbar}\Gamma[\Phi]}\right] = \int \DD{\phi}  
\exp \left[ -\frac{1}{\hbar} \left( S - \int \dd{^4x}  \frac{\delta \Gamma [\Phi]}{ \delta \Phi_i}  (\phi_i - \Phi_i)  \right) \right],
\end{equation}
obtained combining the definition \eqref{pi-quantum-effective-action-definition} and the expression for the generating functional \eqref{pi-generating-functional}.

The effective action $\Gamma $ can be expanded in powers of $\hbar$, reading, at least as a formal expansion, 
\begin{equation}\label{pi-hbar-expansion-effective-action}
{\Gamma} =  \sum_L \hbar^L \Gamma_{(L)}
=  S + \hbar \Gamma_{(1)} + O\left(\hbar^2\right)
\end{equation}
where $\Gamma_{(L)}$ is the $L$-loop contribution to the effective action. In particular, $S = \Gamma_{(0)}$ is the classical tree-level action and $\Gamma_{(1)}$ is the first quantum correction. A similar computation shows that the generating functional $E$ can be computed with the tree level diagrams generated by $\Gamma$, and according to \eqref{pi-hbar-expansion-effective-action} this is enough to take all the quantum corrections into account, at least perturbatively.

The nomenclature of `loop' expansion originates from the diagrammatic approach to quantum field theory.
Consider indeed any connected Green's function. Differentiating \eqref{pi-quantum-effective-action-definition} with respect to the sources one can express all the connected Green's functions using the vertex functions and propagators.
Green's functions can then be represented with Feynman diagrams, where the propagator is represented by a line and vertex functions by vertices.

 The relation between the number of vertices (including those connected to the sources) $V$, the number of propagators $I$  and the number of internal momenta $L$ is $L = I - V + 1$; since every vertex brings a factor of $\hbar^{-1}$ and every propagator comes with $\hbar$, the full diagram is of order $\hbar^{I-V} = \hbar^{L-1}$. Comparing with \eqref{pi-hbar-expansion-effective-action}, we can therefore recognise $\Gamma_{(n)}$ to be the $n$-loop vertex function.

We have therefore seen that $\Gamma$ behaves in a way analogous to the classical action, can be computed in terms of a formal parameter $\hbar$ parametrizing the quantum effects and generates the one-particle irreducible Green's functions. These remarkable properties justify the name `quantum effective action'; more detailed discussions can be found in \cite{WeinbergII}.

\subsection{The case of gauge theories}

In this work we are mainly interested in studying gauge theories, whose  generating functional obtained from a na\"ive extension of \eqref{pi-generating-functional} is known to be ill-defined.

In general the theory under consideration can be invariant under the action of some symmetry group $\GroupName{G}$, so that every field $\phi_i$ or $\psi_j$ transforms in a given representation of the group itself according to
\begin{equation}\label{group-transformation}
\begin{split}
\delta \phi_i(x) & = \omega\: \phi_i(x)= \omega^a \: T_{(i)}^a \phi_i(x),
\\
\delta \psi_j(x) & = \omega\: \psi_j(x)= \omega^a \: T_{(j)}^a \psi_j(x)
\end{split}
\end{equation}
where $T_{(i)}$ ($T_{(j)}$) are the generators of the representation under which $\phi_i$ ($\psi_j$) transforms and $\omega = \omega^a T^a_{(i,j)}$ is an element of the Lie algebra of $\GroupName{G}$.

This means that, generally speaking, every field is actually a multiplet of fields, enumerated by an internal extra index. In order to make the symmetry local and consider position-dependent transformations parametrised by  Lie~algebra-valued fields $\omega(x)$ in \eqref{group-transformation}, a gauge field connection must be introduced  to ensure covariance of derivative terms under gauge transformation. Explicitly, the minimal coupling prescription consists in the replacement 
\begin{equation}
\partial_\mu \rightarrow \covD_\mu = \partial_\mu + A_\mu,
\end{equation}
where $A_\mu$ is a Lie-algebra valued field acting on the representation in which the fields are transforming according to
\begin{equation}\label{gauge-transf-A}
\delta A_\mu = -\partial_\mu \omega + [\omega,A_\mu] = -\covD_\mu \omega
\end{equation}
where the covariant derivative is in the adjoint representation.
After that, a kinetic term for $A_\mu$ can be considered in order to make gauge fields dynamical. The conventional kinetic term is
\begin{equation}\label{kinetic-term-F}
\Lagr_A = \frac{1}{2g^2} \tr F_{\mu\nu} F_{\mu\nu} 
= -  \frac{1}{4 g^2} F^a_{\mu\nu} F^a_{\mu\nu}
\end{equation}
with
\begin{equation}
F_{\mu\nu} = [\covD_\mu,\covD_\nu] = \partial_\mu A_\nu - \partial_\nu A_\mu + [A_\mu, A_\nu]
\end{equation}
being the \ym{} field strength, or curvature tensor, associated to the covariant derivative $\covD_\mu$. Notice that the covariant derivative and the field strength  transform in the adjoint representation:
\begin{align}
\delta \covD_\mu = [\omega, \covD_\mu],
\hspace{3em}
\delta F_{\mu\nu} = [\omega, F_{\mu\nu} ].
\end{align}

The problem with the definition \emph{\`a la} \eqref{pi-generating-functional} lies in the redundancy of the description of the physical observables. Indeed, physical quantities are invariant under the action of the gauge group: this means that solutions of the equations of motions that differ by a gauge transformation should provide the same observable quantities. The path integral measure in \eqref{pi-generating-functional} ignores this redundancy in the description of the same physical system and the integration is performed over the infinite orbit of the action of the gauge group. A good definition of the path integral measure should therefore single out just one physical representative for the  fields of a given system.

A solution to this issue consists in introducing a gauge-fixing condition $G[A](x)=\theta(x)$, where $G$ is some invertible non-gauge-invariant functional of the gauge field and $\theta$ is a function, and restricting the integration in \eqref{pi-generating-functional} to the fields satisfying such gauge condition. Up to a redefinition  of the normalization of the measure in \eqref{pi-generating-functional}, Faddeev and Popov have shown that a well-defined path integral is
\begin{equation}\label{pi-generating-functional-gauge-intermediate}
Z[J]
	=
\int  \DD{\phi} \DD{A} \det M[A] \: \delta(G-\theta)\exp\left[{-S +  S_{\text{sources}}} \right],
\end{equation}
with
\begin{equation}\label{gauge-fixing-jacobian}
\hat M[A](x,y) = M[A](x) \delta^{(4)}(x-y) = \left. \frac{\delta G[A^\omega] }{\delta \omega }\right|_{\omega=0} ,
\end{equation}
$M$ being a differential operator and $\delta(\, \cdot\,)$ a Dirac delta-like functional. The functional Jacobian determinant $\hat M$ is computed from the variation of the gauge fixing functional $G[A](x)$ with respect to a gauge transformation parametrized by the element $\omega(y)$. Internal indices of such determinant are therefore of the representation in which the gauge fields are.

In the case of the \ym{} field, $M$ has therefore indices in the adjoint representation, and the transformation $A^\omega$ is \eqref{gauge-transf-A}.
A common option for the gauge fixing is $G[A]=\partial_\mu A_\mu$; with these choices, the operator $\hat M$ reads 
\begin{equation}\label{M-lorenz-gauge}
\hat M[A] = \frac{\delta G[A](x)}{\delta \omega(y)} = - \partial_\mu \covD_\mu\delta(x-y) 
\end{equation}
and therefore $M = -\partial_\mu \covD_\mu$, with $\covD_\mu$ in the adjoint representation.

Integrating \eqref{pi-generating-functional-gauge-intermediate} with the respect to the function $\theta$ with a Gaussian weight 
\begin{equation}
\sqrt{ \det H }\ \exp\left\lbrace -\int \dd{^4x} \tr \theta(x) \: H(x) \: \theta(x) \right\rbrace,
\end{equation}
$H(x)$ being a generic differential operator independent of the quantum fields, one gets the well-known expression for the generating functional for gauge theories
\begin{equation}\label{pi-generating-functional-gauge}
Z[J]
	=
\int  \DD{\phi} \DD{A}
 \det M
 \sqrt{ \det H }
 \exp\left[{-\frac{1}{\hbar}( S_\tot  - \Ssrc)}\right].
\end{equation}
with
\begin{equation}\label{Stot}
S_\tot = S + \int \dd{^4 x} \tr G H G.
\end{equation}

Often in diagrammatic computations the determinants are represented by introducing ghost fields in the exponential, effectively modifying the Lagrangian density. We do not need to follow this paradigm, since we are interested only in the renormalization  properties and in Section~1.3 we will explain how to compute determinants directly looking at the form of the operators.

The definitions given in the Section \ref{basic-definitions} and the results presented are naturally extended to the generating functional \eqref{pi-generating-functional-gauge}.
In particular we have the definition
\begin{equation}
\mathcal{A}_\mu^a
	=
\frac{\delta E }{\delta J^{\mu a}}
\end{equation}
which brings us to the definition of the effective action $\Gamma[\Phi]$ as 
\begin{equation}\label{eff-act-gauge}
\begin{split}
&\exp\left[{- \frac{1}{\hbar}\Gamma[\Phi,\mathcal{A}]}\right] =
\\
& \hspace{5em}
\int \DD{\phi}   \DD{A} \det M \: \sqrt{\det H} \:  
\exp \left[ -\frac{1}{\hbar}( S_\tot - \Ssrc ) \right],
\end{split}
\end{equation}
where, with abuse of notation, the source term is intended
\[
\Ssrc = \int \dd{^4x}  \frac{\delta \Gamma [\Phi, \mathcal{A}]}{ \delta \Phi_i}  (\phi_i - \Phi_i) + \int \dd{^4x}  \frac{\delta \Gamma [\Phi, \mathcal{A}]}{ \delta \mathcal{A}^a_\mu }  (A^a_\mu - \mathcal{A}^a_\mu).
\]

\section{One-loop effective action}

In this section we exploit the formalism introduced to explain how it can be used to compute one-loop corrections. We identified the one-loop correction as the first nontrivial term in the expansion of the effective action $\Gamma$ in terms of $\hbar$; we will therefore expand the relevant quantities in order to get such contribution. This approach is discussed in a modern fashion but with different levels of depth in \cite{Avram, WeinbergII, Peskin}.

We will do the explicit computations for the generating functional \eqref{eff-act-scalar} containing only scalar fields, only sketching  the solution for the general case with fermions and gauge fields.

Expanding the effective action according to \eqref{pi-hbar-expansion-effective-action} in the expression \eqref{eff-act-gauge} and shifting the integration variables
\begin{equation}\label{1lopp-generic-expansion-action-shift}
\phi = \Phi + \sqrt{\hbar}\: \varphi,
\end{equation}
one can determine, order by order in $\hbar$, the quantum effective action $\Gamma$. At the lowest order in $\hbar$, this expansion can be understood as a semi-classical perturbation of a classical background $\Phi$, but it also encapsulates higher order corrections that can in principle be computed order by order in $\hbar$.
Notice that
\begin{equation}\label{1lopp-generic-expansion-action}
\begin{split}
S[\Phi + \sqrt{\hbar} \phi  ] 
	& =
S[\Phi]
 + \sqrt{\hbar} \int \dd{^4x} \left. \frac{\delta S}{\delta \phi_i(x)} \right|_{\Phi} \varphi_i(x)
\\
&
\quad   + \hbar \int \dd{^4 x} \dd{^4 y} \left. \frac{\delta^2 S}{\delta \phi_i(x) \delta \phi_j(y)} \right|_{\Phi} \varphi_i(x)  \varphi_j(y) + O(\hbar^{3/2})
\end{split}
\end{equation}
where the functional derivatives of the action are evaluated at $\Phi$. The second term of this expansion is nonzero, since $\Phi$ makes $\Gamma$, and not $S$, stationary, so that ${\delta_\phi S}[\Phi]$ is of order $\hbar$ -- it could be then ignored in this expansion, however it cancels with the contribution from $\delta_\Phi \Gamma$ (yet they contribute to higher order effects).
The left-hand-side of \eqref{eff-act-scalar} indeed gives, truncating the expressions at the relevant power of $\hbar$,
\begin{equation}
\exp\left[ - \frac{1}{\hbar}  S[\Phi] - \: \Gamma_{(1)}[\Phi]  \right]
\end{equation}
while the right-hand-side reads, thanks to the expansion \eqref{1lopp-generic-expansion-action}
\begin{equation}
\int \DD{( \sqrt{\hbar }\phi) }
\exp\left[
	- \frac{1}{\hbar} 
		S[\Phi] - \int \dd{^4 x} \dd{^4 y} \left. \frac{\delta^2 S}{\delta \phi_i(x) \delta \phi_j(y)} \right|_{\Phi} \varphi_i(x)  \varphi_j(y)
\right];
\end{equation}
the constant factor \(\exp[ -S[\Phi]/\hbar] \) simplifies between both sides of the expansion, and remembering the rule for functional integration over bosonic fields 
\begin{equation}
\int \DD{\phi} e^{- \phi \cdot \Delta \phi} = \frac{1}{\sqrt{ \det \Delta}},
\end{equation}
we arrive to the result
\begin{equation}\label{1loop-generic-gauge+matter}
\Gamma_{(1)}
	=
\frac{1}{2}\log \det \hat \Delta_\phi,
\hspace{2.5em}
\text{with}
\hspace{2.5em}
\hat \Delta_\phi:= \left. \frac{\delta^2 S}{\delta \phi_i(x) \delta \phi_j(y)} \right|_{\Phi}.
\end{equation}

The operator obtained differentiating twice the action in \eqref{1loop-generic-gauge+matter} is more easily obtained explicitly expanding the action, or rather the Lagrangian density, about the solution $\Phi$. Notice that $\Phi$ is really, in the absence of sources, a stationary `point' of the quantum effective action $\Gamma$. However, we can use any classical solution, \ie any stationary point of the action $S$, since the difference is of order $\hbar$ and thus resulting in higher-order corrections. For this reason, when choosing a background field configuration we can make considerations purely based on the classical equations of motion.

It is easy to extend this result in order to include gauge fields, whose generating functional is \eqref{pi-generating-functional-gauge}. We assume here that the solution $(\Phi,\mathcal{A})$ provides an operator in \eqref{1loop-generic-gauge+matter} that does not mix fields with different spin. The operator for the gauge field is obtained differentiating the total action $S_\tot$ given in \eqref{Stot}; gauge fields are just bosonic fields so they contribute with another factor \(\sim 1/\sqrt{\Delta_A}\). The presence of a gauge symmetry, as we discussed, modifies the generating functional to \eqref{eff-act-gauge}, adding the contributions $\det M$ and $ \sqrt{ \det H} $. The former  does not contain any factor of $\hbar$, so it can be simply evaluated on the solution $\Phi$, being the difference of higher order (remember the shift \eqref{1lopp-generic-expansion-action-shift}), and we will write $M[\Phi] \equiv M_0$. The latter is independent on the quantum field.

Fermionic contributions are slightly more delicate since in \eqref{1lopp-generic-expansion-action} the functional derivatives should be taken acting first on the right on $\psi_j$ and then on the left on $\bar \psi_j$. This is indeed natural when computing the expansion `by hand', since the rule for functional integration is
\begin{equation}
\int \DD{\psi} \DD{\bar \psi} e^{\bar \psi \cdot \Delta \psi} = {\det \Delta}.
\end{equation}
for complex Grassmann fields.

Explicitly, the one-loop contribution to the quantum effective action for the theory defined by \eqref{pi-generating-functional-gauge} reads, considering also fermionic matter,
\begin{equation}\label{eff-act-1loop-gauge}
\Gamma_{(1)}
	=
\frac{1}{2} \log \frac{ (\det M_0 )^2 \det H \  \prod_j (\det \Delta_{\psi,j})^2 }{\det \Delta_A \ \det \prod_i \Delta_{\phi,i} }.
\end{equation}
Now that we have isolated the one-loop contribution, we set $h=1$.

An important feature of the expansion \eqref{1lopp-generic-expansion-action-shift} is that the effective action $\Gamma$ obtained with this method can be constructed  so that it is invariant under (formal) gauge transformations of the background field. This can be done by choosing a proper gauge-fixing functional -- that might look like quite unusual in the conventional framework.  We stress here that it is not a `true' gauge transformation because the background field is not a dynamical variable, it is just an assigned function. The necessary condition for this invariance is that $H$ transforms covariantly in the adjoint representation under a gauge transformation of the background field.

The shift of the gauge field can be explicitly written
\begin{equation}\label{bfq-shift-gauge-field}
A_\mu^a \to  \mathcal{A}_\mu^a + A_\mu^a 
\end{equation}
and the gauge fixing functional that allows for the desired property is
\begin{equation}\label{bfq-gauge-fixing}
G[A](x) = \covD_\mu A_\mu 
	= (\partial_\mu + \mathcal{A}_\mu) ( A_\mu - \mathcal{A}_\mu ).
\end{equation}
where we used $\covD_\mu$ to denote the covariant derivative over the background field only. This will be done in the rest of the work as well: After the shift of the fields, implicit expressions such as $\covD_\mu$ and $F_{\mu\nu}$ are intended as functions of the background field only.
The Jacobian determinant for such gauge-fixing reads
\begin{equation}
M[A] = - \covD_\mu (\partial_\mu + A_\mu);
\end{equation}
notice that $M$ is computed considering real gauge transformations on $A_\mu$.
After the shift \eqref{bfq-shift-gauge-field}, the gauge-fixing becomes
\begin{equation}
G[A+ \mathcal{A}] = \covD_\mu( A_\mu + \mathcal{A}_\mu)
 = (\partial_\mu + \mathcal{A}_\mu) A_\mu 
\end{equation}
and the Jacobian reads
\begin{equation}
M[A+\mathcal{A}] = - \covD_\mu (\partial_\mu + A_\mu + \mathcal{A}_\mu)
	\equiv - \covD_\mu (\covD_\mu + A_\mu).
\end{equation}

We are interested in considering formal gauge transformations of the background field expressed in infinitesimal form by
\begin{equation}\label{eff-action-gauge-inv-transf-bkg}
\mathcal{A}_\mu \rightarrow \mathcal{A}_\mu -\covD_\mu \omega.
\end{equation} 
We want to prove the invariance of the right-hand-side of \eqref{eff-act-gauge} 
In order to do this, we change the quantum field under integration according to 
\begin{equation}\label{eff-action-gauge-inv-transf-quant}
A_\mu \rightarrow A_\mu + [\omega, A_\mu];
\end{equation}  
we then get that the original shifted  field (according to \eqref{bfq-shift-gauge-field}) transforms as 
\begin{equation}
\delta (A_\mu +\mathcal{A}_\mu) = -\covD_\mu \omega + [\omega, A_\mu] = - \partial_\mu \omega + [\omega, A_\mu + \mathcal{A}_\mu].
\end{equation}
This is precisely a gauge transformation in terms of the field prior to the shift, and therefore the action $S[A+\mathcal{A}]$ is invariant under such transformation. Then, since \eqref{eff-action-gauge-inv-transf-quant} is a gauge transformation in the adjoint representation and \eqref{eff-action-gauge-inv-transf-bkg} implies that $\covD_\mu$ transforms in the adjoint representation too,  $G[A+\mathcal{A}] = \covD_\mu A_\mu  $ transforms covariantly in the adjoint and the same is true for $M$, whose determinant is therefore invariant. The invariance of the source term is apparent provided the sources transform in the same representation of the respective fields. As anticipated, if $H$ transforms in the adjoint representation, its determinant is invariant and therefore such are all terms in $S_\tot$. We conclude this paragraph observing that $M$ in (1.2.13) evaluated at the classical solution reads
\begin{equation}\label{M0}
M_0 = - \covD^2
\end{equation}

We just proved that background field method provides an effective action that is gauge invariant in the background field; this fact strongly constraints the terms that can appear in its expression. In particular we see that the one-loop correction in \eqref{1loop-generic-gauge+matter} is gauge invariant and therefore a limited number of local functions can be present in its expression.

Notice that, since we are dealing with local theories only, the two-point operators $\hat \Delta(x,y)$ are actually diagonal in the spacetime coordinates and can be expressed in the form $\Delta_x\: \delta^{(4)}(x-y)$, being $\Delta_x$ a local differential operator. The determinant of $\hat \Delta$ therefore factorises in the determinant of the differential operator, depending on the background field, and the constant contribution of the delta function that we will ignore for it does not contribute to the divergence. For the integral over commuting variables to be convergent, we require $\hat \Delta$ -- or equivalently the differential operator $\Delta$ -- to be positive, that means 
\begin{equation}
\phi \cdot \hat \Delta \phi 
	=
\int \dd{^4 x} \dd{^4 y} \phi^*(x) \: \hat \Delta(x,y)  \: \phi(y)
	=
\int \dd{^4 x} \phi^*(x) \:  \Delta_x  \: \phi(x)
	\geq
0
\end{equation}
for any function $\phi$. Grassmann variables do not need such requirement since the exponential actually expands to a finite number of terms.

\section{Renormalization}

Let us go back to the expression \eqref{eff-act-1loop-gauge}. In the diagrammatic approach to Quantum Field Theory, the determinants can be evaluated considering loop processes. We will not follow this path, but let us just assume for now that we have some machinery that allows us to evaluate them, and that we similarly get divergent results. Let us also assume that we can parametrize the divergence with an ultraviolet cut-off $\Lambda$, and that the only relevant divergences are the logarithmic ones.

The essence of the renormalization procedure is the redefinition of the coupling constants and wavefunctions in order to reabsorb the  divergent contributions.
Suppose that the action contains a term that can be written in the form
\begin{equation}\label{ren-classical}
S = a_\cl S^0
\end{equation}
with \(a_\cl\) being a (bare) coupling constant and $S^0$ the spacetime integral of some local function of the fields. 
The quantum effective action up to order $\hbar$ is given by the sum of the tree-level contribution \eqref{ren-classical} and the one-loop term \eqref{eff-act-1loop-gauge}; if the divergent contribution coming from the latter has the same structure, namely can be written in the form
\begin{equation}
\Gamma_{(1)}^\Lambda =  - \frac{ \bar \beta }{ 16 \pi^2 }  S^0 \log \frac{\Lambda}{\mu},
\end{equation}
we can realise that the parameters of the bare Lagrangian do not correspond to measurable quantities, and therefore we conclude that, in order for the result to be finite as the physical system is, the bare parameters are actually divergent in such a way that they cancel the divergence.  We introduced the energy scale $\mu$ that has to appear for dimensional reasons.

In the case that no wavefunction renormalization is necessary, the requirement of $S +\Gamma_{(1)}^\Lambda  $ to be finite in the limit $\Lambda \rightarrow \infty$ becomes a redefinition of the bare coupling to depend on \( \Lambda \) in such a way that the renormalized coupling is independent of it, but we allow for a possible dependence  on the scale \( \mu \). In formul\ae{}, we are requiring that the final result
\begin{equation}
S_{\text{ren}} \supseteq
a_\mu S^0 = a_\Lambda S^0 - \frac{ \bar \beta }{ 16 \pi^2 } \ S^0  \log \frac{\Lambda}{\mu}
\subseteq S + \Gamma_{(1)},
\end{equation}
where $a_\Lambda$ is the bare coupling, now depending on $\Lambda$, and $a_\mu$ is the renormalised coupling, is finite in the renormalised action $S_{\text{ren}} $.
This brings us to the relation
\begin{equation}\label{renorm-a}
a_\mu = a_\Lambda  - \frac{ \bar \beta }{ 16 \pi^2 } \log \frac{\Lambda}{\mu}.
\end{equation}
In the case \( a = g^{- 2 } \), we obtain
\begin{equation}\label{running-coupling}
g^{- 2 }_\mu = g^{- 2 }_\Lambda  - \frac{ \bar \beta }{ 16 \pi^2 } \log \frac{\Lambda}{\mu};
\end{equation}
we can also express the running of the coupling by means of the  $\beta$-function
\begin{equation}\label{beta-function-generic}
\beta(g_\mu) = \mu \frac{\partial g_\mu}{\partial \mu} = - \frac{g_\mu^3}{32 \pi^2} \bar \beta.
\end{equation}
From now on, the dependence of the renormalized coupling constant $g_\mu$ on the scale $\mu$ will be understood.
This discussion extends naturally to the case in which different couplings are present.

We have implicitly chosen a minimal subtraction scheme for the divergences: Indeed, one can rescale the bare coupling in \eqref{renorm-a} of also a finite quantity, but this is not of interest for our purposes.

This procedure holds as long as no wavefunction renormalization is necessary; indeed, this turns out to be the case of interest for the present work since wavefunctions for gauge fields do not get renormalised. 
This is a consequence of the fact that the effective action is gauge invariant with respect to the background field. This implies that the one-loop effective action $\Gamma_{(1)}$ is a gauge invariant expression of the background field too, and therefore it must be written as the trace of covariant expressions,  the only available with gauge fields being $\covD_\mu$ and $F_{\mu\nu}$. These are not homogeneous functions of the gauge field $A_\mu$: Rescaling such field of a  $Z_A$, one obtains
\begin{equation}
\covD_\mu \rightarrow \partial_\mu + Z_A A_\mu,
\hspace{2em}
F_{\mu\nu} \rightarrow Z_A (\partial_\mu A_\nu - \partial_\nu A_\mu) + Z_A^2 [A_\mu,A_\nu],
\end{equation}
but in order for the result to be gauge covariant they should be some combination of $\covD_\mu$ or $F_{\mu\nu}$, but this clearly implies $Z_A = 1$.

We will now describe how to explicitly evaluate the divergent contribution for determinants of operators of interest.

\section{The Heat Kernel method}

We now define the Heat Kernel of a differential operator, and describe how this tool can be used to evaluate the determinants that appear in the expression for the one-loop effective action. Mathematical treatment for finding the Heat Kernel coefficients can be found in \cite{Vassilevich, gilkey} while for a more physically motivated procedure  the reader should consult \cite{dewitt, Fradkin:1981iu}.

Let us consider the initial-value problem for the evolution under  $\Delta$, a positive self-adjoint differential operator  of order $r$ defined in $\mathbb{R}^4$, that might carry also internal indices, with a formal time $t$,
\begin{equation}
(\partial_t  + \Delta_x )u(x,t) = 0, \hspace{2.5em} u(x,0) = f(x)
\end{equation}
with $f(x)$ a given function. The solution to this problem can be formally written as
\begin{equation}
u(x,t) = e^{-t \Delta_x} f(x).
\end{equation}
An alternative expression for the solution can be given as the convolution with an operator  $K(t;x,y,\Delta)$
\begin{equation}\label{hk-convol}
u(x,t) = \int \dd{x} K(t;x,y,\Delta) f(y)
\end{equation}
provided that  $K$ satisfies  the differential equation
\begin{equation}\label{hk-definition-heat-equation}
(\1 \partial_t + \Delta)\indices{^i_j}K(t;x,y;\Delta)\indices{^j_k}
	=
0
\end{equation}
with the boundary condition
\begin{equation}\label{hk-definition-boundary-condition}
K(0;x,y;\Delta)\indices{^i_j}
	=
\delta\indices{^i_j} \delta(x-y).
\end{equation}

From \eqref{hk-convol}, writing $f(x) = \braket{x|f}$ where $\ket{x}$ is a set of eigenkets of the position operator, we see that we can express the heat kernel as
\begin{equation}
K(t;x,y;\Delta)
	=
\braket{x | e^{- t \Delta} | y }.
\end{equation}

We can now go a step further towards the definition of the determinant of the operator \( \Delta \) through the relations
\begin{equation}\label{hk-logdet-definition}
\log \det  \Delta = \tilde{\Tr}\, \log  \Delta  = \tilde{\Tr}\,\left[ - \int_0^\infty \frac{dt}{t} e^{- t \Delta} \right]
\end{equation}
where $\tilde \Tr$  is the trace over spacetime indices as well as internal ones.
In order to understand the last equality recall that, being $\Delta$ self-adjoint, it admits a complete basis of eigenfunctions $\{\ket{n}\}_n$, where $f_n(x) := \braket{x |n}$ has eigenvalue $\lambda_n$, \ie
\(
\Delta f_n(x) = \lambda_n f_n(x)
\); the operator in the square brakets then reads
\begin{equation}
\braket{x | \int_0^\infty \dd{t} \frac{e^{-t \Delta }}{t} | n } =
\braket{x  | n }\int_0^\infty \dd{t} \frac{e^{-t \lambda_n }}{t}  
=f_n(x) \log \lambda_n
\end{equation}
using the formal equality $\log \lambda = \int_0^\infty \dd{t} {e^{-t \lambda }}/{t}  $ up to an  infinite constant (indeed, one can `verify' this relation by differentiating $\lambda$).
Computing then the trace over spacetime indices in \eqref{hk-logdet-definition}, we therefore obtain the sought relation
\begin{equation}
\log \det  \Delta =   - \Tr   \int_0^\infty \frac{dt}{t} \braket{x | e^{- t \Delta} | x} = -   \int_0^\infty \frac{dt}{t} \int \dd{^4x}\Tr  K(t;x,x;\Delta).
\end{equation}
This integral is in general divergent over in both limits. Since  $t$ has canonical dimension $ [t] = - r < 0$, and we are interested in studying the ultraviolet behaviour of the theories, we consider only the possible divergence in the lower bound.

An asymptotic expansion for the trace of the heat kernel near $t=0^+$ is known in the general case of a self-adjoint differential operator of order $r$:
\begin{equation}\label{hk-asymptotics}
\Tr K( t ; x , x ; \Delta) 
	\sim_{0^+}
\sum_{k\geq 0} \frac{2}{(4\pi)^2 r} b_k(x) \ t^{{(k-4)}/{r}},
\end{equation}
where the so-called Seeley-deWitt coefficients $b_k$ can be expressed in terms of local invariants computed from the operator $\Delta$. The reason for the strange normalization will be clear later.
We also introduce the definition
\begin{equation}
B_k( \Delta)
	:=
\frac{1}{(4\pi)^2}
\int \dd{^4x}  b_k(x).
\end{equation}

Comparing the asymptotics \eqref{hk-asymptotics}  with \eqref{hk-logdet-definition} it is clear that the integral is indeed divergent at the lower bound. There are many ways to regulate it, such as dimensional or $\zeta$-function regularization; here we will simply introduce an explicit UV cut-off $\Lambda$, so that the divergent contributions read
\begin{equation}
\begin{split}
\left( \log \det \tilde \Delta \right)_\Lambda 
	& = 
-
\frac{2}{r}
\int_{ (\Lambda/\mu)^{-r} } \frac{dt}{t} \sum_{ k \geq 0 } \left( \frac{t}{\mu^r} \right)^{ (k - 4)/r  } B_k(\Delta) \\
	& =
	- 2 B_4 \log \frac{\Lambda}{\mu}
	+ \ldots,
\end{split}
\end{equation}
where we omitted power-law divergences. Notice that the $\log\Lambda$-divergence is given by the $b_4$ coefficient regardless of the order of the differential operator. This indeed gives the divergent contribution that we were looking for, and  therefore one can use the Heat Kernel method that we just described for finding the $\beta$ function.

Considering the effective action in \eqref{eff-act-1loop-gauge}, we can apply elementary properties of the logarithm function and we see that the overall coefficient for the logarithmic divergence in the effective action \eqref{eff-act-1loop-gauge} is
\begin{equation}
B^\tot_4 = \frac{1}{(4\pi)^2}\int \dd{^4 x} b_4^{\tot}(x)
\end{equation}
with
\begin{equation}\label{b4-total}
\begin{split}
b_4^\tot = 
	b_4( \Delta_{A})
	- 2 b_4( M_0)
	-	b_4( H) 
	- 2 \sum_j	b_4( \Delta_{\psi,j}) 
	+ \sum_i 	b_4( \Delta_{\phi,i})
.
\end{split}
\end{equation}
If the decomposition
\begin{equation}\label{ren-1loop-contribution}
B^\tot_4 \equiv \int \dd{^4 x } \frac{b^\tot_4(x) }{ 16 \pi^2 } = \frac{ \bar \beta }{ 16 \pi^2 }   S^0
\end{equation}
holds, then the divergences have the same structure of the terms already present in the Lagrangian, and the renormalization programme that we outlined in the previous Section can be carried out.

After these formal considerations, the algorithm to compute one-loop corrections is clear: We start with any Lagrangian, then we expand it to the second order about a classical solution  and then we extract the operator and we evaluate the coefficient $b_4$. In the following pages we will obtain expression for such coefficient for operators of interest.

\subsection{Determinants: Second order differential operators}

Let us start by analysing convetional two-derivative theories containing only bosonic fields; these kind of operators can be treated almost explicitly and constitute the fundamental building block for evaluating all the other relevant cases.

In order to understand the definition that we gave in the previous paragraphs, we start by considering a trivial case that can be worked out explicitly, that is the case of the scalar Laplacian $\Delta= -\partial_\mu\partial_\mu$ (the minus is added to ensure the positivity). It is easy to verify that the solution to the heat equation \eqref{hk-definition-heat-equation} is the family of Gaussian functions
\begin{equation}\label{hk-laplacian}
K \left( t;x,y; \Delta \right)
	=
\frac{1}{(4 \pi t)^{2}} \exp\left[ - \frac{(x-y)^2}{4t} \right].
\end{equation}
The expansion \eqref{hk-asymptotics} is then trivially given by $b_0 (x) = 1$ and $b_n(x) = 0$ for all other coefficients with $n\geq 1.$ 

Let us move on to a more general case. The most general positive, self-adjoint, second order differential operator is
\begin{equation}\label{generic-second-order-diff-op}
\Delta_2 
	=
	- \covD^2 + X
\end{equation}
where $\covD_\mu$ is some covariant derivative\footnote{Notice that it might not be the obvious one: The expansion of the action could give also a contribution of the form $c_\mu \covD_\mu $ for some functions $c_\mu$, but such contribution can be reabsorbed introducing a new covariant derivative and redefining $X$.} and $X\indices{^i_j}(x)$ is a function, possibly dependent on the background field.

Considering the explicit solution \eqref{hk-laplacian}, one could try to consider a power-law correction of the form
\begin{equation}\label{hk-delta2-full-solution}
K \left( t;x,y; \Delta \right)
	=
\frac{1}{(4 \pi t)^{2}} \exp\left[ - \frac{(x-y)^2}{4t} \right]
\sum_{n \geq 0}
	a_{n}(x,y) t^{n}
;
\end{equation} indeed, plugging this into the heat equation for the operator $\Delta$ we find a set of recursive differential equations between the coefficients that, in principle, one can solve. The algebra is quite lengthy but in principle doable $n$ by $n$;  comparing with the asymptotic \eqref{hk-asymptotics}, after some work one gets
\begin{align}
 b_0(x) &  = \Tr a_0(x,x) = \Tr \1
\\
 b_1(x) &  = 0 
\\
 b_2(x) &  = \Tr  a_1(x,x) = \Tr X
\\
 b_3(x) &  = 0 
\\ \label{b4-second-order}
 b_4(x) & = \Tr a_2(x,x) = \Tr \left[ \frac{1}{12} F_{\mu\nu} F_{\mu\nu}  + \frac{1}{2} X^2 \right], 
\end{align}
where \(F_{\mu\nu} = [ \covD_\mu , \covD_\nu ] \) is the curvature tensor associated to the covariant derivative in the representation under which the fields transform. For our purposes, $F_{\mu\nu}$ will be the gauge field strength tensor.

As we previously mentioned, we can also study  other kind of operators starting with the second order case. The key relation to study other operators is
\begin{equation}\label{hk-determinant-composition}
\det \left( \Delta_1\Delta_2\right)
	=
\det \Delta_1 \cdot \det \Delta_2,
\end{equation}
valid for any self-adjoint operators $\Delta_{1,2}$. Comparing the expansion \eqref{hk-asymptotics} and the relation \eqref{hk-determinant-composition}, we get a relation between the $b_4$  coefficients
\begin{equation}\label{hk-a4-composition}
b_4 \left( \Delta_1 \Delta_2 \right)
	=
b_4 \left( \Delta_1 ) + b_4( \Delta_2 \right)
\end{equation}
as it follows by applying the definitions and comparing the terms proportional to $t^0$.

\subsection{Determinants: First order differential operators}

An immediate application of \eqref{hk-a4-composition} is the computation of the $b_4$ coefficient for a first order differential operator, that is the case relevant when dealing with spinor fields.

The kind of first-order self-adjoint differential operators acting on a two-dimensional Weyl spinor we will be interested in are
\begin{equation}
\Delta_1 :=  i \bar \sigma^{\nu\; \dot \alpha \beta}\covD_\nu,
 \hspace{4em}
\bar \Delta_{1} := - i 	\sigma\indices{^{\nu}_{ \alpha \dot \beta} } \covD_\nu.
\end{equation}  
Consider now the composition \( \Delta_{1+\bar 1} = \Delta_1 \cdot \bar \Delta_{1} \), that is, remembering \(\covD_\mu \covD_\nu =  \covD_{(\mu} \covD_{\nu)} + \frac{1}{2} F_{\mu\nu}\), and the definition \eqref{notation-gener-lorentz-spinor}
\begin{equation}
\left(\Delta_{ 1 + \bar 1 }\right)^{\dot\alpha}_{\dot \beta}
=
 - \delta^{\dot\alpha}_{\dot \beta} \covD^2
			+ \frac{1}{2} \bar \sigma\indices{^{\rho \nu\; \dot \alpha}_{\dot\beta}} F_{\rho \nu}
\end{equation}
that has the structure of \eqref{generic-second-order-diff-op}, and therefore \eqref{b4-second-order} applies.
Conversely, we then have
\(
\det \Delta_1 = \det \bar \Delta_{1}
\)
since \( \det \varepsilon = -1\) and the relation between $ \sigma^\mu $ and $ \bar \sigma^\mu $ is \eqref{notation-sigma-matrices}, implying 
\begin{equation}
b_4 \left( \Delta_1  \right)
=
b_4 \left( \bar \Delta_{1} \right)
=
\frac{1}{2} b_4 \left( \Delta_{1+\bar 1} \right) .
\end{equation}
Performing the computation one gets
\begin{equation}\label{b4-first-order}
\begin{split}
b_4(\Delta_1)
& = \frac{1}{2} b_4(\Delta_{1+\bar 1})
\\
& =	\frac{1}{2}
	\Tr \left[
		+ \frac{1}{12} F_{\mu\nu} F_{\mu\nu}
		+  \frac{1}{2} \left( \frac{1}{2} \bar \sigma\indices{^{\rho \nu\; \dot \alpha}_{\dot\beta}} F_{\rho \nu} \right)^2
	\right]
\\
& =
-	\frac{1}{6}
	\tr \left[
		  F_{\mu\nu} F_{\mu\nu}
	\right]
\end{split}
\end{equation}
where the identity \eqref{identity-spinor_trace-double-sigmamn-V} has been used.

\subsection{Determinants: Fourth order differential operators}

The kind of operators that we will be dealing with has the structure
\begin{equation}\label{fourt-order-self-adjoint}
\Delta_{4,\text{sf}} = \covD^4 + \covD_\mu  \mathcal{V}_{\mu\nu} \covD_\nu +  \mathcal{N}_\mu \covD_\mu + \covD_\mu \mathcal{N}_\mu  + \mathcal{U};
\end{equation}
where the derivatives act on everything at their right and the matrices of the gauge-covariant coefficients satisfy 
\begin{equation} \label{fourth-order-symmetry-requirement}
\mathcal{V}_{\mu\nu} = \mathcal{V}_{\nu\mu}, 
	\qquad
\mathcal{V}^T_{\mu\nu} = \mathcal{V}_{\mu\nu},
	\qquad
\mathcal{N}^T_{\mu} = - \mathcal{N}_{\mu},
	\qquad
\mathcal{U}^T = \mathcal{U},
\end{equation}
where superscript \( T \) indicates the transpose with respect to internal indices.
This is the most general self-adjoint fourth-order operator without a term cubic in the covariant derivative. This is an important feature because this is the form of an operator that is obtained when composing two second order differential operators with the structure \eqref{generic-second-order-diff-op}.

The requirement that the coefficient $b_4$ is a scalar and a local expression of mass dimension $4$ implies that it is the trace of some linear combination of \( F_{\mu\nu}F_{\mu\nu} \), \(  \mathcal{V}_{\mu\nu} \mathcal{V}_{\mu\nu} \), \( ( \mathcal{V}_{\mu\mu} )^2 \) and $\mathcal{U}$. Other possible invariants such as  \( \covD_\mu \mathcal{N}_\mu \) are total derivatives vanishing when integrated to get the trace in \eqref{hk-logdet-definition}. 
This observation, and the comparison with the result for the composition suitably chosen second order operators using the rule \eqref{hk-a4-composition}, provides enough information to reconstruct
\begin{equation}
b_4(\Delta_{4,\text{sf}}) = \tr \left[
	  \frac{1}{6} F_{\mu\nu}F_{\mu\nu}  
	+ \frac{1}{24}\mathcal V_{\mu\nu} \mathcal V_{\mu\nu}  
	+ \frac{1}{48} \mathcal V^2 
	- \mathcal U 
	\right].
\end{equation}
Also in this case the algebra is straightforward but lengthy; the computation is technical and we will not reproduce it here.

Notice that \eqref{fourt-order-self-adjoint} is not the natural expression for the operator that one obtains expanding a Lagrangian up to the quadratic order in the fluctuations. In that case, the natural form for the operator is
\begin{equation}\label{hk-fourth-order-generic}
\Delta_4 = \covD^4 + V_{\mu\nu} \covD_\mu  \covD_\nu +  2N_\mu \covD_\mu + U;
\end{equation}
with, again,
\begin{equation}\label{hk-fourth-order-generic-coefficeints}
{V}_{\mu\nu} = {V}_{\nu\mu}, 
	\qquad
{V}^T_{\mu\nu} = {V}_{\mu\nu},
	\qquad
{N}^T_{\mu} = - {N}_{\mu},
	\qquad
{U}^T = {U}.
\end{equation}
However, it is immediate to relate the coefficients with those of \eqref{fourt-order-self-adjoint}; it is easy to find that the difference is only given in terms of total derivatives, that do not contribute after the integration performed in \eqref{hk-logdet-definition}. It is therefore justified to use in our computations the expression
\begin{equation}\label{b4-coeff-4order}
b_4(\Delta_{4}) = \Tr \left[
	  \frac{1}{6} F_{\mu\nu} F^{\mu\nu}  
	+ \frac{1}{24} V_{\mu\nu} V_{\mu\nu}  
	+ \frac{1}{48}  V^2 
	-  U 
	\right],
\end{equation}
slightly more immediate given an operator of the form \eqref{hk-fourth-order-generic}.

As implicit in what we just said, $N_\mu$  does not enter into the computation, hence we will not consider its contribution in the expansion \eqref{hk-fourth-order-generic}.

This method of computation allowed \cite{Fradkin:1981iu} to obtain for the first time this result.
In \cite{Gusynin:1988zt}, the same formula was derived using an asymptotic expansion as described for the second order case.
The interested reader can find the coefficient for a generic fourth order differential operator, with the $\sim \covD^3$ term, in \cite[p.~54]{Barvinsky:1985an}, however we will not need that result.

\subsection{Determinants: Third order differential operators}

This case is relevant for the spinor field operator with higher derivatives. The self-adjoint version of the relevant operator  is
\begin{equation}\label{3rd-order-operator-generic}
(\Delta_{3})_{\alpha \dot \beta} 
=
i \covD_\mu \sigma^\rho_{\alpha \dot \beta} \covD_\rho \covD_\mu
+
\frac{i}{2} K^{\mu}_{\alpha \dot \beta} \covD_\mu
+
\frac{i}{2} \covD_\mu K^{\mu}_{\alpha \dot \beta} 
+
B_{\alpha \dot \beta}
\end{equation}
being $K^\mu$ and $B$ matrices with respect to internal indices and hermitian for all the indices. The determinant of such an operator can be obtained by the previous ones via composition with a suitable first order operator. Considering self-adjoint operators, defining \( \Delta_{3+1} = \Delta_3 \cdot \Delta_1 \), 
given the leading symbol of \eqref{3rd-order-operator-generic}, a suitable choice is
\(
	\Delta_1 =  i \bar \sigma^{\nu\; \dot \alpha \beta}\covD_\nu
\), whose determinant was evaluated in \eqref{b4-first-order}.
Using \eqref{hk-a4-composition} once more, we can then find
\begin{equation}\label{b4-coeff-3order-implicit}
\begin{split}
b_4(\Delta_3) & = b_4(\Delta_{3+1}) - b_4(\Delta_1) \\
& =  b_4(\Delta_{3+1}) 
+	
\frac{1}{6}
	\tr \left[
		  F_{\mu\nu} F_{\mu\nu}
	\right]
\end{split}
\end{equation}

The explicit evaluation of the operator \(\Delta_{3+1}\) does not yield an insightful  general formula because of the presence of the sigma matrices; it will be directly evaluated in the case of interest.
However, a few comments are in order to make the computation slightly easier. Notice that, after the composition with the first order operator, the coefficient $B$ in \eqref{3rd-order-operator-generic} becomes the coefficient  \(N_\mu\) of \eqref{b4-coeff-4order}; since the latter effectively does not enter into the computation, we can discard the $B$-term in \eqref{3rd-order-operator-generic}, that will be systematically ignored from now on. Then, we can see that the operator \eqref{3rd-order-operator-generic} can be rewritten in the more natural  form
\begin{equation}\label{3rd-order-operator}
\Delta_{3}
=
i \sigma^\rho_{\alpha \dot \beta} ( \covD_\mu )^2 \covD_\rho 
+
i \tilde K^{\mu}_{\alpha \dot \beta} \covD_\mu
\end{equation}
that, after the composition with $\Delta_1$, gives the same operator $\Delta_{3+1}$ of \eqref{3rd-order-operator-generic}.

This kind of technique has employed for the first time to compute the determinant of higher-derivative spinor operators in \cite{Fradkin:1981jc}.

\section{Examples}

As a warm-up, before dealing with the higher-derivative  theories, we are now going to compute one-loop correction to the gauge sector of conventional \ym{} theories. In this way we will also show how to apply the rather abstract formalism discussed so far to concrete Lagrangians.
 We will consider the pure \ym{} case first, and then we will consider matter fields as well, specialising the final result to the $N=1$, $2$, $4$ supersymmetric extension.

We are interested in the renormalization of the \ym{} coupling $g$ that appears in front of the kinetic term for the gauge field \eqref{kinetic-term-F}. Since the quantum effective action is symmetric with respect to gauge transformations of the background field, all contributions must be (traces of) combinations of $F^{\mu\nu}$ and possibly $\covD_\mu$. As we discussed, these quantities are not homogeneous on the background field and no other internal parameter is present, so there cannot be any wavefunction renormalization. The procedure outline in Section~1.3 therefore applies.

\subsection{Pure Yang-Mills}

We consider the Yang-Mills Lagrangian in euclidean spacetime
\begin{equation}\label{L-YM-abstract}
\mathcal{L}_\YM 
	=
	- \frac{1}{2 g^2} \tr	F_{\mu\nu} F_{\mu\nu}
	=
	\frac{1}{4 g^2} F_{\mu\nu}^a F_{\mu\nu}^a
\end{equation}
where
\(
F_{\mu\nu} = [\covD_\mu, \covD_\nu]
\)
is the field strength tensor for the covariant derivative \( \covD_\mu = \partial_\mu + A_\mu \). We will use the gauge fixing functional defined in \eqref{bfq-gauge-fixing} to ensure formal gauge invariance of the effective action; we postpone the choice of the integration weight in order to show how it can be made to simplify the Lagrangian density.

We now proceed expanding the Lagrangian around the background field configuration $B_\mu^a$, shifting the quantum field
\begin{equation}\label{ym-shift-field}
A_\mu \rightarrow A_\mu + B_\mu.
\end{equation}
We will make the quantum field $A_\mu$ explicit in all expressions; $\covD_\mu$ and $F_{\mu\nu}$ in the expanded Lagrangian are intended as functions of $B_\mu$ only.
The covariant derivative and the field strength tensor transform according to
\begin{equation}\label{ym-shift-covD}
\covD_\mu  \rightarrow \covD_\mu + A_\mu,	 	
\end{equation}
and
\begin{equation}\label{ym-shift-Fmunu}
F_{ \mu \nu}  \rightarrow F_{\mu\nu} +  \covD_\mu A_\nu - \covD_\nu A_\mu + [A_\mu,A_\nu].
\end{equation}

In order to determine the operator for the quantum fluctuation, we have to expand the Lagrangian keeping only quadratic contributions in $A_\mu$. Following the previous formul\ae{}, we get 
\begin{equation}
\begin{split}
\left( F_{\mu\nu} \right)^2
\rightarrow
&
\left( F_{\mu\nu} +  \covD_\mu A_\nu - \covD_\nu A_\mu + [A_\mu,A_\nu] \right)^2
\\
& \simeq
2 F_{\mu\nu} [A_\mu,A_\nu]  + 
2 ( \covD_\mu A_\nu  )( \covD_\mu A_\nu ) - 2 ( \covD_\mu A_\nu  )( \covD_\nu A_\mu).
\end{split}
\end{equation}
Performing the trace over gauge indices the previous expression reads
\begin{equation}
\begin{split}
\left( F^a_{\mu\nu} \right)^2
& \rightarrow
2 F^a_{\mu\nu} A^b_\mu A^c_\nu f^{abc} + 
2 ( \covD_\mu A^a_\nu  )( \covD_\mu A^a_\nu ) 
- 2 ( \covD_\mu A^a_\nu  )( \covD_\nu A^a_\mu).
\end{split}
\end{equation}
and integrating by parts, dropping total derivatives,
\begin{equation}\label{YM-operator}
\left( F^a_{\mu\nu} \right)^2
\rightarrow
2  A^a_\mu   \left[
  	- (\covD^2)^{ab} \delta_{\mu\nu}  
  	+ (\covD_\mu \covD_\nu)^{ab} 
	- 2 F^m_{\mu\nu} f^{amb} 
  	\right]
 A^b_\nu.
\end{equation}
It is convenient, in order to slightly simplify the expressions in the higher-derivative case, to adopt a compact notation; using the fact that the fields are in the adjoint representation, we can suppress the indices and write
\begin{equation}\label{YM-operator-implicit-indices}
\left( F^a_{\mu\nu} \right)^2
\rightarrow
2  A_\mu  \cdot \left[
  	- (\covD^2) \delta_{\mu\nu}  
  	+ (\covD_\mu \covD_\nu)
	- 2 F_{\mu\nu} 
  	\right]
 A_\nu.
\end{equation}

The quadratic sector in the expanded Lagrangian density therefore reads
\begin{equation}\label{L-YMquadr}
\mathcal{L}_{\YM, A^2 } = \frac{1}{2 g^2} 
 A^a_\mu   \left[
  	- (\covD^2) \delta_{\mu\nu}  
  	+ (\covD_\mu \covD_\nu) 
	- 2 F_{\mu\nu}
  	\right]^{ab}
 A^b_\nu.
\end{equation}

It is now instructive to choose the gauge fixing functional $G$ and integration weight $H$.   As discussed, the convenient gauge choice in the framework of the background field quantization is
\(
{G}[A + \mathcal{A}] = \covD_\mu A_\mu
\),
with $H=  1/g^2.$ In this case \( \det H \) is trivial because it is independent of the fields and therefore can be reabsorbed in the overall normalization of the path integral measure; had we used any different weight we would have needed to take into account its contribution as well. The gauge fixing term can be written as, integrating by parts,
\begin{equation}
\int {G} H {G}
	=
- \int d^4x \frac{1}{2g^2} A_\mu^a \left( \covD_\mu \covD_\nu \right)^{ab} A^b_\nu.
\end{equation}
The contribution of $\det M_0$ must be taken into account, since, as computed in \eqref{gauge-fixing-jacobian}, $M_0= - \covD^2$ clearly depends on the background field.

We are now ready to obtain the operator related to the gauge field, that is, following \eqref{Stot} or equivalently the total Lagrangian density \( \Lagr_{\YM,A^2} - GHG \)
\begin{align}\label{ym-operator-ecplicit}
(\Delta_A)_{\mu\nu}  = 	 	
  	- (\covD^2) \delta_{\mu\nu}  
  	- 2 F_{\mu\nu} ;
\end{align}
or, in terms of operators in the adjoint representation,
\begin{equation}\label{ym-operator}
\Delta_A =
  	- \covD^2
	- 2 F  .
\end{equation}

The coefficients of the determinants are ready to be evaluated; using \eqref{b4-second-order}, the gauge-fixed operator for the \ym{} field \eqref{ym-operator} contributes with
\begin{equation}\label{b4-ym-operator}
\begin{split}
b_4( \Delta_A ) 
& = 
\Tr \left[ \frac{1}{12} F_{\mu\nu} F_{\mu\nu} \delta_{\alpha \beta} + \frac{1}{2} \cdot 4 \:  F_{\alpha\mu} F_{\mu \beta} \right]  
\\
& = 
\tr \left[ \frac{1}{3} F_{\mu\nu} F_{\mu\nu} - 2  F_{\mu\nu} F_{\mu \nu} \right]
\\
&
= \frac{5}{3} C_2 F^a_{ \mu \nu }F^a_{ \mu \nu }
\end{split}
\end{equation}
where we used \eqref{notation-trFmunuFmunu}; the Jacobian $M_0$ contributes with (notice the different index structure)
\begin{equation}\label{b4-M0-operator}
b_4( M_0 )  =  \tr \left[ \frac{1}{12} F_{\mu\nu} F_{\mu\nu} \right] = - \frac{1}{12} C_2 ( F^a_{ \mu \nu } )^2 .
\end{equation}

The total coefficient introduced in \eqref{b4-total} (with the convention $b_4(H) = 0$) is easily evaluated and gives
\begin{equation}
b_4^{\tot} = \frac{11}{6} C_2 F^2
\end{equation}
and comparing with \eqref{running-coupling} and \eqref{beta-function-generic}, we get
\begin{equation}
\bar \beta = \frac{22}{3} C_2
\end{equation}
yielding the familiar result
\begin{equation}
g^{- 2 }_\mu = g^{- 2 }_\Lambda  - \frac{ {22 C_2}/{3}  }{ 16 \pi^2 } \log \frac{\Lambda}{\mu}.
\end{equation}
The $\beta$ function, according to \eqref{beta-function-generic}, reads
\begin{equation}
\beta_\YM = - \frac{g^3}{16 \pi^2} \cdot\frac{11}{3} C_2
\end{equation}
in agreement with the literature, as in \cite{Ram, WeinbergII}.
The interpretation of this result is that the coupling $g$ diminishes with increasing scale $\mu$, and the first-order solution suggests $g \rightarrow 0$ as $\mu \rightarrow \infty$, that is asymptotic freedom. In contrast, for finite scale $\mu$ above a low-energy threshold, $g$ remains finite and the gauge field is self-interacting.

\subsection{Yang-Mills fields with matter }

The Euclidean Lagrangian describing the theory is
\begin{align}
\Lagr & = 
\Lagr_\YM + \Lagr_{\phi} + \Lagr_{\psi}
\end{align}
with $\Lagr_\YM$ defined in \eqref{L-YM-abstract} and
\begin{align}
\Lagr_{\phi} &= \left( \covD_\mu \phi_i \right)  \left( \covD_\mu \phi_i \right),\\
\Lagr_{\psi} &=  \psi_j \sigma^\mu \covD_\mu \bar \psi_j ;
\end{align}
again, we employ the gauge fixing condition
\(
{G}[A] = \covD_\mu ( A_\mu - \mathcal{A}_\mu )
\)
weighted with $ H =  1/g^2 $, whose determinant drops from the computation. The contraction of internal indices is understood. As we are going to comment, this is the most general Lagrangian density for the renormalization of the gauge coupling; it is interesting to notice that, since the matter fields appear just with their kinetic term, the $ \beta $ function is determined knowing only the field content of the theory.

As we have already mentioned, the gauge invariance over the background field implies that it is enough to compute the divergence proportional to $F^2$. In order to do this, we can expand the Lagrangian about a solution with vanishing matter fields.  A consequence of this is that, even if we added a gauge invariant potential $V$ (\eg a Yukawa coupling)  it would not contribute to the term that is quadratic in the fluctuations, and therefore it would be ineffective for our purposes. 
In formul\ae{}, the the expansion reads
\begin{equation}\label{bkg-with-matter}
A_\mu \rightarrow A_\mu + B_\mu ,
\hspace{3.5em}
\phi_i \rightarrow \phi_i,
\hspace{3.5em}
\psi_j \rightarrow \psi_j.
\end{equation}
The shift of the gauge field is therefore again \eqref{ym-shift-field}, and we will simply call the fluctuation of the matter field without changing symbol.

The expansion is therefore relatively simple. The operator of the gauge fields was already obtained in \eqref{ym-operator};
as far as matter fields are concerned, the Lagrangian is already quadratic in them, and therefore it is enough to evaluate the covariant derivatives on the background field. Up to total derivatives the the operator for scalar fields is obtained simply by integrating by parts,
\begin{equation}
\Delta_{\phi,i} = 
	- \covD^2
\end{equation}
and that of spinors is obtained simply stripping spinor the fields, \ie
\begin{equation}
(\Delta_{\psi,j})_{\alpha\beta} = 
	i \sigma^\mu_{\alpha\beta} \covD_\mu
\end{equation}

The relevant Seeley-deWitt coefficients for the \ym{} field and the gauge fixing contribution have already been evaluated in \eqref{b4-ym-operator} and \eqref{b4-M0-operator}. 
Applying \eqref{b4-second-order} and \eqref{b4-first-order} we can immediately evaluate the coefficients for the matter fields, obtaining
\begin{align}\label{b4-scalar-usual-der}
b_4( \Delta_{\phi,i} ) & = - \frac{1 }{12}  C_{\phi,i} F^a_{\mu\nu} F^a_{\mu\nu}  
\end{align}
and
\begin{align}\label{b4-scalar-higher-der}
b_4( \Delta_{\psi,j} ) & = \frac{1}{6} C_{\psi,j} F_{\mu\nu}^a F_{\mu\nu}^a  .
\end{align}
The total coefficient reads
\begin{align}
b^{\tot}_4 & = 
	b_4( \Delta_\YM ) 
	- 2 b_4( M|_0 ) 
	+  \sum_i b_4( \Delta_{\phi,i} )
	- 2 \sum_j b_4( \Delta_{\psi,j} ) \\
& =
	F_{\mu\nu}^a F_{\mu\nu}^a  \left(
	\frac{11}{6} C_2 
	-   \frac{1 }{12} \sum_i C_{\phi,i}
	-   \frac{1}{3} \sum_j C_{\psi,j}
	\right)
\equiv \frac{1}{4} F_{\mu\nu}^a F_{\mu\nu}^a  \bar \beta
\end{align}
and leads to
\begin{align}\label{beta-ym-w-matter-generic}
\beta(g) 
= - \frac{g^3}{32 \pi^2} \bar \beta
= - \frac{g^3}{8 \pi^2} \left(
	\frac{11}{6} C_2 
	- \frac{1 }{12} \sum_i C_{\phi,i}
	- \frac{1}{3} \sum_j C_{\psi,j}
	\right).
\end{align}

It is now interesting to compute the beta function in some relevant cases studied in the literature.

Let us consider Quantum Chromodynamics (QCD) with $n_c$ colors. It is a theory of $n_f$ Dirac fermions in the same representation of the gauge group. There are therefore no scalars; then, $C_{\psi,j} = C_\psi$ that factorizes, and considering the doubling of the Weyl components in each Dirac spinor one gets $\sum_j C_j = 2 n_f C_j$. \eqref{beta-ym-w-matter-generic} therefore reads
\begin{align}
\beta(g) = \mu \frac{d g}{d	\mu} 
= - \frac{g_\mu^3}{16 \pi^2} \left(
	\frac{11}{3} C_2 
	-  \frac{4}{3} n_f C_{\psi}
	\right).
\end{align}
For an $\GroupName{SU}$-invariant theory, $C_2 = C_2(\GroupName{SU}(n_c)) = n_c$.
This is the classical result usually computed with diagrammatic techniques; the coupling $g$ is asymptotically free for small number of fermions; on the other hand for big $n_f$, depending on the representation in which the spinors are, the one-loop result suggests the presence of a UV Landau pole.

\subsection{Super-Yang-Mills}

Let us specialise \eqref{beta-ym-w-matter-generic} to the supersymmetric case. Some details on supersymmetric theories in four specetime dimensions are given in Appendix B. In terms of $N=1$ supermultiplets, at our disposal we have the chiral multiplet (one complex scalar and one Weyl fermion) and the vector multiplet (one gauge field and one Weyl fermion); systems with extended supersymmetry can be represented using $N=1$ supermultiplets.

\subsubsection{$\boldsymbol { N = 1} $ \sym}
This is simply the vector multiplet: no scalar fields are present in this model, so formally $\sum C_{\phi,i} = 0 $; other than the gauge field, there is one Weyl fermion in the adjoint representation, so that $C_{\psi} = C_2$ and the final result is
\begin{align}
\beta(g) = - \frac{g^3}{8 \pi^2} \left(
	\frac{11}{6} 
	- \frac{1}{3}
	\right)C_2 
= - \frac{g^3}{16 \pi^2} 3C_2 .
\end{align}

\subsubsection{$\boldsymbol { N = 2 } $ \sym}
The field content is one complex scalar, two Weyl fermions and the gauge field, hence in terms of $N=1$ superfields, this theory contains exactly the vector multiplet and the chiral multiplet. Being all the fields  in the adjoint representation,  $\sum C_{\psi,j} = 2 C_2 $ and since the complex field splits into real and imaginary components, $\sum C_{\phi,i} = 2 C_2$. Therefore,
\begin{align}
\beta(g) = - \frac{g^3}{8 \pi^2} \left(
	\frac{11}{6}
	- \frac{1 }{6} 
	- \frac{2}{3}
	\right) C_2 
= - \frac{g^3}{8 \pi^2} C_2  .
\end{align}

\subsubsection{$\boldsymbol { N = 4} $ \sym}
This theory describes three complex scalars, four Weyl fermions and the gauge field; that are three chiral multiplets and the vector rmultiplet.
As usual, all fields are in the adjoint representation, so that $\sum C_\psi = 4 C_2 $ and $\sum C_\phi = 6 C_2$ obtaining the celebrated result
\begin{align}
\beta(g) = - \frac{g^3}{8 \pi^2} \left(
	\frac{11}{6}
	- \frac{1}{2}
	-  \frac{4}{3}
	\right) C_2 
	= 0.
\end{align}

\chapter{Higher-derivative gauge theory}

This Chapter is organised as follows. In Section 2.1 we define the higher-derivative gauge theory that we want to consider and study some of its properties; then, in Section~2.2 we explicitly compute the one-loop $\beta$ function for the \ym{} coupling. In Section~2.3 we consider also the presence of matter fields, allowing for a higher-derivative Lagrangian for them as well.

\section{Pure-gauge higher derivative theory}

Let us consider a pure-gauge theory invariant under the action of a semisimple gauge group $\GroupName{G}$; for the notation and the definition of the mathematical structures, see Appendix~A. The theory that we are considering describes a set of gauge fields $A_\mu^a$, where $a=1,\ldots,\dim\mathrm{Lie}\, \GroupName{G}$ is the `color' index.

The higher-derivative extension of the \ym{} Euclidean Lagrangian density that we are going to consider is
\begin{equation}\label{lagr-hdym}
\Lagr^\HD_{\YM}
	=
- \frac{1}{2 g^2} \tr  F_{\mu\nu} F_{\mu\nu} 
	- \frac{1}{ m^2 g^2 }\tr
		\left[
			 \left( \covD_\mu F_{\mu\nu} \right)^2
			+ \gamma   \: F_{\mu\nu}
				\left[ F_{\mu\lambda},
				F_{\nu\lambda} \right] 
			\right],
\end{equation}
where $\covD_\mu$ is the covariant derivative of the gauge field $A_\mu$ and, as usual,
\(
F_{\mu\nu} 
	=
[\covD_\mu, \covD_\nu] 
\); $ \gamma$ is a real dimensionless parameter, whereas $m$ has dimension of mass, as the name suggests. 
In components, the Lagrangian density \eqref{lagr-hdym} reads
\begin{equation}\label{lagr-hdym-comp}
\Lagr^\HD_{\YM}
	=
\frac{1}{4 g^2}  F^a_{\mu\nu} F^a_{\mu\nu}
	+ \frac{1}{ 2 m^2 g^2 }
		\left[
			 \left( \covD_\mu F^a_{\mu\nu} \right)^2
			+ \gamma  f^{abc}  F^a_{\mu\nu}
				F^b_{\mu\lambda}
				F^c_{\nu\lambda}  
			\right],
\end{equation}

In \eqref{lagr-hdym} we added to the \ym{} Lagrangian density the most general contribution that can be written from terms of mass dimension six  preserving gauge invariance.
Such new terms enjoy interesting properties. This can be understood observing that $(\covD_\mu F_{\mu\nu})^2$ is a kinetic-like term, because it contains contributions that are quadratic in the fields. Indeed, this contribution is not studied in perturbation theory: It can be resummed in an additional term modifying the propagator, that, after a proper gauge fixing, for high momenta scales with $p^{-4}$. Such UV behaviour results in an improved scaling of loop integrals that is a first hint to the fact that this kind of theories have, in general, better renormalization properties than usual-derivative theories.
As we are going to discuss, despite having a parameter with negative mass dimension (\ie $1/m^2$), the scaling of the propagator at high energies is such that the theory turns out to be renormalizable.


\subsection{Degrees of freedom}

In this subsection we will comment on the structure of degrees of freedom of the theory \eqref{lagr-hdym}. Since the kinetic term is unconventional, we first specify what we actually mean when we consider the degrees of freedom. We will distinguish between `off-shell' and `on-shell'  degrees of freedom. The former is the number of independent real fields that are used to define the Lagrangian density, regardless of the dynamics; the latter is half the number of initial real data necessary to specify a solution of the equation of motions. These definitions are motivated by the comparison with ordinary-derivative theories.

The theory describes a vector field with gauge invariance, hence it has $ 4 - 1= 3 $ off-shell degrees of freedom for each color index. 

Taking the dynamics into consideration, the Lagrangian density contains terms with up to four derivatives, so that the equations of motions derived from it are fourth-order partial differential equations.  The component $A^0$  of the gauge field is still non-dynamical, since, as in the usual \ym{} case, $F^{00}=0$.  The number of on-shell degrees of freedom is therefore $ 2 \cdot 3 - 1 = 5 $ for each gauge index.

For comparison, recall that the conventional Yang-Mills action $\sim  \tr F_{\mu\nu}F_{\mu\nu}$ propagates $3 - 1=2$ on-shell degrees of freedom for color index. The additional number of degrees of freedom in $\Lagr_\YM^\HD$ can be made manifest rewriting the Lagrangian density in a suitable fashion; as we are going to show, the degrees of freedom can be realized through a conventional Yang-Mills field interacting with a massive vector field conveying the $3$ missing degrees of freedom.

For simplicity, and since we are interested only in the structure of the degrees of freedom, in our discussion we will consider only the extended kinetic term
\begin{equation}\label{lagr-hdym-kinetc}
\left. \Lagr^\HD_{\YM} \right|_{\text{kin}}
	=
-\frac{1}{2 g^2} \tr  F_{\mu\nu} F_{\mu\nu} 
	- \frac{1}{ m^2 g^2 }
			\tr \left( \covD_\mu F_{\mu\nu} \right)^2
\end{equation}
Let us introduce an auxiliary field $A'_\mu$  transforming in the adjoint representation of the gauge group. Let $\covD_\mu$ and $F_{\mu\nu}$ denote the covariant derivative and the field strength tensor of the field $A_\mu$. The Lagrangian density
\begin{equation}
\Lagr =
	-\tr \left[
			\frac{1}{2} F_{\mu\nu} F_{\mu\nu} 
			- m^2 A'_\mu A'_\mu
			- 2 F_{\mu\nu} \covD_\mu A'_\nu
		\right]
\end{equation}
gives back the original \eqref{lagr-hdym-kinetc} on equations of motion of the auxiliary field 
\begin{equation}
  m^2 A'_\mu = \covD_\mu F_{\mu\nu} .
\end{equation} Shifting $ A_\mu \rightarrow A_\mu + A'_\mu $, the auxiliary field becomes dynamical and the mixing $\sim \tr F \covD A'$ exactly cancels:
\begin{equation}\label{lagr-HD-ym-splitted-dof}
\begin{split}
\Lagr
& =
	-\tr \bigg[
			\frac{1}{2} \left(
				F_{\mu\nu} + \covD_\mu A'_\nu - \covD_\nu A'_\mu + [A'_\mu,A'_\nu]
				\right)^2 
			- m^2 A'_\mu A'_\mu
\\
&
\hspace{3em}
			- 2 \left(
				F_{\mu\nu} + \covD_\mu A'_\nu - \covD_\nu A'_\mu + [A'_\mu,A'_\nu]
				\right)
				(\covD_\mu  A'_\nu + [A'_\mu,A'_\nu]) 
		\bigg]
\\
& 
	 =
	-\tr \bigg[
		\frac{1}{2}	F_{\mu\nu}F_{\mu\nu} 
		- \frac{1}{2} \left( \covD_\mu A'_\nu - \covD_\nu A'_\mu \right)^2 
		- m^2 A'_\mu A'_\mu 
		+ \text{interactions} 
		\bigg]
\end{split}
\end{equation} 
where in the last equality we focused on the kinetic term.

Notice that, in this decomposition, both the kinetic and the mass term for the massive fields have the wrong sign. Already at the classical level this is known to be a problem because the equations of motion allow for modes growing exponentially at infinity, that upon canonical quantization result in negative norm states in the Hilbert state space.

These aspects of the theory could lead to a violation of causality; this problem can be solved by imposing future boundary conditions, or modifying the contour of integration in the definition of the propagator so that the particles associated to negative-norm decay in every physical process. It can also be shown that the theory with $\gamma = 0$, at least in the non-abelian case or at low energies (smaller than the scale dictated by $m$) is perturbatively unitary. Anyway, dealing with  these problems goes much beyond the aim of the present work, since we are interested in the one-loop renormalization properties, therefore we will not investigate further these issues.

\subsection{Renormalizability}

We are now going to discuss the properties of renormalizability of the theory \eqref{lagr-hdym}, showing that, as anticipated, it turns out to be renormalizable despite having a parameter with negative mass dimension. We will do so by performing an explicit power counting of the divergences in Feynman diagrams and without considering the background field approach. Since this approach is somehow tangential to the main topics of the present work, we will outline the main steps in the derivation without all details and explicit expressions. Notice that we are not using the background field technique outlined in the previous Chapter.

We choose the gauge fixing to be
\begin{equation}
G[A] = \partial_\mu A_\mu,
\end{equation}
integrated with weight 
\begin{equation}
 H = - \frac{1}{m^2 g^2}\partial_\mu \partial_\mu + \frac{1}{g^2},
\end{equation}
so that the Lagrangian density picks up the contribution
\begin{equation}\label{hdym-renorm-gauge-fixing}
\Delta_G \Lagr
	=
	-
\frac{1}{2}
\tr \left[ 
	\partial_\mu A_\mu 
	\left( 1 - \frac{1}{m^2}\partial_\nu \partial_\nu \right)
	\partial_\rho A_\rho
\right]
\end{equation}
Notice that the first term, \ie the `$1$'  in the expression of $H$, is the same Lorenz gauge-fixing contribution in the case of the \ym{} Lagrangian cancelling the contribution $\partial_\mu A_\nu \, \partial_\nu A_\mu$ from the expansion of $F_{\mu\nu}F_{\mu\nu}$; the second cancels the analogous term originating from $(\partial_\mu F_{\mu\nu})^2 \sim \partial_\mu \partial_\mu A_\nu \, \partial_\rho  \partial_\rho A_\nu  $.

The factor $\det H$ is just a numerical contribution independent of the fields, therefore it can be reabsorbed in the definition of the integration measure.
This choice of $G$ gives the Jacobian contribution \eqref{M-lorenz-gauge}
\begin{equation}
M = - \partial_\mu  \covD_\mu 
.
\end{equation}
In order to evaluate the determinant in a diagrammatic framework, we represent it by introducing (anticommuting) Lorentz-scalar ghost fields  $c$ and  $\bar c$ via
\begin{equation}
\det \hat M = \int \DD{c} \DD{\bar c} 
	\exp \left[ {\int \dd{^4x}  \bar c \ M \ c } \right]
\end{equation}
that effectively adds a new contribution to the Lagrangian density
\begin{equation}
\Delta_c \Lagr
	=
\tr \left[ - \bar c \ \partial_\mu \covD_\mu c \right].
\end{equation}

Considering the kinetic term \eqref{lagr-hdym-kinetc} with the gauge fixing contribution \eqref{hdym-renorm-gauge-fixing}, it is apparent that at high momenta $p$ the Feynman propagator for the gauge field scales as
\begin{equation}
D^{ab}_{\mu\nu}(p) \sim  \frac{ \delta^{ab} \delta_{\mu\nu} }{(p^2)^2},
\end{equation}
while the ghost fields have the usual propagator behaving at high energies like
\begin{equation}
D^{ab}(p) \sim  \frac{ \delta^{ab} }{{p}^2}.
\end{equation}

Let us consider a  Feynman diagram with $n_A$ internal gauge lines, $n_c$ internal ghost lines, $V_n$ vertices with exactly $n$ gauge field lines and $V_c$ ghost vertices. From the interaction terms in \eqref{lagr-hdym}, it is clear that  $n$ takes the values $3$, $4$, $5$ and $6$. 
The number of loops in the diagram is 
\begin{equation}
L = n_A + n_c - \sum_{n=3}^6 V_n - V_c + 1;
\end{equation}
the number of external particles is
\begin{equation}
E_A
	=
\sum_{n=3}^6 n V_n + V_c - 2n_A,
\hspace{4em}
E_c
	=
2V_c - 2n_c.
\end{equation}
Combining these results the superficial degree of divergence is
\begin{equation}
\begin{split}
d 
	& =
4L - 4n_A - 2 n_c + \sum_{n=3}^6 (6-n) V_n + V_c
\\
	& =
6 - 2L - E_A - 2E_c
\end{split}
\end{equation}
This result is enough to prove the renormalizability of the theory. Indeed, at one-loop level the only candidates for a non-renormalizable divergent result are the two-point function of the gauge field and the ghost two-point function. The former is ruled out by gauge invariance; the latter, in principle logarithmically divergent, is actually finite since one derivative acts on one external ghost line.
All other diagrams with two or more loops are finite. This proves that the theory is super-renormalizable, in the sense that only a finite number of diagrams are divergent.

\section{One-loop renormalization of the gauge sector}

In this Section we will compute the one-loop quantum correction to the gauge coupling $g$. We follow the path outlined in the first Chapter using the background field method. As discussed for the usual \ym{} theory in Section~1.5, we have to compute the divergence proportional to $\tr F_{\mu\nu} F_{\mu\nu} $, and therefore a clever choice of background solution is one with vanishing matter fields. The expansion therefore reads, as in \eqref{ym-shift-field},
\begin{equation}
	A_\mu \rightarrow A_\mu + B_\mu 
\end{equation}
with the same notation as before; in particular the covariant derivative $\covD_\mu$ and the field strength tensor $F_{\mu\nu}$ are expressed in terms of the background field only. Their expressions have been given in \eqref{ym-shift-covD} and \eqref{ym-shift-Fmunu}.

We now expand the Lagrangian density about the classical solution. 
The expansion of the  kinetic term $\sim \tr F_{\mu\nu} F_{\mu\nu}$ can be straightforwardly read from the \ym{} case studied in Section~1.5.1; up to a total derivative, \eqref{YM-operator-implicit-indices} reads 
\begin{equation}
\label{LW-ym-expansion}
\left( F^a_{\mu\nu} \right)^2
\rightarrow
2  A_\mu \cdot   \left[
  	- ( \covD^2 ) \delta_{ \alpha \beta }  
	- 2 F_{ \mu\nu }
  	\right]
 A_\nu 
-
2  (\covD_\mu A_\mu ) \cdot (\covD_\nu A_\nu) 
\end{equation}
where all the fields are written in the adjoint representation.

The expansion of the other two terms is quite lengthy and technical; the results are
\begin{align}
\label{LW-DF2-expansion}
\left(\covD_\mu F_{\mu \beta}^a \right)^2  
& \rightarrow 
 	A_\alpha \cdot \delta_{\alpha \beta} \covD^4 A_\beta
 	+ (\covD_\mu A_\mu) \cdot \covD^2 (\covD_\nu A_\nu )  \\\nonumber
& \hspace{6em}
 	+ 4\ A_\alpha \cdot F_{ \alpha \beta} \covD^2 A_\beta
 	+ 4\ A_\alpha \cdot F_{ \alpha \mu} F_{\mu \beta } A_\beta
\\\nonumber
\\
\label{LW-FFF-expansion}
f^{abc} F_{\mu\nu}^a F_{\mu\lambda}^b F_{\nu\lambda}^c
& 
\rightarrow
A_{\alpha} \cdot
	\left[
		3 F_{ \mu \nu } \delta_{ \alpha \beta }  
		+  3 F_{ \alpha \beta} \delta_{ \mu \nu } 
		- 6 F_{ \mu \beta }\delta_{ \alpha \nu } 
	\right]
	 \covD_{ \mu } \covD_{ \nu } A_{\beta}.
\end{align}
These formulae are derived in Appendix~C; in the rest of this Section we manipulate these expressions in order to get the final operator and perform the computation of the divergence.

Putting together the contributions \eqref{LW-DF2-expansion}, \eqref{LW-FFF-expansion} and \eqref{LW-ym-expansion}, the part of the Lagrangian that is quadratic in the fluctuation reads
\begin{equation}\label{LW-expansion-intermeriate}
\begin{split}
{\Lagr_{\YM,A^2}^\HD}
	= 		\frac{1}{2 g^2 m^2 } A^a_\alpha   &
			\bigg[
				\delta_{\alpha \beta} \covD^4
  				+ W^{\mu\nu}_{\alpha\beta} \covD_{ \mu } \covD_{ \nu } 
  				+ U_{\alpha\beta}
 			\bigg]^{ab}
			 A^b_\beta \\
 	& - \frac{1}{2 g^2 m^2}( \covD_\mu A_\mu )^a
 			\left[
 				- \covD^2 +  m^2	 
 			\right]^{ab}
 			( \covD_\nu A_\nu )^b,
\end{split}
\end{equation}
where
\begin{align}
\label{V}
W^{\mu\nu}_{\alpha\beta} & =  \left[
  					\left(
  						(4 + 3 \gamma) F_{ \alpha \beta} 
	  					- m^2 \delta_{\alpha \beta} 
	  				\right) \delta_{\mu\nu}
  					- 6 \gamma F_{(\mu|  \beta }\delta_{|\nu ) \alpha }
  				\right],
  				\\
\label{U}
U_{\alpha \beta} & = \frac{3}{2} \gamma
 					F_{ \mu \nu } F_{ \mu \nu  }\delta_{ \alpha \beta }
	  			- 2 m^2 F_{\alpha \beta }
				+ \left(
				 		4 + 3 \gamma
				 	\right) F_{ \alpha \mu} F_{\mu \beta }.
\end{align}

With this result, we are now ready to proceed to the quantization of the theory and to the determination of the divergences in the one-loop effective action.
The gauge fixing functional that we will use to perform the computation is again 
\begin{equation}
G\left[A+\mathcal{A}\right](x) = \covD_\mu A_\mu.
\end{equation}
The form of the expanded Lagrangian density \eqref{LW-expansion-intermeriate} suggests to use the integration weight functional
\begin{equation}
H = -\frac{1}{g^2 m^2}\covD^2 + \frac{1}{g^2} ,
\end{equation}
that cancels exactly the terms written in the second line of \eqref{LW-expansion-intermeriate}. The `$1$' is the same contribution for the \ym{} case with the background field approach; the Laplacian cancels the analogous term originating from $(\covD F)^2$. As we anticipated before, the choice that we made for the functional $G$ and the operator $H$ is motivated by the fact that they make the Lagrangian density simpler. Indeed, it exactly cancels against the term in the second line of the expanded Lagrangian \eqref{LW-expansion-intermeriate}. 
The contribution of the Jacobian determinant of the gauge fixing functional has already been evaluated in \eqref{M0}, hence here we recall that the relevant operator is
\begin{equation}
M_0
	=
-\covD_\mu \covD_\mu .
\end{equation}

Considering he Lagrangian density (2.2.5),
the quantity in brackets in the first line of  closely resembles an operator of the form \eqref{hk-fourth-order-generic}. In order to apply the procedure described in Section~1.2 and 1.3 to compute the divergences and the $\beta$ function, we have to make sure that the operators involved are self-adjoint, up to total derivatives: we have to define the coefficients with the symmetries described in \eqref{hk-fourth-order-generic-coefficeints}.

Let us consider the coefficients of dimension two, \ie
\begin{equation}\label{LW-operator-D2-prov-coeffs}
W^{\mu \nu}_{\alpha\beta} =
			\left[
  				(4 + 3 \gamma) F_{ \alpha \beta} 
	  			- m^2 \delta_{\alpha \beta} 
	  			\right] \delta^{\mu\nu}
  				- 6 \gamma F\indices{^{(\mu}_\beta }\delta^{\nu )}_{ \alpha };
\end{equation}
some indices have been raised for notational purposes but without any meaning since we are in Euclidean spacetime. We have to implement symmetry with respect to the transposition  of the \( A_\mu \)'s;  this corresponds to swapping \emph{simultaneously} both the gauge indices and the spacetime indices -- with the notation of  \eqref{LW-expansion-intermeriate}, the couples \( (a,\alpha) \) and \( (b,\beta ) \). This operation has different consequences on the three terms of $W^{\mu \nu} $. Indeed, the first and the third terms are antisymmetric under the exchange of the gauge indices, whereas the second one is symmetric; hence, the spacetime indices must be respectively antisymmetrized and symmetrized.

Therefore the coefficients of dimensions two for the self-adjoint operator read
\begin{equation}\label{LW-V-final}
\left(V^{\mu \nu}\right)_{\alpha \beta}
 	= 
 	\left[
  	(4 + 3 \gamma) F_{ \alpha \beta} 
	- m^2 \delta_{\alpha \beta} 
	\right] \delta^{\mu\nu}
  	- 6 \gamma F\indices{^{(\mu}_{[ \beta } } \delta^{\nu )}_{{ \alpha] } }.
\end{equation}
The non derivative term, on the other hand, is already symmetric, and hence
\eqref{U}
is the correct coefficient.

 The relevant operator for the gauge field therefore reads
\begin{equation}
\Delta_A = \covD^4 + V^{\mu\nu} \covD_\mu \covD_\nu + U
\end{equation}
with $V^{\mu\nu}$ and $U$ given in \eqref{LW-V-final} and \eqref{U}, and we can proceed to the evaluation of its Seeley-deWitt coefficient.

\subsection{Divergences}

We are now ready to perform the computation of the divergences of one-loop quantum effective action.

Comparing with the general expressions for $\Gamma_{(1)}$, we can see that our total coefficient \eqref{b4-total} reads the same of Yang-Mills Lagrangian
\begin{equation}\label{b4-tot-hdym}
b_4^\tot
	=
b_4(\Delta_A)
- b_4(H)
- 2 b_4(M_0).
\end{equation}

Let us now evaluate the first contribution, $b_4(\Delta_A)$, according to \eqref{b4-coeff-4order}.
 We start by computing the trace on `external' indices of \( V^{\mu \nu} \)
\begin{equation}
\begin{split}
V_{\alpha \beta} \equiv \left( V^{\mu\mu} \right)_{\alpha \beta } 
	& =
		4 (4 + 3 \gamma) F_{ \alpha \beta} 
		- 4  m^2 \delta_{\alpha \beta} 
		- 6 \gamma F\indices{^{\mu}_{[ \beta } } \delta^{\mu }_{{ \alpha] } }\\*
	& = 
		(16 + 6 \gamma) F_{ \alpha \beta} 
		- 4  m^2 \delta_{\alpha \beta} .
\end{split}
\end{equation}
Then the the relevant contributions read
\begin{align}
\begin{split}
\Tr V^2 
	& 	= \tr
			\left[ 
				(16+6\gamma)^2 F_{ \alpha \beta } F_{ \beta \alpha }
				+ 4 \cdot 16 m^2 \ 1_{\text{ad}}^G
			\right] \\
	&	= 4(8 + 3 \gamma)^2 C_2 F^a_{\mu\nu}F^a_{\mu\nu} +  64 m^4 \ d_\ad 
\end{split}
	\\ \nonumber \\
\begin{split}
\Tr V^{\mu \nu} V^{\mu \nu}		
	&	= \tr 
			\big[ 
				4 (4+3\gamma)^2 F_{ \alpha \beta } F_{ \beta \alpha }
				+ 36 \gamma^2 F_{\alpha \beta  } F_{ \beta  \alpha}
			\\
			& \qquad\qquad
				- 12\gamma (4 + 3 \gamma ) F_{\alpha \beta} F_{\beta \alpha}
				+ 16 m^4 \  1_{\text{ad}}^{\GroupName{G}} 
			\big]\\
	&	= (36 \gamma^2 + 64 \gamma + 48)C_2 F^a_{\mu\nu}F^a_{\mu\nu}  +  16 m^4 \ d_{\ad}
\end{split}
	\\ \nonumber \\
\begin{split}
\Tr U	
	&	= \tr
			\left[
				6 \gamma F_{ \mu \nu } F_{ \mu \nu }
				+ (4 + 3 \gamma ) F_{\alpha \mu} F_{ \mu \alpha}
			\right] \\
	& = ( 4 - 3 \gamma)C_2 F^a_{\mu\nu}F^a_{\mu\nu}  
\end{split}
\end{align}
where we used
\begin{equation}
F\indices{^{\mu}_\beta }\delta^{\nu }_{ \alpha }
F\indices{^{(\mu}_{ [ \alpha } }\delta^{\nu )}_{\beta ]  }
	= 
F\indices{^{\mu}_\beta }
F\indices{^{(\mu}_{ [ \alpha } }\delta^{ \alpha )}_{\beta ]  }
	=
- F_{ \mu \beta } F_{\mu \beta}
\end{equation}
and \eqref{notation-trFmunuFmunu}. \(d_\ad = \tr 1_{\text{ad}}^{\GroupName{G}} = \dim \text{Lie}\ \GroupName{G}\) is the dimension of the adjoint representation, \ie the number of colors.

We can now sum up the contributions with the correct weight in order to get the coefficient the fourth order operator, obtaining
\begin{align}\label{b4-operator-hdym}
b_4(\Delta_{A})
 =\phantom{.} &		
 	 \left( \frac{9}{4} \gamma^2 + 9 \gamma + \frac{10}{3} \right) C_2 F^a_{\mu\nu}F^a_{\mu\nu} 
 	+ 2 m^4 d_\ad
\end{align}
The gauge fixing contribution $M$ has been evaluated in \eqref{b4-M0-operator}
\begin{equation}
b_4 ({ M } ) = - \frac{1}{12} C_2 F^a_{\mu\nu}F^a_{\mu\nu}    ;
\end{equation} 
the integration weight $H$ contributes, using \eqref{b4-second-order}, with
\begin{align}\label{b4-H-HD-operator}
b_4 (H) & = - \frac{1}{12} C_2 F^a_{\mu\nu}F^a_{\mu\nu}    + \frac{m^4}{2} d_\ad.
\end{align}

The total coefficient \eqref{b4-tot-hdym} reads
\begin{equation}\label{b4-tot-hd-gauge}
b_4^\tot  =  \left( \frac{9}{4} \gamma^2 + 9 \gamma + \frac{43}{12} \right) C_2 F^a_{\mu\nu}F^a_{\mu\nu} + \frac{3}{2} m^4 d_\ad.
\end{equation}
A remarkable feature of this result is that the the one-loop effective action contains only a contribution proportional to usual the Yang-Mills  kinetic; this implies that only a renormalization of $g$ is necessary. The second contribution is just a constant independent of the fields contributing to the vacuum energy.

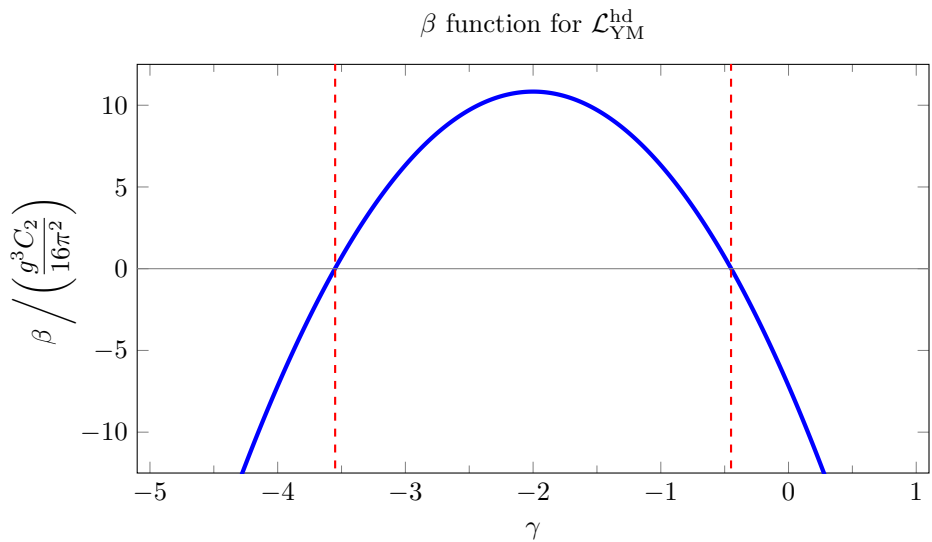
\begin{figure*}[!ht]
\centering
\begin{tikzpicture}
\begin{axis}[ 
	title = $\beta$ function for $\Lagr^\HD_\YM$, 
	xlabel = $\gamma$,
	ylabel = $\beta \left/ \left( \dfrac{g^3 C_2 }{ 16\pi^2 }\right) \right. $ ,
	xmax=1.1,
	ymax=12.5,
	xmin=-5.1,
	ymin=-12.5, 
	samples=500,
	minor x tick num = 1,
	width=12cm,
	height=7cm
	]
  \addplot[blue, ultra thick] (x,-4.5*x*x - 18*x - 43/6);
  \addplot[dashed, red,  thick, domain=-20:20] (-3.55,x);
  \addplot[dashed, red,  thick, domain=-20:20] (-0.45,x);
  \addplot[gray, domain= -10:10] {0};
\end{axis}
\end{tikzpicture}
 \captionof{figure}{$\beta$ as a function of the parameter $\gamma$. External to vertical lines: $\beta<0$, \ie asymptotic freedom; internal: $\beta >0$, \ie UV Landau pole. }
 \label{FIG_plot-beta}
\end{figure*}

We can then identify, according to \eqref{ren-1loop-contribution} and \eqref{running-coupling}, the required renormalization
\begin{equation}
g^{- 2 }_\mu = g^{- 2 }_\Lambda  - \frac{ \bar \beta }{ 16 \pi^2 } \log \frac{\Lambda}{\mu},
\end{equation}
where
\begin{equation}
\bar \beta  = \left( 9 \gamma + 36 \gamma + \frac{43}{3}\right)C_2
\end{equation}
that gives, applying \eqref{beta-function-generic}, the beta function for the \ym{} coupling $g$
\begin{equation}\label{beta-hd-ym}
\beta(g)  = - \frac{g^3}{32\pi^2}\bar \beta =  - \frac{g^3 C_2}{16 \pi^2} \left( \frac{9}{2} \gamma^2 + 18 \gamma + \frac{43}{6}\right);
\end{equation}
Figure~\ref{FIG_plot-beta} shows a plot of $\beta$ as a function of the parameter $\gamma$.

A few comments on the one loop result are in order. As already mentioned, the divergent contribution turns out to be proportional to the term $\tr F^2$ only, so that $g$ gets renormalized but the product $gm$ and $\gamma$ do not. This implies that $m$ runs in the opposite way than $g$.

Then, the $\beta$ function is independent of $m$, but depends on the parameter $\gamma$ that can be fine-tuned to make the former zero; in this case the flow of the renormalization group is trivial and conformal symmetry is not broken at quantum level. The values of $\gamma$ that make the $\beta$ function vanish are 
\begin{equation}
\gamma_{\pm} = -2 \pm \sqrt{\frac{65}{27}} \approx -3.55  \text{,}\ -0.45;
\end{equation}
we can therefore distinguish the two regimes
\( 
	\gamma_- < \gamma < \gamma_+
\)
where $\beta < 0$, and 
\(
\gamma < \gamma_- \vee \gamma > \gamma_+
\)
that gives
\(
\beta > 0
\). We also observe that $\beta$ is maximized when \( \gamma = -2 \) and its value is
\(
\beta(g)  =  {65 g^3 C_2} / {93 \pi^2}  
\)
but this does not seem to have any deeper implication.
These considerations are actually meaningful since, as mentioned, $\gamma$ does not run under renormalization.

In the first case the theory is asymptotically free, that is the same situation of the usual Yang-Mills theory; the coupling decreases towards zero with increasing scale, and according to this one loop result has an infrared Landau pole. Conversely, in  the second case, the coupling $g$ grows with energy and the theory has a UV Landau pole.

It is also interesting to notice that for $\gamma = 0 $ the $\beta$ function differs from the Yang-Mills case. Indeed, the coupling runs roughly twice as fast; this is nothing but a consequence of the additional degrees of freedom in the Lagrangian discussed in \eqref{lagr-HD-ym-splitted-dof}.

In the literature two different values for the beta function \eqref{beta-hd-ym} were reported. In \cite{Fradkin:1981iu} the computation techniques discussed here was employed, but the reported result is different. In its preprint \cite{Fradkin:preprint}  few more details are given, and some inconsistencies can be found already observing the symmetry properties of the coefficients of the derived operators.
On the other hand, \cite{Grinstein:2008qq} and \cite{Schuster}, using a conventional diagrammatic approach to the computation, obtained a result equivalent to ours, confirming the computation.

\section{Gauge theory with matter fields}

In this Section we consider the higher-derivative gauge theory with Lagrangian \eqref{lagr-hdym} coupled to matter fields. We will consider both (real) scalar and Weyl fermionic fields, studying the renormalization properties of the gauge sector only.

We are interested in the renormalization properties of the gauge sector, hence the computation we are going to do here is in fact similar to that we did for the usual \ym{} field in Section 1.5.2; for the same reasons we need just the divergence proportional to $\tr F_{ \mu \nu } F_{ \mu \nu } $  and therefore we choose a background with vanishing matter fields and non-zero gauge fields $B_\mu$. The expansion therefore will be performed as in \eqref{bkg-with-matter} that we repeat here for clarity, 
\begin{equation}
A_\mu \rightarrow A_\mu + B_\mu, 
\hspace{3.5em} 
\phi_i \rightarrow \phi_i,
\hspace{3.5em}
\psi_j \rightarrow \psi_j,
\end{equation}
with the same notation as before, that is $\phi$, $\psi$ and $A_\mu$ are the quantum fluctuation. Again, $\covD_\mu$ and $F_{\mu\nu}$ are expressed in terms of the background field only.


The structure of the Lagrangian that we are going to consider is then
\begin{equation}
\Lagr =
	\Lagr_{\YM}^\HD + \Lagr_\phi + \Lagr_\psi,
\end{equation}
where \(\Lagr_{\YM}^\HD\) is \eqref{lagr-hdym}; \( \Lagr_\phi \) and \( \Lagr_\psi \) are the matter field Lagrangian densities with at most quadratic contributions in matter fields and with minimal coupling with gauge fields.
We are ignoring interactions between matter fields because they would be terms at least of third order in the fields themselves, and for the background about which we expand the Lagrangian they do not contribute to the terms quadratic in the fluctuations.

Considering usual-derivative matter fields, power counting arguments analogous to those performed in Section 2.1.2 lead to the conclusion that such theories are renormalizable if all the coefficients of the terms containing $\phi$ or $\psi$ have mass dimension grater or equal than zero. For higher-derivative matter of the type we are considering here, the mass dimension of any coefficient can be greater or equal than $-2$ without spoiling renormalizability. Going through these results would be long and not very instructive since it is just matter of playing with identities about graph topology, therefore we just accept this result and explicitly verify the renormalizability of the gauge sector under one-loop corrections.

We quantize the theory with the same gauge-fixing functional used above, \ie $G[A + \mathcal{A}]= \covD_\mu A_\mu$, and we will consider the same integration weight \( H = 1/g^2 - \covD^2/g^2m^2 \). 
As a consequence, the heat kernel coefficients  $b_4$ for the gauge fixing terms and for the gauge field are unaffected by the matter sector, and therefore are again those computed in \eqref{b4-M0-operator}, \eqref{b4-H-HD-operator} and \eqref{b4-operator-hdym}.

Considerations similar to those outlined for the \ym field coupled matter, bring us to the conclusion that, applying \eqref{eff-act-1loop-gauge}, we have to compute the total coefficient
\begin{equation}\label{b4-total-most-general}
b_4^\tot = 
b_4(\Delta_A)
- 2 b_4(M_0)
- b_4(H)
+ \sum_i b_4(\Delta_{\phi,i})
-2 \sum_j b_4(\Delta_{\psi,j}).
\end{equation}


In the following we analyse the cases of usual-derivative matter and higher derivative matter. Mixed combinations could also be considered, but the number of free parameters makes the situation quite involved. For simplicity we will consider massless matter only.

\subsection{Usual derivative matter fields}

First we briefly cover the somewhat trivial theory describing higher-derivative gauge fields coupled with usual-derivative matter.

Given the restrictions that we imposed on the terms that can be present in the Lagrangian density, the contribution for scalar fields can only read
\begin{equation}
\Lagr_\phi = 
	\left( \covD_\mu \phi_i \right) \left( \covD_\mu \phi_i \right) 
\end{equation}
and that of the spinor field is
\begin{equation}
\Lagr_\psi = 
	i \psi_j \sigma^\mu \covD_\mu \bar \psi_j.
\end{equation}
$i$ and $j$ are generic indices enumerating the matter fields $\phi_i$ and $\psi_j$; no  structure related to the gauge symmetry is associated with these indices, that are summed when repeated. The fields might transform under any representation of the gauge group, that means that $\phi_i$ or $\psi_j$ could carry internal indices, whose contraction is understood.

The Langrangian densities are the same of the matter sectors studied for the usual \ym{} theories in Section~1.5, so that we can directly take the results \eqref{b4-scalar-usual-der} and \eqref{b4-scalar-higher-der}, \ie  
\begin{equation}
b_4( \Delta_{\phi,i} ) = - \frac{1 }{12}  C_{\phi,i} F^a_{\mu\nu} F^a_{\mu\nu} 
\end{equation}
for scalar fields, and
\begin{equation}
b_4( \Delta_{\psi,j} ) = \frac{1}{6} C_{\psi,j} F_{\mu\nu}^a F_{\mu\nu}^a  
\end{equation}
for spinor fields.

The total coefficient \eqref{b4-total-most-general} now reads
\begin{equation}
\begin{split}
b_4^\tot
	& =
F_{\mu\nu}^a F_{\mu\nu}^a
\bigg[
\frac{9}{4} \gamma^2 + 9 \gamma + \frac{43}{12}  
- \frac{1}{12} \sum_i C_{\phi,i}
- \frac{1}{3}  \sum_j C_{\psi,j}
\bigg]
+ \frac{3}{2} m^4 d_\ad
\end{split}
\end{equation}
that again implies a renormalization of the \ym{} coupling $g$ only, and includes the vacuum energy contribution. Such divergence gives the $\beta$ function
\begin{equation}
\beta(g)
	=
- \frac{g^3}{8 \pi^2}
\bigg[
\frac{9}{4} \gamma^2 + 9 \gamma + \frac{43}{12}  
- \frac{1}{12} \sum_i C_{\phi,i}
- \frac{1}{3}  \sum_j C_{\psi,j}
\bigg]
\end{equation}
This result is of course just the same as in the case \ym{} with matter with the difference that the gauge field contribution is given by \eqref{b4-tot-hd-gauge}.
Depending on the value of $\gamma$ and the representation in which the matter fields are, the computed $\beta$ function can cause different renormalization group flow behaviours.


\subsection{Higher derivative matter fields}

Now we extend the Lagrangian densities for the matter fields  in order to include higher derivative contributions.

The most general contribution that we can have in the scalar sector is
\begin{equation}\label{lagr-hd-scalars}
\Lagr^\HD_\phi = 
- \frac{1}{2 g^2} \phi_i \covD^2 \phi_i  
+ \frac{\delta_{1,i}}{2 g^2	m^2}  \phi_i \left[  ( \covD^2 )^2 
						+ \delta_{2,i} F_{\mu\nu} F_{\mu\nu}
						\right] \phi_i
			,
\end{equation} where $\delta_{k,i}$ are generic real coefficients.
This is a quite general expression at least for our purposes: There cannot be any term \( \sim \covD_\mu \covD_\nu \), since no rank-2 symmetric Lorentz-scalar tensor of mass dimension 2 is available to contract those indices;\footnote{Of course a choice could be $m^2 \delta^{\mu\nu}$, but it is actually the ordinary-derivative contribution already taken into account in the Lagrangian density \eqref{lagr-hd-scalars}.} any mixing like $ \phi_i \Sigma_{ij} F_{\mu\nu} F_{\mu\nu} \phi_j $ can be eliminated by diagonalising $\Sigma_{ij}$ since the kinetic term is proportional to $\delta_{ij}$; a contribution of the form $\sim \covD_\mu F_{\mu\nu} \covD_\nu $ is a $N_\mu$ term (in the notation of \eqref{hk-fourth-order-generic}) so it does not contribute to the divergence. 

The operator for (Weyl) spinor fields that we will consider is
\begin{equation}\label{lagr-hd-spinors}
\begin{split}
\Lagr^\HD_\psi =  
& \frac{1}{g^2}  \psi_j i \sigma^\mu \covD_\mu \bar \psi_j \\
& + \frac{\tau_{1,j}}{ g^2 m^2 }  \psi_j \bigg[ 
		i \sigma^\mu  \covD_\mu \bar\sigma^\nu\covD_\nu \sigma^\rho \covD_\rho
		-i  \tau_{2,j} F_{ \mu \tau } \sigma^{\mu \tau}  \sigma^\rho \covD_\rho 
		-i \tau_{3,j} F_{ \mu \nu } \sigma_\nu \covD_\mu 
		\bigg]  \bar \psi_j
\end{split}
\end{equation}
where $\tau_{k,j}$ are generic real coefficients.
This is less general than the Lagrangian for the scalar field: the spinor index structure allow also for other combinations of sigma matrices, and we are ignoring possible mixing between the spinors;  \eqref{lagr-hd-spinors} is a general expression and we dropped total derivatives or ineffective contributions as described for the scalar field.

\subsection{General aspects}

Similarly to the pure-gauge case, the theory defined by 
\( \Lagr =	\Lagr_{\YM}^\HD + \Lagr^\HD_\phi + \Lagr^\HD_\psi \)
enjoys some interesting renormalization properties. In particular we will see that a class of this kind of theories is not affected by the hierarchy problem.


\subsubsection{Degrees of freedom}

As done for the pure gauge Lagrangian, it is interesting to count the degrees of freedom conveyed by these kind of theories and see how they can be thought of in a different way. As outlined in \cite{CONFORMAL, Grinstein:2008qq}  We will consider scalars first, and then spinors; we will focus only on the kinetic term, since we are just considering the structure of the degrees of freedom.
We will consider only one field at time.

\paragraph{Scalar fields.} 
The higher derivative Lagrangian density reads
\begin{equation}
\Lagr_\phi = 
	\frac{1}{2} \partial_\mu \phi \partial_\mu \phi	
	+ \frac{1}{2 m^2} (\partial^2 \phi)^2.
\end{equation}
Off-shell each real scalar field describes, of course, just one degree of freedom; on shell, since the equation of motions are fourth order differential equations, one has to specify four initial data for each real field, corresponding to two degrees of freedom.

These additional degrees of freedom can be made manifest by introducing a real auxiliary field $\phi'$:
\begin{equation}\label{lagr-HD-scalar-splitted-dof-intermediate}
\Lagr_\phi = 
	\frac{1}{2} \partial_\mu \phi \partial_\mu \phi	
	- \frac{1}{2 } m^2 \phi'^2
	- \phi' \square \phi,
\end{equation}
upon equation of motion \( \phi' =  \partial^2  \phi / m^2 \), \eqref{lagr-HD-scalar-splitted-dof-intermediate} gives back the original kinetic term. Shifting the field  \( \phi \rightarrow \phi - \phi' \) in order to make the derivative term diagonal, we obtain
\begin{equation}\label{lagr-HD-scalar-splitted-dof}
\Lagr_\phi = 
	\frac{1}{2} \partial_\mu \phi \partial_\mu \phi	
	- \frac{1}{2} \partial_\mu \phi' \partial_\mu \phi'
	- \frac{1}{2} m^2 \phi'^2.
\end{equation}
In this Lagrangian density, the degrees of freedom are decomposed in a massless scalar field and a massive scalar field (with mass $m$), the latter having an unusual sign in front of its kinetic and mass terms. The number of degrees of freedom is conserved, since in the theory defined by \eqref{lagr-HD-scalar-splitted-dof} each scalar field propagates one degree of freedom. Similarly to the gauge field case, the massive particle is a ghost corresponding to the exponentially growing modes of the higher-derivative  formulation of the theory.

\paragraph{Spinor fields.} Off-shell there are 4 degrees of freedom; the higher derivative equations of motion are third order partial differential equations, so that the number of propagating degrees of freedom is $ 4 \cdot 3 / 2 = 6  $. In order to make such degrees of freedom manifest, we rewrite the Lagrangian as
\begin{equation}
\begin{split}
\Lagr_{\psi} =
&\phantom{.}	\bar \psi i \bar \sigma^\mu \partial_\mu \psi
	- m (\bar{\psi'} \bar{\psi''} + \psi' \psi'')
	+ \bar{\psi'} i \bar \sigma^\mu  \partial_\mu \psi 	+ \bar{\psi} i \bar \sigma^\mu \partial_\mu \psi' 
	- \psi'' i \sigma^\mu \partial_\mu \bar{\psi''}
\end{split}
\end{equation}
where we introduced {two} Weyl spinors \(\psi'\) and \(\bar \psi'' \), whose classical equations of motion read
\begin{align}
i \bar \sigma^\mu \partial_\mu \psi' - m \bar \psi''  = 0
		, \hspace{2em}
i \sigma^\mu \partial_\mu \bar \psi'' +  m \psi'  = 0 .
\end{align} A diagonal kinetic term is obtained by shifting $\psi \rightarrow \psi - \psi' $, so that the Lagrangian now reads
\begin{equation}\label{lagr-HD-spinor-splitted-dof}
\begin{split}
\Lagr_{\psi} & 
	= \bar \psi   i \bar \sigma^\mu \partial_\mu \psi 
	- \bar \psi' i \bar \sigma^\mu \partial_\mu \psi' 
	- \psi'' i \sigma^\mu \partial_\mu \bar{\psi''}
	- m (\bar{\psi'} \bar{\psi''} + \psi' \psi'')
\\
&
	=  \bar \psi   i \bar \sigma^\mu \partial_\mu \psi 
	- \bar \Psi i \slashed{\partial} \Psi
	- m \bar \Psi \Psi
\end{split}
\end{equation}
where we emphasized that the two new spinors group together into a (massive) Dirac spinor $\Psi := (\psi' , \bar \psi'')$ as it is required for charged spinors in order to have a mass term.  Since a Dirac spinor propagates $4$ degrees of freedom, and the Weyl spinor 2, the total number of dynamical degrees of freedom is conserved.
Again, the auxiliary massive field has an extra minus sign in the Lagrangian that classify it as a ghost particle.

\subsubsection{On the hierarchy problem}

A particular class of such theories, moreover, solves the hierarchy problem, because the mass of the scalar grows only logarithmically with the cut-off, while the scalar mass in usual theories gets also quadratic corrections. The theory that we will consider here are those in which the higher derivative contribution is only on the kinetic term, so that the scalar field has a Lagrangian of the form 
\begin{equation}\label{hd1l-lagr-scalar-hierarchy}
\tilde \Lagr_{\phi} \sim (\covD_\mu \phi) (\covD_\mu \phi) - \frac{1}{m^2}  (\covD^2 \phi) (\covD^2 \phi) + V(\phi) 
\end{equation}
where the potential satisfies the usual renormalization prescriptions for conventional-derivative theories, \ie it is a polynomial of degree at most four. We ignore spinor contributions for simplicity, even though it is easy to see that Yukawa couplings would not change the result.

This remarkable feature is again a consequence of the improved UV behaviour of the propagator, scaling with $p^{-4}$ instead of $p^{-2}$. The proof of this fact relies on an explicit power counting argument on Feynman diagrams, that we will not go through.
From \cite{Grinstein:2008qq} we quote that the superficial degree of divergence is, denoting $E_X$ the number of external lines for the field $X$,
\begin{equation}
d = 6 - 2 L - E_A - 2 E_c - E_\phi,
\end{equation}
so that the only possible quadratic divergence for the two-point function of the scalar field is at one loop.
We will now explicitly show the vanishing of the coefficient of $\Lambda^2$ in the one-loop effective action in the context of the formalism that we are using in this thesis.

We need to compute the divergences for the coefficient of $\phi^2$ and the possible wavefunction renormalization; we therefore need to choose a background with non-vanishing scalar $\phi_{\text{bkg}}$, while fermion and gauge fields can be set to  zero. The operator obtained expanding the fields will be a fourth-order differential operator of the form \eqref{hk-fourth-order-generic}; the power divergence is weighted with the heat kernel coefficient $a_1$ (equivalently $ b_2$) that can be computed using the techniques described in Section 1.2 and turns out to be proportional to $V_{\mu\mu}$. But let us check what could contribute to $V_{\mu\nu}$: We would need a rank-two symmetric tensor with mass dimension 2 containing $\phi_{\text{bkg}}^2$, but the Lagrangian density \eqref{hd1l-lagr-scalar-hierarchy} does not allow for it, since it requires a contribution with the structure $\phi^2 \covD \phi \covD \phi$ to appear. For the same reason, since no divergence proportional to $(\covD \phi)^2$ can be present, the wavefunction does not get renormalized, at least at one-loop level, and this shows that there is no mass counterterm with power-law dependence on the cut-off.

\subsection{One-loop renormalization}
As  mentioned above, being the Lagrangian densities \eqref{lagr-hd-scalars} and \eqref{lagr-hd-spinors} already quadratic in the matter fields, the expansion about the background solution is immediate.

\paragraph{Scalar fields.} The operator relevant to the scalar field reads
\begin{align}
\Delta_{\phi,i}
	& = ( \covD^2 )^2 
						+ \frac{m^2}{\delta_{1,i}}\delta^{\mu\nu} \covD_\mu\covD_\nu
						+ \delta_{3,i} F_{\mu\nu} F_{\mu\nu}
\end{align}
 The $b_4$ coefficient is ready to be evaluated using \eqref{b4-coeff-4order}, and the result is
\begin{equation}\label{b4-hd-scalar}
b_4(\Delta_{\phi,i})
	=
	- \left( \frac{1}{6} + \delta_{3,i} \right)C_{\phi,i} F^a_{\mu\nu} F^a_{\mu\nu}
	+ \frac{ m^4 }{ 2 ( \delta_{1,i} )^2 } d_{\phi_i}.
\end{equation}
where
\(
d_{\phi_i}
	=
\tr \1_{\phi,i}
\)
is the dimension of the representation of the scalar $\phi_i$.

\paragraph{Spinor fields.} The operator for the spinor field reads
\begin{align}
\Delta_{\psi,j}
	& =
	i \sigma^\mu  \bar\sigma^\nu \sigma^\rho \covD_\mu\covD_\nu  \covD_\rho
	- i \tau_{2,j} F_{ \mu \tau } \sigma^{\mu \tau}  \sigma^\rho \covD_\rho 
	- i \tau_{3,j} F_{ \mu \nu } \sigma_\nu \covD_\mu
	+ \frac{m^2}{\tau_{1,j}} i \sigma^\mu \covD_\mu  
\end{align}
In order to compute its contribution to the divergence, we apply the procedure described in \eqref{b4-coeff-3order-implicit}. The Seeley-deWitt coefficient therefore reads
\begin{equation}\label{b4-hd-spinor}
\begin{split}
b_4(\Delta_{\psi,j})
& =
\left[
-\frac{1}{2}	
+ \tau_{2,j}
- \frac{1}{2} \tau_{2,j} \tau_{3,j}
- \frac{1}{2}	(\tau_{2,j})^2
- \frac{1}{4} (\tau_{3,j})^2
\right]
\tr F_{\mu\nu} F_{\mu\nu}
+ \frac{m^4}{2(\tau_{1,j})^2} d_{\psi_j}
\end{split}
\end{equation}
where \(d_{\psi,j} = \tr 1_{\psi,j}\) is the dimension of the representation of the field $\psi_j.$
The first term is confirmed from the immediate calculation when the $\tau$'s are all zero, since the operator factorizes $ \Delta_\psi = \Delta_1 \bar \Delta_{1} \Delta_1 $, and therefore $b_4(\Delta_\psi) = 3 b_4(\Delta_1) = -\frac{1}{2} \tr F_{\mu\nu} F_{\mu\nu} $.

\subsubsection{Computation of $ b_4(\Delta_\psi)$}

In this section we go through the steps that bring to \eqref{b4-hd-spinor}. For simplicity of notation, we understand the index $j$.

Considering  the leading symbol in the operator the right choice of first-order differential operator to compose it with is \( \Delta_1 = - i \bar \sigma^\mu \covD_\mu \), and we get
\begin{equation}
\begin{split}
\Delta_{\psi+1} 
	& =
\Delta_\psi \cdot \Delta_1
	=
(\covD^2)^2
+ 
V^{\mu\nu} \covD_{\mu} \covD_{\nu}
+ U
\end{split}
\end{equation}
with coefficients
\begin{align}
V^{\rho \kappa} & =
	\left(-1 + \frac{\tau_2}{2}\right) 
	F_{\mu\nu} \sigma^{\mu} \bar \sigma^\nu \delta^{\rho\kappa}
	- \tau_3 F\indices{^{(\rho|}_\nu }\sigma^\nu \bar\sigma^{|\kappa)}
	+ \frac{m^2}{\tau_{1}} \delta^{ \rho \kappa }
	\\
U & =
\frac{1}{4} (1 - \tau_2) F_{\mu\nu} F_{\rho \tau} \sigma^\mu \bar \sigma^\nu \sigma^\rho \bar \sigma^\tau
-  \frac{\tau_3}{2} F_{\mu\nu} F_{\mu\rho} \sigma^\nu \bar \sigma^\kappa
+ \frac{m^2}{2 \tau_{1}} F_{\mu\nu} \sigma^\mu \bar \sigma^\nu
\end{align}
where symmetric and antisymmetric parts of the product of two covariant derivatives have been separated in order to ensure the correct symmetry properties.

As a starting point we perform the trace of $V$ over external indices
\begin{equation}
V = V^{\mu\mu}
	=
\left(-4 + 2{\tau_2} + \tau_3 \right) 
	F_{\mu\nu} \sigma^{\mu} \bar \sigma^\nu 
	+ 4 \frac{m^2}{\tau_{1}} 
\end{equation}
and now we have all the necessary ingredients to compute the relevant traces.
By using \eqref{identity-spinor_trace-double-sigmamn} we get
\begin{align}
\begin{split}
\Tr V^2
	& = \left(-4 + 2{\tau_2} + \tau_3 \right)^2
	\tr F_{\mu\nu} F_{\rho\kappa} \
	\str \sigma^{\rho}\sigma^{\mu} \bar \sigma^\nu \bar \sigma^\kappa
\\
	& = -4 \left( 16 + 4({\tau_2})^2 + (\tau_3)^2 - 16 \tau_2 + 4 \tau_2\tau_3 -8 \tau_3 \right) \tr F_{\mu\nu} F_{\mu\nu}
+ 16 \frac{m^2}{\tau_{1}} d_{\psi}
\end{split}	
\end{align}
and using \eqref{identity-4}
\begin{align}
\Tr U 
	& = \frac{1}{4} (1 - \tau_2) \tr F_{\mu\nu} F_{\rho \tau} \str \sigma^\mu \bar \sigma^\nu \sigma^\rho \bar \sigma^\tau   +
 \frac{\tau_3}{2} 2 \tr F_{\mu\nu} F_{\mu\nu}
\\\nonumber
& = - (1 - \tau_2 - {\tau_3}) \tr F_{\mu\nu} F_{\mu \nu}.
\end{align}
\(\Tr V^{\mu\nu} V^{\mu\nu}\) is a trickier for its spinor index structure.
Expanding the contraction one gets
\begin{equation}
\begin{split}
\Tr V^{\mu\nu}V^{\mu\nu}
&	=
4 \left(-1 + \frac{\tau_2}{2}\right)^2
\tr F_{\mu\nu} F_{\rho\kappa} \str \sigma^\mu \bar \sigma^\nu \sigma^\rho \bar \sigma^\kappa
\\
& \qquad
+ (\tau_3)^2 \tr F\indices{_{\rho\nu }} F\indices{^{(\rho|}_\mu } 
\str \sigma^\nu \bar\sigma^{\lambda} \sigma^\mu \bar\sigma^{|\lambda)}
\\
& \qquad
+ \tau_3  \left(-2 + {\tau_2} \right) 
	\tr F_{\mu\nu} F_{\rho \nu}
	\str \sigma^\mu \bar \sigma^\nu \sigma^\rho \bar \sigma ^\nu
\\
& \qquad
+ \frac{m^2}{\tau_{1}} d_{\psi}
\end{split}
\end{equation}
The first term is of the kind discussed above.
The third term, using again \eqref{identity-6}, reads
\begin{equation}
\tau_3  \left(-2 + {\tau_2} \right) 
	\tr F_{\mu\nu} F_{\rho \nu}
	\str \sigma^\mu \bar \sigma^\rho 
= 
- 4 \tau_3  \left(-2 + {\tau_2} \right)  \tr  F_{\mu\nu} F_{\mu \nu}.
\end{equation}
Splitting the symmetrization in the second one we get
\begin{equation}
\frac{(\tau_3)^2}{2} \tr F\indices{_{\rho\nu }} F\indices{_{ \rho }_\mu } 
\str \sigma^\nu \bar\sigma^{\lambda} \sigma^\mu \bar\sigma^{ \lambda }
+
\frac{(\tau_3)^2}{2}  \tr F\indices{_{\rho\nu }} F\indices{_{\lambda}_\mu } 
 \tr \bar\sigma^{\rho} \sigma^\nu \bar\sigma^{\lambda} \sigma^\mu ,
\end{equation}
where we also used the cyclicity of trace. They can be evaluated with the same techniques already seen; indeed we get
\begin{equation}
- 2 {(\tau_3)^2} \tr F\indices{_{\mu\nu }} F\indices{_{ \mu }_\nu } 
- 2  {(\tau_3)^2} \tr F\indices{_{\mu\nu }} F\indices{_{\mu}_\nu }  
= - 4 (\tau_3)^2 \tr F\indices{_{\mu\nu }} F\indices{_{\mu}_\nu }  .
\end{equation}
The result is
\begin{equation}
\begin{split}
\Tr V^{\mu\nu}V^{\mu\nu}
&	=
-4 \left[
	4 
	-4 \tau_2  
	-2 \tau_3 
	+ \tau_2 \tau_3
	+ (\tau_2)^2
	+ (\tau_3)^2  
\right] 
\tr  F_{\mu\nu} F_{\mu \nu} 
+ 4 \frac{m^4}{(\tau_{1})^2}d_{\psi} .
\end{split}
\end{equation}

Summing those contributions together according to \eqref{b4-coeff-4order} we arrive at the partial result
\begin{equation}
\begin{split}
b_4(\Delta_{\psi+1})
& =
\left[
-\frac{2}{3}	
+ \tau_2
- \frac{1}{2} \tau_2 \tau_3
- \frac{1}{2}	(\tau_2)^2
- \frac{1}{4} (\tau_3)^2
\right]
\tr F_{\mu\nu}  F_{\mu\nu}
+  \frac{m^4}{2(\tau_{1})^2}d_{\psi} 
;
\end{split}
\end{equation}
the coefficient for the spinor operator is obtained subtracting the coefficient for $\Delta_1$, and the result is \eqref{b4-hd-spinor}.

\subsubsection{$\boldsymbol \beta$ function}

In order to evaluate the divergence we just apply \eqref{b4-total-most-general} with  \eqref{b4-hd-scalar} and \eqref{b4-hd-spinor}. 
The resulting expression is long and not insightful.
Again, the divergence is proportional to the \ym{} contribution only, so that $g$ gets renormalized but $\gamma$, $gm$, $\tau$'s and $\delta$'s do not run. The $\beta$ function reads 
\begin{equation}\label{beta-hd-ym-matter}
\begin{split}
\beta  = 
 - \frac{g^3}{16 \pi^2}
& \bigg[
\left( \frac{9}{2} \gamma^2 
+ 18 \gamma + \frac{43}{6}\right) C_2
- \sum_i	 \left( \frac{1}{3} + 2 \delta_{3,i} \right) C_{\phi,i}
\\
& \quad
+ \sum_j 2 \left(
- 1	
+ 2 \tau_{2,j}
-  \tau_{2,j} \tau_{3,j}
-  (\tau_{2,j})^2
- \frac{1}{2} (\tau_{3,j})^2
\right) C_{\psi,j}
\bigg]
\end{split}
\end{equation}
The renormalization group for this theory is rich and variable depending on the values of the parameters, and a complete discussion is impossible. As in the pure-gauge case \eqref{beta-hd-ym}, there is the possibility of a fine-tuning of the constant in order to have a vanishing $\beta$ function; this is consistent because the expression \eqref{beta-hd-ym-matter} is an invariant of the flow of the renormalization group.

A diagrammatic computation for a similar system has been performed in \cite{Grinstein:2008qq}.
The gauge field and scalar contributions agree with those obtained here; the same is true for the spinor contribution with $\tau_2=0$.

	\newpage
	\pagestyle{plain}

\chapter{Supersymmetric higher-derivative theory }

\pagestyle{fancy}
	\fancyhead{} 
	\fancyhead[LE]{\scshape \leftmark}
	\fancyhead[RO]{\scshape \rightmark}
	\fancyfoot[CE,CO]{\thepage}
	\fancyfoot[LO,RO]{ }
	\fancyfoot[LE,RE]{ }
	\renewcommand{\headrulewidth}{0.4pt}
	\renewcommand{\footrulewidth}{0.4pt}

In this chapter we are going to discuss the supersymmetric generalization of the higher-derivative \ym{} Lagrangian \eqref{lagr-hdym}. 

In the first Section we will give some introductory observations about the problem, and we will motivate the need of formulating the theory in six spacetime dimension and then dimensionally reduce it. In Section~3.2 supersymmetry in six dimension is briefly discussed, and the formalism of harmonic superspace is presented, being the natural framework in which six dimensional supersymmetry is realised. After that, gauge theories are discussed and the higher derivative action for the vector multiplet is formulated.
In Section~3.3, the dimensional reduction of the theory is performed, and in the following two Sections the $d=4$, $N=1$ and $2$ supersymmetric higher-derivative Yang-Mills Lagrangians are formulated and the $\beta$ function of the gauge coupling is computed. In Section~3.6 the linearised $d=4$, $N=4$ Lagrangian is discussed and the $\beta$ function evaluated. Section~3.7 is devoted to some concluding comment.

The basic facts and notation about supersymmetric theories in four spacetime dimensions are given in Appendices~A.

\section{Introductory considerations}

The Lagrangian \eqref{lagr-hdym} that we are considering contains higher derivative of the gauge field; since supersymmetry mixes the fields in the Lagrangian, we expect the matter fields to have higher-derivative contributions as well.
What may be less obvious is that also the auxiliary fields get an higher derivative contribution, therefore becoming dynamical, as we are going to motivate in a few paragraphs.

To start with, let us consider for simplicity the case of an Abelian symmetry group, namely Maxwell theory. This case yields a somewhat trivial supersymmetric extension, but it points out many relevant features and serves as a `base case' against which we will verify more advanced techniques.
The higher-derivative Maxwell Lagrangian can be obtained from \eqref{lagr-hdym-comp}  and reads
\begin{equation}\label{lagr-hdmaxwell}
\Lagr_{\text{M}} = 
\frac{1}{4 g^2} F_{\mu\nu} F_{\mu\nu} 
- \frac{1}{4 m^2 g^2}  F_{\mu\nu}  \square  F_{\mu\nu}.
\end{equation}
where we also used \eqref{CovDF-dentity} and integrated by parts to rewrite the higher derivative term in this fashion. The term from the commutator of course vanishes being the theory Abelian.
We already know the $N=1$ supersymmetric Lagrangian containing the first contribution, namely the Lagrangian for super-Maxwell theory,
\begin{equation}
\Lagr_{\text{sM}} = 
\frac{1}{2 g^2} \left( \frac{1}{2} F_{\mu\nu} F_{\mu\nu} 
+2 i \bar \psi \bar \sigma^\mu \partial_\mu \psi - D^2  \right),
\end{equation}
whose action can be expressed in terms of a superspace integral as
\begin{equation}\label{act-supermaxwell-superspace}
S_{\text{sM}} = 
\frac{1}{4}
\int \dd{^4x} \dd{^2 \theta}\left[ W W + \text{hc}
\right],
\end{equation}
$W$ being the superfield strength.

Let us now notice that the higher derivative contribution in \eqref{lagr-hdmaxwell} consists in a simple insertion of a $ - \square = - \partial_\mu \partial_\mu$ inside the ordinary kinetic term. This suggests to insert such operator between the two factors of $W$ in the superspace action \eqref{act-supermaxwell-superspace}
\begin{equation}
S'_{\text{sM}} = 
- \frac{1}{4}
\int \dd{^4x} \dd{^2 \theta}\left[ W \square W +  \text{hc} 
\right].
\end{equation}
Since $W$ and $\square$ are all gauge-invariant operator, this action is gauge invariant.
In terms of component fields the Lagrangian density corresponding to $S_{\text{sM}} + S'_{\text{sM}} $ reads
\begin{equation}\label{lagr-N=1-supermax}
\begin{split}
\Lagr^{N=1}_{\HD\text{sM}}
=
&
\frac{1}{2 g^2} \left( \frac{1}{2} F_{\mu\nu} F_{\mu\nu} 
+2 i \bar \psi \bar \sigma^\mu \partial_\mu \psi - D^2  \right)
\\
&\quad + \frac{1}{2 m^2 g^2} \left(- \frac{1}{2} F_{\mu\nu} \square F_{\mu\nu} 
- 2 i \bar \psi \bar \sigma^\mu \square \partial_\mu \psi +  D\square D
\right),
\end{split}
\end{equation}
In \cite{Gama:2011ws}  this model was used to construct a higher-derivative extension of QED, also adding matter fields.

The Lagrangian \eqref{lagr-N=1-supermax} is quite interesting and points out a number of features that must be taken into account in writing a supersymmetric theory for the higher-derivative \ym{} field. Though, it is still trivial in its dynamics, since it is a free theory also at the higher-derivative level. 

As anticipated, the auxiliary field becomes dynamical. According to the computation of the degrees of freedom described in the previous Chapter, \eqref{lagr-N=1-supermax} describes 5 ($A_{\mu}$) + 1 ($D$) bosonic and 6 ($\psi$) fermionic degrees of freedom, and hence the propagation of $D$ is indeed necessary to make the number of bosonic and fermionic degrees of freedom equal on-shell. Its kinetic term, however, has the extra minus sign that indicates its ghost nature.\footnote{Remember that the euclidean Lagrangian density is the energy.}
Notice that the presence of unphysical degrees of freedom was already recognised in the formulation of the higher-derivative theories in terms of two-derivative fields; this kind of multiplet structure of the ghost fields will be discussed later in some detail.

The non-Abelian generalization of this result is highly nontrivial for many reasons. We used the identity \eqref{CovDF-dentity} that in the case of a non-Abelian gauge group adds also another contribution. Moreover, the partial derivative, and as a consequence $\square$, is not a gauge invariant object any more; nor is the \ym{} superfield $W_\alpha$, and it is not clear how a super-invariant can be constructed.

In the case of Maxwell theory the term proportional to $\gamma$ in \eqref{lagr-hdym} vanishes since the gauge group is Abelian. However, we can argue that in general for a supersymmetric theory there cannot arise any also in the non-Abelian gauge symmetry. This is a consequence of the superfield formulation of the theory: The vector superfield $W^\alpha$ cannot construct a scalar Lagrangian density that is cubic in itself. We will therefore restrict to the case $\gamma = 0$.

We want to find the $N=1$, $2$ and $4$ extension of \eqref{lagr-hdym}, or at least to evaluate the one-loop $\beta$ function for such theories.
Instead of trying to find a supersymmetric formulation in four spacetime dimension, we will obtain the corresponding theory in $5+1$ spacetime dimensions and then dimensionally reduce it to four dimensions. This is motivated by many reasons.

The unextended supersymmetry algebra in six spacetime dimensions is indeed very similar to the $N=2$ algebra in four spacetime dimension; this comes, of course, with no surprise, because the dimensional reduction of the former yields automatically the latter. In this context, working in six spacetime dimensions simplifies the multiplet that we have to deal with, because it consists only of the gauge field, of one spinor and of the auxiliary fields; the rest of the $N=2$, $d=4$ supermultiplet is  generated via dimensional reduction.
We will also see that studying $N=2$ higher derivative \sym{} will provide enough information to compute, at least at one-loop, the  $\beta$ function of the $\beta$ function for the extended $N=4$  case.

\section{Preliminaries}

In this Section we will give some generalities of the six dimensional spinors and supersymmetry algebra. We will then introduce a superspace that represents in a manifest way the group structure associated with the relevant fields, namely the harmonic superspace. This will allow us to formulate the gauge theory in six dimensions, then to be dimensionally reduced to four spacetime dimension.
Appendix~B gives a review of supersymmetry in four spacetime dimensions

As will turn out in the computation, the construction of a higher derivative Lagrangian in $d=6$ is much more straightforward than the formulation of the usual \ym{} action. This could be related to the fact that in six dimensions such theories might be more natural, since the coupling of the higher derivative term happens to be dimensionless (while in \eqref{lagr-hdym} it has dimension $-2$).

Six dimensional Minkovsky spacetime is the natural generalisation of the four dimensional one. It is described by coordinates $x^M$, $M=0,1,\ldots,5$; here we will be working with the mostly positive metric $\eta^{MN} = \text{diag}({}-{}+{}+{}+{}+{}+{})$, and Wick-rotate the Lagrangian at the end, in order to get a formally convergent 
functional integral to proceed as discussed in Chapter~1. As mentioned in the Introduction, though, this is not a well-defined operation and we use it as a formal tool to simplify the notation, but such subtleties are not of interest for this work.

For completeness and future reference, the covariant derivative and the field strength tensor are 
\begin{equation}\label{6dim-covd-fmunu}
\covD_M = \partial_M + A_M
\hspace{5em}
F_{MN} = [\covD_M, \covD_N ].
\end{equation}
The Laplacian is denoted with $\tilde \covD^2 = \covD^M \covD_M$, the tilde serving to distinguish it from the four dimensional one.
Gauge transformations act on the gauge fields as
\begin{equation}\label{6dim-covd-fmunu-gaugetrans}
\delta_\omega A_M = - \covD_M \omega
\hspace{5em}
\delta_\omega F_{MN} = [\omega, F_{MN}].
\end{equation}

\subsection{Six-dimensional spinors}

In this Section we recall some facts about spinors in the six-dimensional Minkovsky spacetime $\mathbb{R}^{1,5}$. A complete treatment of the spinor representation in six (and other) dimensions can be found in \cite{Kugo:1982bn,VanProeyen:1999ni}.

In order to describe spinor fields we start by considering the relevant Clifford algebra $\GroupName{Cl}(1,5)$ defined by the anticommutation relation
\begin{equation}
\left\{
\Gamma^M, \Gamma^{N}
\right\}
=
2 \eta^{MN} 
\end{equation}
for $M,N = 0,1,\ldots,5$. $\Gamma_M$ can be represented using $ 8 \times 8 $ complex matrices.
In full analogy with four dimensional spacetime, 
\begin{equation}
\Gamma^\dagger_0 = - \Gamma_0,
\qquad\qquad\qquad
\Gamma^\dagger_I = \Gamma_I,
\qquad\qquad
I = 1,\ldots,5,
\end{equation}
and
\begin{equation}
\Gamma_M^\dagger = - \Gamma_0 \Gamma_M \Gamma_0^{-1}.
\end{equation}
We also define the matrix $\Gamma^7$ as
\begin{equation}
\Gamma^7 = \Gamma^0 \Gamma^1 \cdots \Gamma^5
\end{equation}
such that \( (\Gamma^7)^2=\1\),
by means of which we can construct the chiral projectors
\begin{equation}
P_{\pm} = \frac{1}{2} \left( \1 \pm \Gamma^7  \right).
\end{equation}
As in four dimensional spacetime these are good definitions because $\Gamma^7$ anticommutes with all other $\Gamma^M$ matrices.

As it is well known, one can choose different representations for the $\Gamma^M$ matrices; since $\Gamma^{M*}$ still belongs to the same Clifford algebra (being the metric real), it always exists a matrix $B$ realising the change of basis, that is
\begin{equation}
\Gamma^M = -  B^{-1} \Gamma^{M *}  B
\end{equation}
with $ B$ such that
\(  B^*  B = -\1 \).
In six spacetime dimensions the complex conjugation does not mix the two chiral projections; in other words, spinors do not change chirality under complex conjugation, nor with any other covariant operation. This is an important difference with the $d=4$ spinor representations.

All these considerations imply that the spinor representation is completely reducible, since chiral constraints can be applied covariantly, but no Majorana condition can be imposed, since it would require $B^*B = \1$. Dirac spinors have $8$ complex components; Weyl spinors have $4$ complex components.
From now on, we only consider Weyl spinors $\Psi^a$.

It is convenient to consider, instead of one chiral spinor, a couple of spinors $\Psi_i$ with $i=1,2$ and  $\GroupName{SU}(2)$ index, with the additional pseudo-Majorana 
   constraint
\begin{equation}\label{pMW-condition}
	\overline{\Psi^a_i}
		:=
	- C^a_{\dot{b}} (\Psi^b_i)^*
		=
	\Psi^{ai},
\qquad\qquad
	C^a_{\dot{b}}
		=
	\begin{pmatrix}
		0 & \mathbbm{1} \\
		-\mathbbm{1} & 0
	\end{pmatrix};
\end{equation}where $*$ is the complex conjugation and $C^*C=-\1$ ($C$ it is the restriction of $B$ to Weyl spinors), implying  that \( \overline{\overline{\Psi}^a } = - \Psi^a \) for any spinor $\Psi^a$.

This arises from two motivations. First, it allows us not to consider dotted indices, making explicit the fact that the two chiralities are not mixed under conjugation; second, after dimensional reduction it becomes the $R$-symmetry of the $N=2$, $d=4$ superPoincar\`e algebra. For this reason we will refer also to the $\GroupName{SU}(2)$ symmetry introduced in \eqref{pMW-condition} as $R$-symmetry. Notice that the pseudo-Majorana condition does not modify the number of degrees of freedom; that is, a pseudo-Majorana-Weyl spinor still contains, off-shell, 4 complex degrees of freedom.

A convenient realization of $\Gamma$ matrices is the chiral\footnote{The chiral character of this representation can be seen realising that \( \Gamma_1 \sim \sigma^1 \otimes \ldots \), \(\Gamma_I \sim \sigma^2 \otimes \ldots \) and therefore \(  \Gamma_7 \sim \sigma^1 (\sigma^2)^5 \otimes \ldots \sim \sigma^3 \otimes \ldots \).   } one
\begin{equation}\label{conjug-gamma-6d}
(\Gamma_M)_{AB} =
	\begin{pmatrix}
	0 & (\Sigma_M)_{a  b}  \\
	(\tilde \Sigma_M)^{ a b} & 0 
	\end{pmatrix},
\end{equation}
where
\begin{equation}
(\Sigma_M)_{a  b} = (\1, \Sigma^I)
\qquad\qquad
(\tilde \Sigma_M)^{ a b} = (\1, - \Sigma^I)
\end{equation}
with  $\Sigma^M_{ab}$ are the 6d antisymmetric Weyl matrices that, splitting the indices as $a \sim (\alpha,\dot{\alpha})$, read
\begin{equation*}
	\Sigma^\mu_{ab}
			=
		\begin{pmatrix}
			0 & \sigma^\mu_{\alpha  \dot{\beta} } \\
			-\bar \sigma^\mu_{\dot \alpha \beta} & 0
		\end{pmatrix},
\qquad
	\Sigma^4_{ab}
		=
	\begin{pmatrix}
		-i\varepsilon_{\alpha \beta} & 0 \\
		0 & +i \varepsilon_{\dot \alpha \dot \beta}
	\end{pmatrix},
\qquad
	\Sigma^5_{ab}
		=
	\begin{pmatrix}
		-\varepsilon_{\alpha \beta} & 0 \\
		0 & -\varepsilon_{\dot \alpha \dot \beta}
	\end{pmatrix};
\end{equation*}
the usual hermitian matrices are obtained contracting \( \Sigma^M_{ab} \) with \( C^b_{\dot{b}} \):
\begin{equation}
(\Sigma_M)_{ab} = -  C\indices{^{\dot b}_b} (\Sigma_M)_{a \dot b} .
\end{equation}

Then,
\begin{equation}
\tilde \Sigma^{M\, ab} = \frac{1}{2} \varepsilon^{abcd} \Sigma^M_{cd},
\end{equation}  where \(\varepsilon^{1234}=\varepsilon_{1234}=+1\), so that
\begin{equation*}
	\tilde \Sigma^{\mu \; ab}
			=
		\begin{pmatrix}
			0 & - \sigma^{\mu \; \alpha  \dot{\beta} } \\
			\bar \sigma^{ \mu \; \dot \alpha \beta} & 0
		\end{pmatrix},
\qquad\qquad
	\tilde \Sigma^{4 \; ab}
		=
	\begin{pmatrix}
		-i\varepsilon^{\alpha \beta} & 0 \\
		0 & i \varepsilon^{\dot \alpha \dot \beta}
	\end{pmatrix},
\qquad\qquad
	\tilde \Sigma^{5 \; ab}
		=
	\begin{pmatrix}
		\varepsilon^{\alpha \beta} & 0 \\
		0 & \varepsilon^{\dot \alpha \dot \beta}
	\end{pmatrix}.
\end{equation*}

The generators of the Lorentz group in the $(1,0)$ spinor representation are 
\begin{equation*}
	\left( \Sigma^{MN}\right)^b_a =
	\left( \tilde \Sigma^{[M} \Sigma^{N]} \right)^b_a =
	\frac{1}{2} \left( \tilde \Sigma^{M} \Sigma^{N}
		- \tilde \Sigma^{N} \Sigma^{M}  \right)^b_a
\end{equation*}
\ie\hspace{-0.5em}\footnote{In \( \Sigma\indices{^{\mu\nu \;} ^{\alpha}  _{\beta} } \) the extra minus sign is due to the conventions for contracting spinor indices in $\sigma^{\mu\nu}$.}
\begin{align*}
	\Sigma\indices{^{\mu\nu \; }^a_b}
			=
		\begin{pmatrix}
			- \sigma\indices{^{\mu\nu \;} ^\alpha  _\beta } & 0 \\
			0 & \bar\sigma\indices{^{\mu\nu \;} ^{\dot \alpha} _{\dot\beta}} 
		\end{pmatrix},
& \qquad
	\Sigma\indices{^{\mu 4 \; }^a_b}
		=
	\begin{pmatrix}
		0 & i \sigma\indices{^{\mu\; } ^\alpha _{\dot \beta}} \\
		i \bar \sigma\indices{^{\mu\; } ^{\dot \alpha} _{\beta}} & 0
	\end{pmatrix},
\\
	\Sigma\indices{^{\mu 5 \; }^a_b}
		=
	\begin{pmatrix}
		0 & - \sigma\indices{^{\mu\; } ^\alpha _{\dot \beta}} \\
		\bar \sigma\indices{^{\mu\; } ^{\dot \alpha} _{\beta}} & 0
	\end{pmatrix},
& \qquad
	\Sigma\indices{^{4 5 \; }^a_b}
		=
	\begin{pmatrix}
		i \delta^\alpha_\beta & 0 \\
		0 & -i \delta^{\dot \alpha}_{\dot \beta} 
	\end{pmatrix}.
\end{align*}

We conclude this review of spinors in six spacetime dimensions by underlying that the nontrivial aspects of this spinor algebra are the the fact that in the right-hand-side of \eqref{conjug-gamma-6d} there is a minus sign (instead of a plus) and the fact that $B^*B = -\1$ instead of $\1$. Such signs are determined by the signature of the metric, and they can be proven to be the only consistent combination in $5+1$ dimension. We could have as well  obtained these results by studying the representations of the universal covering group of $\GroupName{SO}(1,5)$, that is \(\GroupName{Spin}(1,5) \approx \GroupName{SL}(2,\mathbb{H}) \approx \GroupName{SU*}(4) \),\footnote{The notation $\GroupName{SU*}(4)$ identifies the subgroup of $\GroupName{SU}(4)$ to which \( \GroupName{SL}(2,\mathbb{H}) \) is injected through the conventional representation of quaternions via \( \GroupName{SU}(2) \) matrices.} that is equivalent to the approach outlined here.


\subsection{Six-dimensional supersymmetry and harmonic superspace}

The six-dimensional supersymmetry algebra can be obtained as the immediate extension of the four-dimensional algebra \eqref{susy1}--\eqref{pedissequo}, by extending the Lorentz indices and imposing the pseudo-Majorana condition \eqref{pMW-condition}.
The commutator of supersymmetry generators then reads, following \cite[p.\ 32]{Galperin:book} and \cite{Howe:1983fr}
\begin{equation}\label{6d-susy-alg}
	\{Q_{a}^i, Q_{b}^j\} = 2 \varepsilon^{ij} \Sigma^M_{ab} P_M,
%
%
\end{equation}

In full analogy with the four dimensional case, one can study the representations of the six-dimensional supersymmetry algebra. Since we are employing an explicit $\GroupName{SL}(2,\mathbb{C})\oplus \GroupName{SL}(2,\mathbb{C})$ notation, we will not go through all the derivation again, since the steps are basically the same up to a doubling of the indices. We briefly mention that, since we are studying massless representations, we can diagonalise the momentum operator to $P_M = (-E, 0, 0, E, 0, 0)$ and from this  follows that only one supercharge is nontrivially realised, and raises the helicity of $1/2$. The vector multiplet therefore is made of the gauge boson and the massless spinor, with both helicities because of CPT invariance. Notice that, off-shell, the spinor carries $8$ real degrees of freedom, while the vector boson only $6-1=5$ (one is killed by gauge invariance), therefore we expect $3$ bosonic real auxiliary fields to appear.

The algebra \eqref{6d-susy-alg} can be realised in the conventional superspace, called the real superspace \( \mathbb{R}^{1,5|1} \), where points are described with coordinates
\begin{equation}
\left( x^M, \theta^a_i \right) 
\end{equation} 
 where $x^M$ are (commuting) spacetime coordinates and $\theta^a_i$ are complex Grassmann numbers satisfying the pseudo-Majorana condition \eqref{pMW-condition}.
In this superspace the supersymmetry algebra is realised through the transformations parametrised by the pseudo-Majorana-Weyl spinor $\xi$
\begin{equation}
\delta x^M 
= i \left(
\xi^{ia} \, \Sigma^M_{ab} \, \theta^b_i
- \theta^{ia} \, \Sigma^M_{ab} \, \xi^b_i
\right),
\hspace{3em}
\delta\theta^a_i =  \xi^a_i. 
\end{equation}
In the real superspace we then introduce the spinor covariant derivatives
\begin{equation}\label{spinor-covD-usual}
\covDs_a^k  = \partial_a^k - i \theta^{bk} \partial_{ab}
\end{equation}
where the partial derivatives are defined as
\begin{equation}
\partial_{ab} 
= \frac{1}{2} \Sigma^M_{ab} \partial_M 
\qquad\qquad
\partial_a^k \, \theta^b_i = \delta^k_i \, \delta^b_a.
\end{equation}
We also introduce the notation, valid in general for vector indices,
\begin{equation}
x^M = \frac{1}{2}\Sigma^M_{ab} x^{ab}.
\end{equation}
The spinor derivatives satisfy the algebra
\begin{equation}
\left\{
\covDs^k_a   ,  \covDs^j_b
\right\}
	=
- 2 i \varepsilon^{kj} \partial_{ab}.
\end{equation}
The covariant derivatives (anti)commute with supersymmetry generators and therefore are used to construct objects that transform covariantly under supersymmetry. However this is of little help in this case, in contrast with the unextended supersymmetry in four dimensions, because it is not possible to formulate off-shell supersymmetric theories in six dimensional spacetime in this superspace, as follows from a degree-of-freedom counting argument as discussed in \cite{Howe:1983fr,Galperin:book}.
This no-go theorem drove physicists to look for other kind of superspaces. As we will see, the loophole to this result is to find a superspace that allows for the introduction of an infinite number of auxiliary fields.

There are also other choices of coordinates that one can make. In particular we will deal with the \emph{harmonic} coordinates, that we are now going to construct. For a complete discussion of the construction, of the properties and the superfieds that can be introduced in four and six dimension the reader should check \cite{Galperin:1984av, Galperin:book, Howe:1985ar}.

Let us introduce the bosonic variables $u^\pm_i$, $i$ being a $\GroupName{SU}(2)_R$ index, such that $u^-_i = (u^{+i})^*$ and
\begin{equation}\label{u-defining-property-1}
u^{+i} u^-_i = 1,
\end{equation}
\begin{equation}\label{u-defining-property-2}
u^{+i} u^+_i = 0 = u^{-i} u^-_i.
\end{equation}

They are charged with respect to a $\GroupName{U}(1)_R$ subgroup of $\GroupName{SU}(2)_R$ with charges $\pm 1$ according to the notation.
These properties allow us to define, for any spinor $\Psi^i_a$, its $\GroupName{U}(1)_R$ projections $\Psi^{\pm} = \Psi^i u^\pm_i$ and to decompose it according to
\begin{equation}
\psi^i = u^{+i} \Psi^- - u^{-i} \Psi^+.
\end{equation}

The harmonic variables effectively correspond to assigning an  \( \GroupName{SU}(2) \) matrix
\begin{equation}\label{matrix-u+u-}
\begin{pmatrix}
u^+_1 	&	u^-_1	\\
u_2^+	&	u^-_2	
\end{pmatrix}
\end{equation}
where the requirement of unit determinant is exactly the condition \eqref{u-defining-property-1}.
The action of $\GroupName{U}(1)$ can be then rewritten as
\begin{equation}\label{matrix-u+u--U1action}
\begin{pmatrix}
u^+_1 e^{i\psi}	&	u^-_1 e^{- i\psi}	\\
u_2^+ e^{i\psi}	&	u^-_2 e^{- i\psi}	
\end{pmatrix}
=
\begin{pmatrix}
u^+_1 	&	u^-_1	\\
u_2^+	&	u^-_2	
\end{pmatrix}
e^{i \psi \sigma^3}
\end{equation}
where $\sigma^3$ is the third Pauli matrix and \( e^{i \psi \sigma^3} \in \GroupName{U}(1)_R \). Not all $\GroupName{SU}(2)$ matrices can be represented as \eqref{matrix-u+u-}, while \eqref{matrix-u+u--U1action} parametrises the whole group; 
 we are effectively considering the coset 
 $\GroupName{SU}(2)_R / \GroupName{U}(1)_R$, that is diffeomorphic to the two sphere $\mathbb{S}^2$. This also proves that $\GroupName{U}(1)_R$ is the only component of $ \GroupName{SU}(2)_R $ linearly realised, as a consequence of coset theory.

We also introduce the following integration rules for the harmonic variables
\begin{equation}\label{integratioon-harmonics}
\int \dd{u} 1 = 1,
\hspace{3em}
\int \dd{u} u^+_{(i_1} \cdots u^+_{i_n} u^-_{j_1} \cdots u^-_{j_m)} = 0,
\end{equation}
with the second one holding when $n + m \neq 0 $.

There are different kind of conjugations that we can define over the harmonic variables. We already considered complex conjugation $*$, that affects both the $\GroupName{SU}(2)_R$ and the $\GroupName{U}(1)_R$ indices; we now define the conjugation $\overline{\phantom{x}}$, that affects only the latter:
\begin{equation}
\overline{u^{+i}} = u^{-i}
\qquad
\overline{u^{-}_{i}} = -u^+_{i}.
\end{equation}
This conjugation preserves the defining properties of \eqref{u-defining-property-1} and  \eqref{u-defining-property-2} and squares to minus the identity.
Then, we can define the conjugation $\sim$, that is the composition of $*$ and $\overline{\phantom{x}}$; on scalars and Grassmann numbers it is the complex conjugation, whereas on the harmonic variables it takes the value
\begin{equation}
\widetilde{u^{\pm}_{i}} = u^{\pm i}
;
\end{equation}
it therefore squares to the identity for complex and Grassmann numbers, and to minus the identity for harmonic variables.
This implies that for an harmonic function of charge $q$, 
\begin{equation}
\widetilde{\widetilde{ f^{(q)}}} = (-)^q f^{(q)}.
\end{equation}
In the particular case of even $q=2k$, we have that we can impose the condition
\begin{equation}
\widetilde{f^{2k}} = f^{(2k)}
\end{equation}
or equivalently
\begin{equation}
\overline{f^{i_1 \cdots i_{2k}}} = \varepsilon_{i_1 j_1} \cdots \varepsilon_{i_{2k} j_{2k}} f^{{i_1 \cdots i_{2k}}}.
\end{equation}

We are now ready to introduce the \emph{real harmonic superspace}  $\HR$, that is the space described with the set of coordinates
\begin{equation}
(z,u) = \left( x^M, \theta^a_i, u^{\pm i} \right).
\end{equation}
This superspace is therefore diffeomorphic to $\mathbb{R}^{1,5|1} \times \mathbb{S}^2$, where only part of the $R$-symmetry (the $\GroupName{U}(1)_R$ subgroup) is now linearly realised on the coordinates, while in the conventional superspace the whole $\GroupName{SU}(2)_R$ is linearly realised. In the superspace  $\HR$ we still have the spinor covariant derivatives \eqref{spinor-covD-usual}; in addition we can introduce derivatives (automatically covariant, since the harmonics do not have spinor indices) for the harmonic variables,
\begin{equation}
\partial^{++} = u^{+i} \frac{\partial}{\partial u^{-i}},
\hspace{2em}
\partial^{--} = u^{-i} \frac{\partial }{\partial u^{+i}},
\hspace{2em}
D^0 = u^{+i} \frac{\partial}{\partial u^{+i}} - u^{-i} \frac{\partial}{\partial u^{-i}}.
\end{equation}
Notice that a superfield of definite $\GroupName{U}(1)$ charge is eigenvector of $D^0$ with eigenvalue the charge.

An alternative coordinate system on this superspace is represented by the `analytic' coordinates,
\begin{equation}
( Z_A , u) = (x^M_A, \theta^{\pm a}, u^\pm_i) 
\end{equation}
where
\begin{equation}
\theta^{\pm }_a = \theta^i_a u_i^\pm
\qquad
x^M_A = x^M + \frac{i}{2} \theta^a_k\, \Sigma^M_{ab} \, \theta^b_l \, u^{+(k|} u^{-|l)}.
\end{equation}

Another superspace that can be now introduced, whose importance will be clear soon, is the \emph{Grassmann-analytic} (G-analytic for short) \emph{superspace} $\HR^+$, that is spanned by the coordinates
\begin{equation}
(\zeta , u ) = \left( x^M_A, \theta^{+ a},  u^\pm_i \right) ,
\end{equation}
namely we dropped dependence on the coordinate $\theta^-$.
This new superspace turns out to be very useful because the restriction to it is supersimmetric invariant, as it happens in four spacetime dimensions with the chiral superspace. This can be seen by noticing that the supersymmetry transformations can be realised as 
\begin{equation}
\delta x^M_A
= -2 i \left(
\xi^{ia} \, \Sigma^M_{ab} \, \theta^{+b}
- \theta^{+a} \, \Sigma^M_{ab} \, \xi^{ib}
\right)u^-_i,
\hspace{1.5em}
\delta\theta^a_i =  \xi^{ia} u^+_i,
\hspace{1.5em}
\delta u^\pm_i=  0
,
\end{equation}
where again $\xi$ is the parameter of the supersymmetry transformation. It is clear that such transformation leave the G-analytic subspace invariant.

With respect to analytic coordinates, the covariant derivatives in the analytic basis become
\begin{align}
\label{spinor-covD-6-analytic-basis}
&	\covDs^+_a 	
 =
	\partial_{-a},
\\
\qquad
&	\covDs^{-}_a 	
= 
	- \partial_{+a} - 2 i \theta^{-k} \partial_{ab},
\\
\qquad
&	\covDs^0	
= 
	u^{+i} \frac{\partial}{\partial u^{+i}} 
	- u^{-i} \frac{\partial}{\partial u^{-i}}
	+ \theta^{+a} \partial_{+a} 
	- \theta^{-a} \partial_{-a},
\\
&	\covDs^{++}	
=
	\partial^{++} 
	+ i \theta^{+a} \theta^{+b} \partial_{ab} 
	+ \theta^{+a} \partial_{-a},
\\
\qquad
&	\covDs^{--}
=
	\partial^{--} 
	+ i \theta^{-a} \theta^{-b} \partial_{ab} 
	+ \theta^{-a} \partial_{+a},
\end{align}
where
\( \partial_{\pm a} \theta^{\pm b} = \delta^b_a. \)
The following commutation relations hold:
\begin{equation}
\left\{ \covDs^+_a , \covDs^-_b \right\} = 2i \partial_{ab},
\qquad
[ \covDs^{++} , \covDs^{--} ] = \covDs^0,
\end{equation}
\begin{equation}\label{casa}
[\covDs^{--}, \covDs^-_a] = \covDs^+_a,
\qquad
[ \covDs^{--} , \covDs^+_a ] = \covDs^-_a,
\end{equation}
\begin{equation}\label{covDs++-commutator-1}
[\covDs^{++} , \covDs^+_a ] = 0 =  [ \covDs^{--} , \covDs^-_a ],
\end{equation}
as one can verify by direct computation.
Notice, however, that in the usual (non-analytic) $x^M$ coordinate we can write, with abuse of notation, $\covDs^{\pm}_a = u^{\pm}_i \covDs^i_a$.

We also introduce the notation
\begin{equation}
\left( \covDs^\pm \right)^4
	=
- \frac{1}{24}
	\varepsilon^{abcd} 
	\covDs_a^\pm \covDs_b^\pm \covDs_c^\pm \covDs_d^\pm
\end{equation}
and the integration measure
\begin{equation}
d \zeta^{(-4)} = d^6 x_A \left( \covDs^- \right)^4.
\end{equation}
Notice that with these definitions the only nonvanishing superspace integral is
\begin{equation}\label{integr-zeta(-4)}
\int d \zeta^{(-4)} (\theta^+)^4 = \int d^6 x_A .
\end{equation}

Consider now a superfield $\Phi$.
A first constraint that we can impose  is Grassmann-analiticity, that is said to be satisfied  if a field depends only on the variables
\(
\left( \zeta, u \right)
	=
\left( x^M_A, \theta^{+a}, u^{\pm i} \right),
\)
\ie it is independent of $\theta^{-a}$.  Notice that the constraint can be written in differential form as
\begin{equation}
\covDs^+_a \Phi = 0
\end{equation}
by virtue of \eqref{spinor-covD-6-analytic-basis}. This is analogous to the chiral constraint in the four dimensional superspace.

Since $\GroupName{U}(1)_R$ is linearly realised, superfields transform in representations of it, and this fact allows us to classify them according to their charge. Any covariant function $f^{(q)}$ with charge $q>0$ can then be decomposed in series of harmonics as
\begin{equation}
f^{(q)} = \sum_n^\infty f^{(i_1 \ldots i_{n+q} j_1 \ldots j_n )}
u^+_{i_1} \cdots u^+_{i_{n+q}} u^-_{i_1} \cdots  u^-_{i_n},
\end{equation}
an analogous decomposition holding for $q < 0$. Every $f^{\cdots}$ transforms in an irreducible representation of $\GroupName{SU}(2)_R$. Therefore, given a superfield of definite $\GroupName{U}(1)$ charge, it can be expanded according to the previous formula; this allows for the infinite number of auxiliary fields that we mentioned before.

We have now introduced all the superspace technology that we need in order to work in six dimensional supersymmetry.

\subsection{Gauge theory in harmonic superspace}

\subsubsection{General framework}

We want to consider fields that are not necessarily G-analytic, that is they are functions $\Phi$ of all the real superspace coordinates $(x^M,\theta^a_i)$. Notice that also this subspace is closed under the supersymmetry transformations considered above.  Consider a symmetry transformation of the form
\begin{equation}
\Phi'(x,\theta^i_a) =  e^{i \tau} \Phi(x,\theta^i_a),
\hspace{3em}
\tau = \tau^k \, T^k
\end{equation} 
with $\tau$ real and 
being $T^k$ the generators of the representation of the symmetry group $\GroupName{G}$ under which $\Phi$ transforms. We want to gauge such symmetry transformation allowing for $\tau = \tau(x,\theta^i_a)$, by consistency independent of the harmonic variables $u$, that is 
\begin{equation}
\partial^{++} \tau = 0 ,
\qquad\qquad
\partial^{--} \tau = 0 .
\end{equation}

Let us start gauging the real-superspace derivatives $\covDs_{a}^i$ and $\partial_{ab}$ by introducing a gauge connection $A(x,\theta^a_i)$
\begin{align}
\covD^i_a = \covDs_{a}^i + A_{a}^i(x,\theta^i_a),
\qquad\quad
\covD_{ab} = \partial_{ab} + A_{ab}(x,\theta^i_a).
\end{align}
The field strength tensor in superspace is then constructed by taking (anti)commutators of such covariant derivatives
\begin{equation}
F_{IJ} = \left[ \covD_I, \covD_J \right\} - t\indices{_I_J^K} \covD_K,
\hspace{3em}
I,J,K = \left( ab \sim M,\ ^i_a \right)
\end{equation}
where $t$ is the  torsion coming from the commutators of the superspace (non-gauged) covariant derivative \eqref{covDs++-commutator-1}. The relevant components read
\begin{align}
\label{6d-gauge-F-with-torsion}
\left\{ \covD^i_a , \covD^j_b  \right\}
	& =
F^{ij}_{ab} 
+ i \varepsilon^{ij} 
	\delta^c_{[a} \delta^d_{b]} \covD^{}_{cd},
\\
\left\{ \covD_{ab}  , \covD_{cd}   \right\}
	& =
F_{ a b \, c d },
\\
\left[ \covD^i_a   ,  \covD^{}_{ b c } \right]
	& =
F^i_{ a \, b c }.
\end{align}

The field strength tensor obeys a number of Bianchi identities, but the representation still contains a big number of fields, as usual in such approach in superspace. Constraints must be imposed on the field strength in order to extract the desired multiplet.

The constraints to impose to describe the six dimensional gauge theory have been studied in the literature, see for instance \cite{Nilsson:1980nz, Koller:1982cs}, and read
\begin{equation}
\label{6d-gauge-constr1}
F_{ab}^{ij} = 0;
\end{equation}
even though we will not work in this superspace we solve the constraint to show the ideas that will be employed in the harmonic superspace.
The constraint is solved by
\begin{equation}
\label{6d-gauge-Aialpha}
A^i_a
	=
e^{-v} \covDs^i_a e^{v}
\end{equation}
for some real superfield $v=v(x^M,\theta^i_a) = v^k \, T^k$.
Notice  that $v$ 
is not uniquely determined: The substitution $e^v \rightarrow e^{\kappa} e^{v}$ yields the same $A^i_a$ as long as $\covDs_a^i \kappa = 0 $; by consistency with the initial restriction to the subspace $\mathbb{R}^{1,5|1}$, this means that $\kappa$ is real and depends only on $x$.
Then, requiring covariance with respect to gauge transformations parametrized by $\tau$, one has the full gauge transformations
\begin{align}
e^{v'}
	& =
e^{\kappa} e^{v}e^{- \tau}
\\
{A_a^i} '
	& =
e^{\tau} A_a^i e^{-\tau} 
- 
( \covDs_a^i e^{\tau}) e^{-\tau},
\end{align}
where to verify the latter it might be useful to recall that $ \covDs
e^{-\tau} = - e^{-\tau} \left( \covDs
e^\tau \right) e^{-\tau} $.

\subsubsection{Harmonic superspace}

We are particularly interested in G-analytic superfields. The fundamental object that contains the degrees of freedom of the gauge theory is the connection connection $V^{++}$ for the derivative $\covDs^{++}$, so that
\begin{equation}\label{covds++}
\covD^{++} = \covDs^{++} + V^{++} 
\end{equation}
transforms covariantly under gauge transformations. This is achieved by requiring
\begin{equation}\label{v++-gage-intiial}
\left( V^{++} \right)' = e^{\lambda} \covDs^{++} e^{-\lambda} + e^{\lambda} V^{++} e^{-\lambda}
\end{equation}
begin $\lambda$ the gauge transformation parameter. We shall require that $\lambda$ is a $\sim$-even Grassmann-analytic superfield, that is $\tilde{\lambda}=\lambda$ and $\covDs^+_\alpha \lambda = 0$.

To understand this we would need to introduce matter multiplet and discuss gauging of internal symmetries. We will not go into this in detail since it would require a discussion much broader than the aim of this work, and we will just sketch here the main results. Indeed the matter multiplet can be embedded into a complex G-analytic superfield $\phi^{(1)}(\zeta,u)$ with $\GroupName{U}(1)$ charge $1$ whose action reads
\begin{equation}
S = \int \dd{u} \dd{\zeta^{(-4)}} {\tilde\phi}^{(1),\ell} \covDs^{++} \phi^{(1),\ell},
\end{equation}
being $\ell$ and internal index.
Gauging the invariance under phase transformations $\phi' = e^{ \lambda} \phi$, with $\sim$-even $\lambda = \lambda^k T^k $ being $T^k$ the generators $ \GroupName{G}$, this action is no longer invariant; introducing the $V^{++}$ connection satisfying \eqref{covds++} we can recover it. The G-analyticity of $\lambda$ comes requiring the consistency of the G-analyticity of $\phi^{(1)}$ under gauge transformations, that is a reasonable request if we recall that supersymmetry preserves it too. Notice that in the absence of \ym{} field, the derivative $\covDs^{++}$ preserves G-analyticity thanks to the commutators \eqref{covDs++-commutator-1}.
We therefore understand that all what is needed to encode \ym{} degrees of freedom is the connection $V^{++}$, that we are going to study in some greater detail.

Contracting  \eqref{6d-gauge-constr1} with the harmonics $u^+_i$ and $u^+_j$ we can rewrite it in the form
\begin{equation}
\left\{ 
\covD^+_a , \covD^+_b
\right\}
= 0
\end{equation}
whose solution reads
\begin{equation}
\label{6d-gauge-A+alpha}
A^+_a = e^{-b} \covDs^+_a e^b
\end{equation}
being now $b(x,\theta,u)$ dependent also on $u^\pm_i$ and transforming according to
\begin{equation}
e^{b'}
	=
e^{ \lambda} e^b e^{-\tau}
\end{equation}
where $\tau$ is again a generic superfield independent of $u$ while \(\lambda\) is now G-analytic, namely $\covDs^+_a \lambda =0 $.
Since the covariant derivative and $\lambda$ are $\sim$-even, we also require that $\widetilde{ b }= b$.

The meaning of the superfield $b$ can be understood as follows. A covariantly G-analytic superfield in the real superspace \( \Phi(x,\theta^i,u) \) (so that $\covD^+_a \Phi = 0$) undergoes the gauge transformation \( \Phi \rightarrow e^{\tau} \Phi  \); the field 
\begin{equation}\label{phi-Phi}
\phi := e^b \Phi,
\end{equation}
on the other hand, is G-analytic because
\begin{equation}
0 = \covD^+_a \Phi = e^{-b} \covDs^+_a \phi
\end{equation} 
 and transforms with $\lambda$ only:
\begin{equation}
\phi' =
	e^{b'}\Phi' =  
e^{ \lambda} e^b e^{-\tau} e^{\tau} \Phi = e^{\lambda} \phi,
\end{equation}
that also preserves G-analyticity of $\phi$.

In the analytic superspace we have also two other  derivatives: $\covDs^{++}$ and $\covDs^{--}$. Since we have more derivatives to make gauge-covariant, correspondingly we have to impose other constraints to get rid of the extra components of the superfield strength tensor. We know that $\covDs^{++}$ respects G-analyticity, that is it commutes with \( \covDs^+_a \),  we require this to be true for the gauge-covariant derivatives too:
\begin{equation}\label{6-gauge-commut-cvd++.covd+alpha}
\left[ \covD^{++}  ,  \covD^+_a  \right]
	=
0.
\end{equation}
We require \eqref{6-gauge-commut-cvd++.covd+alpha} to be valid for both the   gauge transformations, generated by $\tau$ and $\lambda$. Considering $\phi$-type fields, in the notation of \eqref{phi-Phi}, we get covariance under ($\lambda$-)gauge transformations if
\begin{equation}\label{6d-gauge-V++-def}
V^{++} = e^{b} \covDs^{++} e^{-b},
\end{equation}
transforming as
\begin{equation}\label{6d-gauge-V++-gauge-transf}
\left( V^{++} \right)' 
	=
e^{\lambda} \covDs^{++} e^{-\lambda} + e^{\lambda} V^{++} e^{-\lambda}
	=
e^{\lambda} \covD^{++} e^{-\lambda} 
\end{equation}
that is indeed our desired result (cf.\ \eqref{v++-gage-intiial}). Notice that $V^{++}$  that is a $\sim$-even function, as consequence of such being the $\covDs^{++}$ and $b$.
We know that for the $\tau$-gauge transformations $\covDs^{++}$ does not need any connection, as a consequence of $\covDs^{++} \tau = 0$; using this fact and plugging \eqref{6d-gauge-A+alpha} into the constraint \eqref{6-gauge-commut-cvd++.covd+alpha} one finds
\begin{equation}
0 = \covDs^{++} A^+_a 
= - e^{-b} \left[  \covDs^+_a \left(  e^b \covDs^{++} e^{-b}   \right)  \right] e^b
\end{equation}
that by comparison with \eqref{6d-gauge-V++-def} is a G-analyticity constraint for $V^{++}$, namely
\begin{equation}
\covDs^+_a V^{++} = 0.
\end{equation}

We  have just arrived to some conclusion. Indeed, if we accept that all the physical degrees of freedom of the vector multiplet are encoded in $V^{++}$ (\ie $A_\mu$ with gauge invariance and $\Psi^i$), through \eqref{6d-gauge-V++-def} can be determined $b$ and from it one can then compute $A^+_\alpha$ using \eqref{6d-gauge-A+alpha}, obtaining therefore the covariant derivative in $\HR^+$. However, in order to reconstruct the whole $A^i_a$, we still need to find $A^-_a$.

The superfield $V^{++}$ can be used to build a \sym{} action but the procedure is quite involved. We will directly study the higher derivative case, that is remarkably simpler. In order to do that, we still have to introduce some definitions. First of all, let us promote the commutation relation \eqref{covDs++-commutator-1} to the gauge-covariant derivatives,  by requiring
\begin{equation}
\label{6d-gauge-V--V++-relation}
\left[
	\covD^{++} ,  \covD^{--}
\right]
	= \covDs^{0};
\end{equation}
notice that $\covDs^0$ is already covariant since it only gives the $\GroupName{U}(1)$-weight and we work only with fields of definite charge.
This is obviously true for the $\tau$-gauge transformations, since $\covDs^{++} \tau = 0 = \covDs^{--} \tau $; to extend it to the $\lambda$-gauge transformations we have to introduce a connection $V^{--}$ for $\covDs^{--}$. \eqref{6d-gauge-V--V++-relation} turns then to an implicit definition of $V^{--}$ in terms of $V^{++}$, since it reads
\begin{equation}\label{6d-gauge-V--V++-differential-relation}
\covDs^{++} V^{--} - \covDs^{--} V^{++} + \left[ V^{++} , V^{--} \right] = 0.
\end{equation}
Notice that we can as well express the $V^{--}$ connection in terms of the superfield $b$ as we did for $V^{++}$  
\begin{equation}\label{6d-gauge-V---def}
V^{--} = e^{b} \covDs^{--} e^{-b}
\end{equation}
from which its gauge transformation property follows
\begin{equation}\label{6d-gauge-V---gauge-transf}
\left( V^{--} \right)' 
	=
e^{\lambda} \covDs^{--} e^{-\lambda} + e^{\lambda} V^{--} e^{-\lambda}
	=
e^{\lambda} \covD^{--} e^{-\lambda}.
\end{equation}
With $V^{--}$ we can reconstruct the connection $A^-_a = - \covDs^+_a V^{--}$ imposing the anti-commutation relation
\begin{equation}
\left\{
\covD^{--}, \covD^+_a
\right\} = \covD^-_a,
\end{equation}
that is the covariant generalization of \eqref{casa}.
We have therefore expressed all the gauge structure in terms of the field $V^{++}$, as promised, since we can reconstruct the whole $A^i_a$.

Following \cite{Ivanov:2005qf}, using \( V^{--} \) one can define the covariant spinor superfield strength
\begin{equation}
W^{+\alpha}
	=
- \frac{1}{6}
\varepsilon^{ \alpha \beta \gamma \delta } 
\covDs^+_\beta \covDs^+_\gamma \covDs^+_\delta V^{--}
\end{equation}
from which we can also build
\begin{equation}
F^{++} = \frac{1}{4} \covDs^+_a W^{+a} \equiv (\covDs^+)^4 V^{--}.
\end{equation}
Notice that as a consequence of the gauge transformation rule for $V^{--}$, \eqref{6d-gauge-V---gauge-transf}, and since $\lambda$ is G-analytic, $F^{++}$ transforms in the adjoint representation of the group too, \ie
\begin{equation}
(F^{++})' = e^\lambda F^{++} e^{-\lambda}.
\end{equation}

Before proposing an action, let us spend a couple of words about dimensions. The harmonics $u$ are dimensionless as a consequence of the defining  properties \eqref{u-defining-property-1} and \eqref{u-defining-property-2}; this implies that the derivatives have dimension $[\covDs^\pm_\alpha] = - [ \theta^\alpha_i ] = 1/2$ and $[\covDs^{++}] = 0 = [\covDs^{--}]$.  Then, from \eqref{6d-gauge-V++-def} and \eqref{6d-gauge-V---def} $V^{\pm\pm}$ are dimensionless too, and therefore $[F^{++}] = 2$. For what we said, follows that $[\dd{\zeta^{(-4)}}] = 4$.

A possible action, gauge and Lorentz invariant, is then
\begin{equation}\label{hd-action-superspace}
S = \frac{1}{2 f } \int \dd{\zeta^{-4}} \dd{u} 
	\Tr \left[ \left(  F^{++} \right)^2 \right]
\end{equation}
with $f$ a dimensionless constant.

\subsection{Action in terms of components fields}

The gauge symmetry transformation \eqref{6d-gauge-V++-gauge-transf} allows us to get rid of all contributions from higher harmonics, choosing the so-called Wess-Zumino gauge. The infinitesimal form of the gauge transformation reads \( \delta V^{++} = - \covD^{++} \lambda \); using it, with some work one can show that is possible to cast $V^{++}$ in the form
\begin{equation}\label{wz-gauge}
V^{++}_{\WZ}(x,\theta^+)
	=
- \theta^{ + a } \theta^{ + b } A_{a b }(x) 
+ 2 \sqrt{2} (\theta^+)^3_a \Psi^{ - a}(x)
- 3 (\theta^+)^4 D^{--}(x)
\end{equation}
where there is still the usual gauge invariance for $A_{ab}$ (\ie $A_M$), and we used the notation
\begin{align}
&\left(
	\theta^+
\right)^3_d
	=
\frac{1}{6} \varepsilon_{a b c d} 
\theta^{ + a } \theta^{ + b } \theta^{ + c },
\quad
&\left(
	\theta^+
\right)^4
	=
- \frac{1}{24} \varepsilon_{a b c d} \theta^{ + a } \theta^{ + b } \theta^{ + c } \theta^{ + \delta },
\end{align}
and the fields read
\begin{equation}
\Psi^{-a} = \Psi^{ai} u^-_i,
\qquad\qquad
D^{--} = D^{ik}u^-_i u^-_k.
\end{equation}
We will not go into this in detail, the interested reader can check \cite{Galperin:book}, but we motivate this result. Both $V^{++}$ and $\lambda$ are a analytic superfield, so they depend on the variables $(x_A^\mu, \theta^+_a, u^\pm_i)$; they therefore admit a finite expansion in powers of $\theta^+$, in which the coefficients are functions of $x$ and $u$; the $\GroupName{U}(1)$ weight of the coefficients is fixed requiring that the superfields have weight respectively $+2$ and $0$. In this way, comparing the harmonic expansions (3.2.44), one can check, explicitly at least in the Abelian case, that using $\delta V^{++} = - \covDs^{++} \lambda $ it is possible to get rid of all component but those in \eqref{wz-gauge}, fixing $\lambda$ up to the function $\lambda(x) = \lambda |_{u=0=\theta^+}$ that generates usual gauge transformation for $A_{ab}$. Indeed, it is immediate to verify that $\covDs^{++} \lambda(x) = \theta^{+a} \, \theta^{+b} \, \partial_{ab} \lambda (x).$

In this expansion the only fields that appear are the gauge field $A^M$, the gluino $\Psi^{ai}$ and the triplet of auxiliary fields $D^{ik}$, so that this is exactly an off-shell representation of the vector supermultiplet. Notice that the fermionic ($4$ complex) and bosonic ($5$ + $3$ real) degrees of freedom indeed agree, as anticipated.

In order to express the action \eqref{hd-action-superspace} in terms of components fields, we have to find the $V^{--}$ superfield associated to $V^{++}_\WZ$, act on it with $(D^+)^4$ in order to find $F^{++}$ and then perform the integration over the Grassmann and harmonic variables. The algebra is quite lengthy but straightforward; we will only outline the steps.

$V^{--}$ is not a G-analytic superfield, so we can decompose it as
\begin{equation}
V^{--} 
	=
v^{--}
+ \theta^{- b} v_{ b }^-
+ \theta^{-c} \theta^{-d} v_{cd}
+ (\theta^-)^3_d v^{+d}
+ (\theta^-)^4 v^{++}
\end{equation}
where the coefficients are functions of \((x_A, \theta^+, u)\), \ie are G-analytic;
noticing that
\begin{equation}
F^{++} =  (\covDs^+)^4 V^{--} = v^{++} 	
\end{equation}
we conclude that we only need to solve for  the lowest  weight component $v^{--}$. 
$F^{++} = v^{++}$, being  G-analytic, can be similarly expanded as
\begin{equation}\label{v++-expansio}
F^{++}
= v^{++} =
\lambda^{++}
+ \theta^{+a} \lambda^+_a
+ \theta^{+a} \theta^{+b} \lambda^+_{ab}
+ (\theta^{+})^3_a \lambda^{-a}
+ (\theta^{+})^4 \lambda^{--};
\end{equation}
 Applying now \eqref{6d-gauge-V--V++-differential-relation} and extracting the equation for the $v^{++}$ component only, we obtain
\begin{equation}
\begin{split}
\covDs^{++} v^{++}
&- \theta^{+a} \theta^{+b} \left[  A_{ab}  , v^{++}  \right]
+ 2 \sqrt{2} (\theta^+)^3_a \left[  \Psi^{-a}  , v^{++}  \right]
- 3 (\theta^+)^4 \left[  \covDs^{--} ,  v^{++} \right]
	=
0;
\end{split}
\end{equation}
inserting the expansion \eqref{v++-expansio} into the previous equation and comparing the coefficients term by term in the Grassmann variables, with a bit of work one can find the solution to be
\begin{align}
	\lambda^{++}
& = 
	- \covDs^{++}
,\\
	\lambda^{+}_a
& = 
	i \sqrt{2} \left( \Sigma^M \right)_{ab} \covD_M \Psi^{+b}
,\\
	\lambda_{ab}
& = 
	\frac{1}{2} \left( \Sigma^M \right)_{ab}
		\left[ i \covD_M D^{+-} - \covD^N F_{NM} \right]
	+ \varepsilon_{abcd} i \left\{ \Psi^{-c},\Psi^{+d}\right\}
,\\
	\lambda^{-a}
& = 
	\sqrt{2} \covD^2 \Psi^{ai} u^-_i 
	- i \sqrt{2} F_{MN} \left( \Sigma^{MN} \right)^a_b \Psi^{bi} u^{ bi } u^-_i
	\\ & \qquad \qquad - \frac{4\sqrt{2}}{3} [\Psi^{ai}, D^l_i] u^-_l
	+ \sqrt{2} \, [\psi^{ai}, D^{kl}]\, u^-_{(i}u^-_ku^+_{l)}
,\\
	\lambda^{--}
& = 
	- \covD^2 D^{--} - 3 \left[  D^{--} , \covD D^{+-} \right]
	- 2i \left\{  \Psi^-, \Sigma^M \covD_M \Psi^- \right\}
.
\end{align}


We are now ready to consider the action \eqref{hd-action-superspace}. First we can integrate the $\theta^{+a}$, that according to \eqref{integr-zeta(-4)} singles out only the $(\theta^+)^4$-component of $F^{++}$, and we get
\begin{equation}\label{step-component field}
S = \frac{1}{2 g^2}
\int \dd{^6 x} \dd{ u }
\left(
	2 \lambda^{++} \lambda^{--}
	-2 \lambda^{+a}  \lambda^{-a}
	- \varepsilon^{abcd} \lambda_{ab} \lambda_{cd}
\right).
\end{equation}
Using \eqref{integratioon-harmonics} and splitting symmetric and antisymmetric part of the product of $u$'s, the following relations are easy to verify,
\begin{equation}
\int \dd{ u } u^+_i u^j_j  = \frac{1}{2} \varepsilon_{ij}
\qquad
\int \dd{ u } u^+_i u^+_j  u^-_k u^-_l
	 =
\frac{1}{6} 
\left( \varepsilon_{ik} \varepsilon_{jl} -  \varepsilon_{il} \varepsilon_{jk}\right)
\end{equation}
now the integration over the harmonics in \eqref{step-component field} can be performed, finally obtaining
\begin{equation}\label{lagr-hd-sym-6d}
\begin{split}
	\Lagr_{\SYM}^{6d,\HD}
		=
	\frac{1}{2f^2}
		\tr \bigg[
	&			 \left( \covD^M F_{ML} \right)^2
				+ i \Psi^i \Sigma^M \covD_M
					\left( \covD \right)^2 \Psi_i
				+ \frac{1}{2} \left( \covD_M D_{ij} \right)^2
	\\
	&		\quad	
				+ D_{ik} D^{kj} D\indices{_j^i}
				+ \left( \Psi^i \Sigma_M \Psi_i \right)^2
	\\
	&		\quad
				- 2 iD_{jk}
					\left( \Psi^j \Sigma^M \covD_M  \Psi^k
						- \covD_M \Psi^j \Sigma^M \Psi^k \right)
	\\
	&		\quad
				+ \frac{i}{2}  \covD_M \Psi^i \Sigma^M \Sigma^{NS} 
					\left[ F_{NS} , \Psi_j \right]
				- 2i \covD^M F_{MN} \Psi^j \Sigma^N \Psi_j
			\bigg].
\end{split}
\end{equation}

\section[Supersymmetric higher-derivative gauge theories in \texorpdfstring{${ d=4 }$}{d=4}]{Supersymmetric higher-derivative gauge theories in \texorpdfstring{$\boldsymbol{ d=4 }$}{d=4}}

In this Section we are going to compute the relevant contributions to the four dimensional Lagrangian density that will allow us to evaluate the one-loop divergence of the supersymmetric extensions of the higher derivative \ym{} theory in four spacetime dimensions.
We will do this performing trivial dimensional reduction to the theory described by \eqref{lagr-hd-sym-6d}.

Already from a superficial inspection of \( \Lagr_{\SYM}^{6d,\HD} \), one could guess that the full dimensionally reduced Lagrangian contains quite a number of terms. Indeed, writing down all of them would take roughly ten pages and is not particularly instructive. Moreover, all we need is the counterterm to the kinetic term $\tr F_{\mu\nu}F_{\mu\nu}$, and most of the terms do not contribute to it in a convenient application of the background field method, similarly to what happened with matter fields in the non supersymmetric case.

In order to do this let us anticipate that we will expand the Lagrangian about a classical background analogous to that considered in \eqref{bkg-with-matter}, namely with a non-vanishing gauge fields only. 
As we will see explicitly, the dimensional reduction transforms the fields as
\begin{align*}
\text{Gauge field } A_M 
&	\longrightarrow 
		\text{gauge field } A_\mu + 
		\text{ complex scalar } \phi = \frac{A + iB }{\sqrt{2} }
\\
\text{Weyl spinor } \Psi^i_a
&	\longrightarrow 
		\text{Weyl spinors } \psi_{ \alpha } + \bar \lambda_{\dot \alpha} 
\\
\text{Auxiliary field } D^{ij}
&	\longrightarrow 
		\text{real + complex auxiliary fields } D + F.	
\end{align*}
where $A$ and $B$ are real scalars, as the notation suggests.
This means that terms at least cubic in the spinors, auxiliary fields and in the induced scalar do not contribute to the one-loop counterterm; moreover in terms quadratic in them and containing gauge fields, the latter is just evaluated at the background value. These considerations allow us to truncate away a lot of contributions.


\subsection[Dimensional reduction of the theory in \texorpdfstring{${ d=6}$}{d=6} ]{Dimensional reduction of the theory in \texorpdfstring{$\boldsymbol{ d=6}$}{d=6} }

\subsubsection{Generalities on trivial dimensional reduction }

We perform here a trivial dimensional reduction of the six dimensional Lagrangian \eqref{lagr-hd-sym-6d}.
By this we mean that we simply drop the dependence on the two coordinates $x^4$ and $x^5$ so that we formally set
\begin{equation}\label{dimensional-reduction-derivative}
\frac{\partial}{\partial x^4} = 0,
\qquad
\frac{\partial}{\partial x^5} = 0
\end{equation}
in \eqref{lagr-hd-sym-6d} and in the definitions in \eqref{6dim-covd-fmunu}. After this simplification, the covariant derivatives in the adjoint representation reads, written in terms of the generator of the fundamental one,
\begin{equation}\label{dimensional-reduction-covD}
\covD_M = 
\left\lbrace
\begin{split}
 & \covD_\mu 	=\partial_\mu + [A_\mu,\#]	\qquad &  M = \mu = 0,1,2,3 \\
 & [A, \#] 									& M = 4 \\			
 & [B, \#]									& M = 5
\end{split}
\right.
\end{equation}
where we denoted $A_{4,5} = (A,B)$.
For future reference we also write the d'Alembertian 
in the adjoint representation
\begin{equation}\label{dimensional-reduction-covDcovD}
\tilde\covD^2 =
\covD_M \covD^M 
	=
	\covD^2 
	+ \left[A, [A ,\#]\right]
	+ \left[B, [B,\#]\right],
\end{equation}
where $\covD^2 = \covD^\mu \covD_\mu.$

Let us apply this to the transformation properties in \eqref{6dim-covd-fmunu-gaugetrans}. For $M = \mu$, \eqref{dimensional-reduction-derivative} is ineffective, and so the first four components of $A^M$  become  the four-dimensional \ym{} potential $A_\mu$. For $M= 4,5$ the derivative drops and what is left is 
\begin{align}
 \delta_\omega A = [\omega, A], \hspace{3em}
 \delta_\omega B = [\omega, B],
\end{align}
that means that $A$ and $B$ are Lorentz-scalar fields transforming under the adjoint representation of the gauge group. 
Therefore, the six-dimensional \ym{} potential $A_{M}$ splits according to 
\begin{equation}
A_M = ( A_\mu , A , B ).
\end{equation}
The six-dimensional field strength  $F_{MN}$ reduces accordingly. Indeed, the components $(M,N) = (\mu, \nu)$ become the four-dimensional tensor $F_{\mu\nu}$. The other components read
\begin{align}
F_{ \mu 4 } & = \covD_\mu A, \\
F_{ \mu 5 } & = \covD_\mu B, \\
F_{ 4 5 } & =  [ A_4 , A_5 ];
\end{align}
the remaining components are related to these by the antisymmetry of $F_{MN}$.

As far as spinors are concerned, the dimensional reduction \( 6d \rightarrow 4d \) is suggested by the manifestly covariant \( \GroupName{SL}(2,\mathbb{C}) \) notation we employed for the $\Sigma$ matrices. Indeed, spinor indices split according to as $a=(\alpha, \dot \alpha)$; the reduction corresponds to restrict the spinor representation to the diagonal subgroup of $\GroupName{SU*}(4)$ 
\begin{equation}
\begin{pmatrix}
A & 0 \\
0 & A^*
\end{pmatrix},
\qquad
A \in \GroupName{SU}(2,\mathbb{C}).
\end{equation}
The symbol $\varepsilon^{abcd}$ factorizes into two-index components according to
\begin{equation}
\varepsilon^{abcd} \rightarrow \varepsilon^{\alpha \beta} \varepsilon^{\dot \gamma \dot \delta}
\end{equation}
and antisymmetry of such tensors.

The spinor  $\Psi^1$ therefore splits into two four-dimensional chiral spinors 
\begin{equation}
(\Psi^1)^a = 
\begin{pmatrix}
\psi^\alpha \\
\bar \lambda ^{\dot{\alpha}}
\end{pmatrix},
\qquad\qquad
(\Psi^2)^a = 
\begin{pmatrix}
\lambda^{\alpha}\\
- \bar \psi^{\dot{\alpha}}
\end{pmatrix}
\end{equation}
where \( \Psi^2 \) is fixed by the pseudo-Majorana constraint.

The tensor of auxiliary fields $D^{ij}$ splits into the auxiliary fields of the vector ($D$, real) and chiral ($F$, complex) supermultiplets:
\begin{equation}\label{dim-red-Dij}
	D^{ij} 
		=
	\begin{pmatrix}
		F^*	&	-D \\
		- D	&	- F	
	\end{pmatrix}.
\end{equation}

The overall coefficient $f$ in four dimensional spaceime has then dimension 2, and we write it in terms of the dimensionless \ym{} coupling $g$ as  $f = m^2 g^2$.

\subsubsection{Dimensional reduction of higher derivative Lagrangian}

Let us now apply the described techniques  to dimensionally reduce the Lagrangian \eqref{lagr-hd-sym-6d}. As mentioned, we will only give those terms that will be relevant for the computation.
In particular, all the terms in the second and third 
 line, 
are, respectively, third order on the auxiliary fields, fourth order on the spinors and third order on spinors and auxiliary fields, so that they do not contribute to the operator relevant for our computation and will not be considered from now on.

The higher derivative kinetic term for the gauge field decomposes as
\begin{equation}\label{dim-red-covdF2}
\begin{split}
\left( \covD^M F_{ML} \right)^2
	& \rightarrow
		\left( \covD^\mu F_{\mu\nu} \right)^2
		+ ( \covD^2 A )^2
		+ ( \covD^2 B )^2
	\\
	&
	\qquad
	- 2 ( \covD^\mu F_{ \mu \nu } )[ A , \covD^\nu A ]
	- 2 ( \covD^\mu F_{ \mu \nu } )[ B , \covD^\nu B ]
\end{split}
\end{equation}
that is the higher derivative kinetic term for $F_{\mu\nu}$ and for the two scalars. It also produces higher order interactions between the fields, that have been discarded. Notice that the scalar contribution can be written in terms of the complex field as
\begin{equation}
( \covD^2 A )^2
+
( \covD^2 B )^2
=
2 ( \covD^2 \phi )^2
\end{equation}

In a somewhat similar way, the higher derivative kinetic term for the spinor field reads
\begin{equation}\label{dim-red-psisigmacovDcovDpsi}
\begin{split}
i \Psi^i \Sigma^M \covD_M \tilde \covD^2 \Psi_i
	& \rightarrow 
		i \left( \Psi^1 \right)^a \Sigma^\mu_{ab} 
			\covD_\mu \left( \covD \right)^2 \left( \Psi^2 \right)^b 
			- i \left( \Psi^2 \right)^a	\Sigma^\mu_{ab} 
			\covD_\mu \left( \covD \right)^2 \left( \Psi^1 \right)^b 
	\\
	& \quad = 
		- i \psi \left\{ \covD_\mu, \covD^2 \right\}
			\sigma^\mu \bar \psi
		- i \bar \lambda \bar \sigma^\mu  
			\left\{ \covD_\mu, \covD^2 \right\}	 \lambda
\end{split}
\end{equation}
the other terms produced by \eqref{dimensional-reduction-covDcovD} and \eqref{dimensional-reduction-covD} do not contribute to the quadratic operator, since they are multiplied by two spinor fields.

The kinetic term for the auxiliary fields becomes
\begin{equation}\label{dim-red-covDDij2}
\begin{split}
\frac{1}{2} \left( \covD_M D_{ij} \right)^2
	& \rightarrow 
		\frac{1}{2} ( \covD_\mu D_{ij}) (\covD^{\mu} D^{ij} )
	\\
	& \quad = 
		- ( \covD_\mu F^* ) ( \covD^\mu F )  
		- ( \covD_\mu D ) ( \covD_\mu D )
\end{split}
\end{equation}
where we again dropped interactions with scalars. We also note here that the auxiliary field $F$ is not canonically normalized

We are now left with only two other contributions.
The first interaction between spinors and gauge fields produces
\begin{equation}\label{dim-red-covDpsisigmasigaFpsi}
\begin{split}
\frac{i}{2} \covD_M \Psi^j \Sigma^M \Sigma^{NS} 
\left[ F_{NS} , \Psi_j \right]
	& \rightarrow 
	\frac{i}{2} \covD_\mu \Psi^1 \Sigma^\mu \Sigma^{\nu\rho} 
	\left[ F_{ \nu \rho } , \Psi^2 \right]
	- \frac{i}{2} \covD_\mu \Psi^2 \Sigma^\mu \Sigma^{\nu\rho} 
		\left[ F_{ \nu \rho } , \Psi^1 \right]
\\
& \quad
		=
		- \frac{i}{2} 
			(\covD_\mu \psi) \sigma^\mu \bar \sigma^{\rho \sigma}
			\left[ F_{ \rho \sigma } , 	 \bar \psi \right]
		- \frac{i}{2}
			(\covD_\mu \bar \psi) \bar \sigma^\mu \sigma^{\rho \sigma}
			\left[ F_{ \rho \sigma } ,  \psi \right]
\\
& \qquad \quad
		- \frac{i}{2} 
			(\covD_\mu \lambda) \sigma^\mu \bar \sigma^{\rho \sigma}
			\left[ F_{ \rho \sigma } , 	 \bar \lambda \right]
		- \frac{i}{2}
			(\covD_\mu \bar \lambda) \bar \sigma^\mu \sigma^{\rho \sigma}
			\left[ F_{ \rho \sigma } ,  \lambda \right]
\end{split}
\end{equation}
where we discarded interactions with scalars.

Finally, the second interaction between spinor and gauge boson reduces as
\begin{equation}\label{dim-red-covDFpsisigmapsi}
\begin{split}
2 i \left( \covD^M F_{MN} \right) \Psi^j \Sigma^N \Psi_j
	& \rightarrow 
		- 2 ( \covD^\mu F_{ \mu \nu } )
			\left(
				 \Psi^1 \Sigma^\nu \Psi^2
				 -
				 \Psi^2 \Sigma^\nu \Psi^1
			\right)
\\
	& \quad
		 = 
		- 2\ ( \covD^\mu F_{ \mu \nu } ) 
			\left[
			 \psi \sigma^\nu \bar \psi 
				+ \bar \psi \bar \sigma^\nu \psi 
				+ \lambda \sigma^\nu \bar \lambda 
				+ \bar \lambda \bar \sigma^\nu \lambda 				
				\right],
\end{split}
\end{equation}
up to the by-now usual higher order contributions.

We are now ready to write the supersymmetric extension to the higher derivative contribution to \sym{}, truncated at second order in the matter fields. In principle we are able to write the case of extended $N=2$ supersymmetry, but first, for clarity and simplicity, we write down the $N=1$ invariant Lagrangian.

The truncated  $N=1$ higher derivative Lagrangian  is then obtained putting together the contributions
\eqref{dim-red-covdF2}-\eqref{dim-red-covDFpsisigmapsi} according to the six-dimensional Lagrangian density \eqref{lagr-hd-sym-6d}, setting to zero the auxiliary field $F$, the spinor $\lambda$ and the scalars $A$ and $B$. 
It therefore reads
\begin{equation}\label{lagr-HDonly-N=1-sym-4d}
\begin{split}
\left.\phantom{\int}\Lagr^{N=1}_{\SYM}\right|_{\HD}
	=
 \frac{1}{m^2 g^2}
 \tr &\bigg[
		\left( \covD^\mu F_{\mu\nu} \right)^2
		- i \psi \left\{ \covD_\mu, \covD^2 \right\}
			\sigma^\mu \bar \psi
		- ( \covD_\mu D ) ( \covD_\mu D )
\\
& \quad
		-\frac{i}{2} 
			(\covD_\mu \psi) \sigma^\mu \bar \sigma^{\rho \sigma}
			\left[ F_{ \rho \sigma } , 	 \bar \psi \right]
		- \frac{i}{2}
			(\covD_\mu \bar \psi) \bar \sigma^\mu \sigma^{\rho \sigma}
			\left[ F_{ \rho \sigma } ,  \psi \right]
\\
& \quad
		- 2 i\ ( \covD^\mu F_{ \mu \nu } ) 
			\left[
			 \psi \sigma^\nu \bar \psi 
				+ \bar \psi \bar \sigma^\nu \psi 
		\right]
\bigg].
\end{split}
\end{equation}
The full $N=1$ higher-derivative \sym{} Lagrangian is obtained  summing  the $N=1$ \sym{} action \eqref{lagr-N1sym-4d} with \eqref{lagr-HDonly-N=1-sym-4d}, that is
\begin{equation}\label{lagr-N=1-HDsym-4d}
\Lagr^{N=1}_{\HD\SYM}
	=
\Lagr^{N=1}_\SYM 
+
\left.\phantom{\int}\hspace{-1em}\Lagr^{N=1}_{\SYM}\right|_{\HD}.
\end{equation}
 By inspecting the contributions to the Lagrangian density \eqref{lagr-N=1-HDsym-4d}, we can recognise the terms related to the na\"ive insertion of $\covD^2$ inside the terms, but also non-trivial contributions dictated by gauge invariance. As anticipated in the introduction to the chapter, the auxiliary field becomes dynamical; by inspecting the sign in front of its kinetic and `mass' term -- that by rescaling the field by $m$ is actually $D^2$ in the usual \sym{} Lagrangian density \eqref{lagr-N1sym-4d} -- we can recognise in it a ghost field. We will come back on this later.

The whole dimensionally reduced Lagrangian gives us the truncated higher-derivative sector of the higher-derivative $N=2$ \sym{} theory. For simplicity of notation we collect  the terms related to higher derivative contributions of the chiral multiplet sector of the full $N=2$  Lagrangian density
\begin{equation}\label{lagr-HDonly-chiral-4d}
\begin{split}
\left.\phantom{\int}\Lagr_{\CH}\right|_{\HD}
	=
\frac{1}{m^2 g^2} \tr & \bigg[
		 2 ( \covD^2 \phi )^2
		- i \lambda \left\{ \covD_\mu, \covD^2 \right\}	\sigma^\mu \bar \lambda
		- ( \covD_\mu F^* ) ( \covD^\mu F )  		
	\\
	& \quad
		- \frac{i}{2} 
			(\covD_\mu \lambda) \sigma^\mu \bar \sigma^{\rho \sigma}
			\left[ F_{ \rho \sigma } , 	 \bar \lambda \right]
		- \frac{i}{2}
			(\covD_\mu \bar \lambda) \bar \sigma^\mu \sigma^{\rho \sigma}
			\left[ F_{ \rho \sigma } ,  \lambda \right]	
	\\
	& \quad
		- 2\ ( \covD^\mu F_{ \mu \nu } )
			\left[
				\lambda \sigma^\nu \bar \lambda 
				+ \bar \lambda \bar \sigma^\nu \lambda 				
				\right]
\bigg].
\end{split}
\end{equation}
The truncated $N=2$ Lagrangian density can then be written
\begin{equation}\label{lagr-N=2-HDsym-4d}
\begin{split}
\Lagr^{ N = 2 }_{ \HD \SYM }
	& =
\Lagr^{N=2}_\SYM 
+ \left.\phantom{\int}\hspace{-1em}\Lagr^{ N = 1 }_{\SYM}\right|_{\HD}
+ \left.\phantom{\int}\hspace{-1em}\Lagr_{ \CH }\right|_{\HD},
\end{split}
\end{equation}
where all the contributions have been just defined.

In order to proceed to a sensible quantization of the theory, namely in order to produce a convergent functional integral as discussed in Chapter~1, we perform a Wick-rotation of the theory. Without facing the details of the procedure, we formally apply the substitution $t \rightarrow i \tau $; the metric then rotates to $\eta^{\mu\nu} \rightarrow \delta^{\mu\nu}$ and the spinor representations are those of $\GroupName{Spin}(4) \approx \GroupName{SU}(2) \times \GroupName{SU}(2) $. As a consequence of the change of the metric, now $\sigma^0 = i \1$. Formally, since we were using a mostly positive metric, we simply cease to distinguish between upper and lower indices, and add an extra minus sign in front of the Lagrangian density, that now equals the energy.

\subsection{Multiplet structure}

In \eqref{lagr-HD-ym-splitted-dof}, \eqref{lagr-HD-scalar-splitted-dof} and \eqref{lagr-HD-spinor-splitted-dof} we have shown that we can rewrite the higher derivative Lagrangian density for massless gauge, scalar and spinor field in terms or a massless field and a massive ghost field, both described by a usual-derivative Lagrangian, the latter being though a ghost.

It is now interesting to check how the degrees of freedom of the involved supermultiplets split within this procedure.
We will analyse the vector and the chiral multiplet, since all other supersymmetric theories of interest for this work can be formulated using them.

We stress here that these  considerations are purely based on the structure of the kinetic term; a more accurate study would be necessary to understand how such grouping relates with supersymmetry transformations.
A hint into this direction was given in \cite{Ferrara:1977mv} for a simple model of the chiral multiplet.

\subsubsection{Vector multiplet}

The vector multiplet contains one vector boson with gauge invariance, one Weyl fermion and one real auxiliary field.

Considering ordinary derivative theories, off shell such fields describe therefore $3_B$, $4_F$ and $1_B$ degrees of freedom; on shell $2_B$, $2_F$ and zero.

Considering higher derivative theories, we have $5+1$ bosonic and $6$ fermionic degrees of freedom on shell. Indeed the multiplet decomposes into massless vector + massless spinor  + massive vector boson + massive fermion + real dynamical auxiliary fields. Off-shell we therefore have $3_B + 4_F$ massless and $3_B + 8_F + 1_B$ massive degrees of freedom; on-shell $2_F + 2_B $ and $3_B + 4_F + 1_B$.

We recognise the structure for the degrees of freedom consisting of the massless vector multiplet without auxiliary field + massive multiplet of ghost fields, the latter consisting of a vector boson + a Dirac fermion and a real dynamical auxiliary field.

The degree-of-freedom counting is summarized in the following table.
 (OD = ordinary derivative; HD = higher derivative; equiv.\ = equivalent defintion in terms of ordinary derivatives only)
\begin{table}[!ht]
\centering
\begin{tabular}{cccccc}
\hline \hline
  &   &   &   & \multicolumn{2}{c}{HD-equiv.} \\ 
field & off-sh.  & on-sh. (OD) & on-sh. (HD) & off-sh. & on-sh. \\  [0.75ex]
$A_\mu$ (+ gauge inv.) & $3$ & $2$ & $5$ & $3+4$ & $2+3$ \\ 
$\psi$ (Weyl) & $4$ & $2$ & $6$ & $4+8$ & $2+4$ \\ 
$D$ (real) &  $1$ & $0$ & $1$ & $0+1$ & $0+1$ \\ 
\hline \hline 
\end{tabular} 
\end{table}

We notice that the number of on shell bosonic and fermionic degrees of freedom coincide separately in the two categories of fields (massless vs.\ massive ghosts), even though they fail to agree off-shell.

\subsubsection{Chiral multiplet}

The chiral multiplet contains one complex scalar, one Weyl fermion and one complex auxiliary field.

Considering ordinary derivative theories, off shell such fields describe respectively $2_B$, $4_F$ and $2_B$ degrees of freedom; on shell $2_B$, $2_F$ and zero.

Considering higher derivative theories, we have $4+2$ bosonic and $6$ fermionic degrees of freedom on shell. Indeed in the ordinary derivative the multiplet decomposes into massless scalar and spinor + massive complex scalar + massive fermion + complex dynamical auxiliary fields. Off-shell we therefore have $2_B + 4_F$ massless and $2_B + 8_F + 2_B$ massive degrees of freedom; on-shell $2_F + 2_B  $ and $2_B + 4_F + 2_B$.

We can identify the degrees of freedom as the chiral multiplet without auxiliary field + massive multiplet of ghost fields, the latter consisting of a complex  scalar field, a Dirac fermion and a complex dynamical auxiliary field.

The degree-of-freedom counting is summarized in the following table.
 (OD = ordinary derivative; HD = higher derivative; equiv.\ = equivalent definition in terms of ordinary derivatives only)
\begin{table}[!ht]
\centering
\begin{tabular}{cccccc}
\hline \hline
  &   &   &   & \multicolumn{2}{c}{HD-equiv.} \\ 
field & off-sh.  & on-sh. (OD) & on-sh. (HD) & off-sh. & on-sh. \\  [0.75ex]
$\phi$ (complex) & 2 & 2 & 4 & 2+2 & 2+2 \\ 
$\psi$ (Weyl) & 4 & 2 & 6 & 4+8 & 2+4 \\ 
$F$ (complex) & 2 & 0 & 2 & 0+2 & 0+2 \\ 
\hline \hline 
\end{tabular} 
\end{table}

We notice, again, that the number of bosonic and fermionic degrees of freedom coincide separately in the two categories of fields (massless vs.\ massive ghosts), though this happens only on shell.

\section[Higher-derivative \texorpdfstring{${ N=1}$}{N=1} \sym{} in \texorpdfstring{${ d=4}$}{d=4}]{Higher-derivative \texorpdfstring{$\boldsymbol{ N=1}$}{N=1} \sym{} in \texorpdfstring{$\boldsymbol{ d=4}$}{d=4}}

In this section we are going to compute the one-loop correction to the gauge coupling for the higher derivative $N=1$ \sym{} theory.

Consider now the $N=1$ supersymmetric Lagrangian density with higher derivative \eqref{lagr-N=1-HDsym-4d}. As done for the other Lagrangian densities that we considered, in order to compute the one-loop beta function we need only the divergent contribution to the \ym{} kinetic term, and this can be obtained by expanding the fields about the background solution
\begin{equation}\label{N=1-bkg-expansion}
A_\mu \rightarrow A_\mu + B_\mu ,
\hspace{3.5em}
\psi \rightarrow \psi,
\hspace{3.5em}
D \rightarrow D,
\end{equation}
where $A_\mu$, $\psi$, $D$ are now the quantum fluctuations, similarly to that we did in  \eqref{bkg-with-matter}.
This means that in the Lagrangian,
the terms that contain only the gauge field can be expanded in the same way that we did for \eqref{lagr-hdym} (with $\gamma = 0$), so we can directly read from \eqref{LW-expansion-intermeriate} the expansion. For this reason it is convenient to choose the same gauge fixing \( G[A+B] = \covD_\mu A_\mu \) with again the integration weight $H = 1/g^2 - \covD^2 /g^2 m^2$ we used there.
The terms containing the spinor and the auxiliary field in \eqref{lagr-HDonly-N=1-sym-4d} are already formally expanded, with the convention that the field $ F_{\rho\sigma} $ is evaluated in the background field configuration.

We are now interested in a decomposition for the Lagrangian to quadratic fluctuations of the type
\begin{equation}
\Lagr^{N=1}_{\HD\SYM,Q^2}
	=
\Lagr^{ N = 1 }_A
+
\Lagr^{ N = 1 }_\psi
+
\Lagr^{ N = 1 }_D
\end{equation}
where each $\Lagr_X$ contains terms up to quadratic order in the fluctuations of the field $X$. In this way we will be able to read the operators for the computation of the effective action and to evaluate their determinants.

As already mentioned, the field content is the vector supermultiplet; it therefore consists of the gauge field, a chiral spinor and a real auxiliary field, that cannot be ignored as we implicitly did for \sym{} theory in Section 1.5.3
since it is now dynamical.

The effective action for this theory is therefore of the type described in \eqref{eff-act-1loop-gauge}, but it is not well defined because of the ghost character of the auxiliary field $D$, whose contribution makes the functional integral divergent. We will ignore this problem and apply \eqref{eff-act-1loop-gauge} anyway; formally this can be thought as the result of an analytic continuation of the auxiliary field $D \rightarrow iD$.\footnote{This is, of course, unacceptable from a superfield perspective, but since we are interested only in renormalization properties and $D$ is not a physical field we will not investigate this any further.}

In order to compute the beta function of the gauge coupling we therefore have to compute the total coefficient
\begin{equation}\label{b4-generic-N=1}
\left.b_4^\tot\right|_{ N = 1 }
	=
b_4(\Delta_A)
+
b_4(\Delta_D)
-
2 b_4(\Delta_\psi)
-
2 b_4(M_0)
-
 b_4(H)
\end{equation}
The Seeley-deWitt coefficients can be evaluated with the  previously exposed techniques. We will compute them in the next Section; here we directly give the result
\begin{equation}\label{b4-N=1}
\left.b_4^\tot\right|_{ N = 1 }
	=
-\frac{7}{4} C_2 F_{\mu\nu}^a F_{\mu\nu}^a
\end{equation}
yielding upon renormalization to the running of $g$ described by the $\beta$ function \eqref{beta-function-generic}
\begin{equation}
\beta_{N=1} = 
\frac{g^3}{32 \pi^2} \cdot \frac{7}{2}
\end{equation}

This $\beta$ function is positive, and corresponds to a theory perturbative at low energy and with a UV Landau pole. This is completely different from what is found for the \ym{} field in its first supersymmetric extension (\ie $N=1$ \sym{}, that has negative beta function).

It is also interesting to notice that in the expression \eqref{b4-N=1} for $\left.b_4^\tot\right|_{ N = 1 }$ the contribution proportional to $m^4$ cancels out, so that there is no cosmological constant contribution from this model, as prescribed by unbroken supersymmetry.

We now turn to the computation and evaluation of the Seeley-DeWitt coefficients for the involved operators. 


\subsection{Gauge field}

Following what we said in the previous paragraphs, we can read the relevant contribution for the gauge field from what we studied in Chapter~2, specialising the result to the case $\gamma = 0$. The relevant Seeley-deWitt coefficient can directly read from \eqref{b4-tot-hd-gauge}, that gives the overall contribution
\begin{equation}
\begin{split}
b_4^{ \text{gauge} } 
& := b_4(\Delta_A) - 2 b_4(M_0) - b_4(H)
 = \frac{43}{12} C_2 F_{\mu\nu}^a F_{\mu\nu}^a + \frac{3}{2} m^2 d_\ad.
\end{split}
\end{equation}
Of course back then this was the total coefficient, whereas here it is just one of the contributions.

\subsection{Spinor field}
Let us define the spinor operator representing the Lagrangian as
\begin{equation}
		\Lagr_{\psi}
	= 
		\frac{1}{2g^2m^2} 
		 \psi \Delta_{\psi} \bar \psi.
\end{equation}
The spinor operator is readily read off the Lagrangian as
\begin{equation}\label{operator_spinor}
\begin{split}
(\Delta_\psi)
  = 
& \ i  \sigma^\mu \covD^2 \covD_\mu 
%
				+ \bigg[
				 i F_{ \rho \mu }  \sigma^\rho	
				- \frac{i}{4} F_{\tau \kappa}
						\sigma^\mu
						\bar \sigma\indices{^{ \tau \kappa}}
				- \frac{i}{4} F_{ \tau \kappa } 
			\sigma\indices{^{ \tau \kappa} }
				\sigma^{\mu}
			 - i m^2  \sigma^\mu
			\bigg] \covD_{\mu} ;
\end{split}
\end{equation}
we ignored the contribution from the term in the last line of \eqref{lagr-HDonly-N=1-sym-4d} that is a boundary term (it is $B$-like in the notation of \eqref{3rd-order-operator-generic}).
Its Seeley-deWitt coefficient reads
\begin{equation}\label{b-4-spinor-n=1}
b_4(\Delta_\psi)
	=
\frac{21}{8} C_2 F_{\mu\nu}^a F_{\mu\nu}^a
+m^2 d_\ad
\end{equation}
as we are now going to verify. The derivation is lengthy and does not provide any particular insight, therefore the uninterested reader can skip it and go directly to the next Section.

\subsubsection{Computation of the determinant for the spinor field}

In order to get the coefficient for the spinor determinant, applying the idea outlined in Section~1.2.4, we compose \eqref{operator_spinor} with \(\bar \Delta_1 = i \bar \sigma^\nu \covD_\nu \). The result is
\begin{equation}
\begin{split}
	\bar \Delta_{\psi+1}
& :=
	\Delta_\psi \cdot \bar \Delta_1
= (\covD^2)^2
+
V_\psi^{\mu\nu} \covD_{\mu} \covD_{\nu}
+
U_\psi
\end{split}
\end{equation}
with coefficients
\begin{equation}\label{N=1-Vmunu}
V_\psi^{\mu\nu}
=
- \frac{3}{4}  F_{\tau\rho} \sigma^{\tau\rho} \delta^{\mu\nu}
+ F\indices{^{(\mu|}_\tau} \sigma^{\tau} \bar \sigma^{|\nu)}
+ \frac{1}{4} F_{\rho\kappa} \sigma^{(\mu|} \bar \sigma^{\rho \kappa} \bar \sigma^{|\nu)}
- m^2 \delta^{\mu\nu}
\end{equation}
and
\begin{equation}\label{Upsi}
\begin{split}
U_\psi
=
& 
\frac{1}{8} F_{\nu\rho} F_{\mu\kappa}
 	 \sigma^{ \mu } \bar \sigma^{\nu\rho} \bar \sigma^{\kappa }
+ \frac{1}{8} F_{\nu\rho} F_{\mu\kappa}
	 \sigma^{ \nu \rho } \sigma^{\mu \kappa}  
- \frac{1}{2} F_{ \mu \rho } F_{ \rho \kappa }
	\sigma^{ \mu } \bar \sigma^{\kappa}
+ \frac{1}{2} m^2 F_{\mu\nu} \sigma^{\mu\nu} 	.
\end{split}
\end{equation}
As in the previous cases, we decomposed the derivatives according to \(\covD_\mu \covD_\nu = \covD_{(\mu}\covD_{\nu)} + \frac{1}{2} F_{\mu\nu} \) in order to ensure the correct symmetries of the coefficients.

The trace of \( U_\psi \) reads
\begin{equation}
\begin{split}
\Tr U_\psi
&	=
\frac{1}{8} 
	\tr F_{ \nu \rho } F_{ \mu \kappa } 
	\, \str \bar \sigma^{\nu\rho} \bar \sigma^{ \kappa \mu }
+ \frac{1}{8}
	\tr F_{ \nu \rho } F_{ \mu \kappa } 
	\, \str \sigma^{\nu\rho} \sigma^{ \kappa \mu }
\\
& \qquad 
- \frac{1}{2}
	\tr F_{ \mu \rho } F_{ \rho \kappa } 
	\, \str \sigma^{\mu}  \sigma^{\kappa} 
\\
& =
- \tr F_{ \mu \nu} F_{ \mu \nu}
\\
& =
C_2  F^a_{ \mu \nu} F^a_{ \mu \nu}.
\end{split}
\end{equation}
It is interesting to notice that the two contributions in the first line of \eqref{Upsi} cancel against each other, since 
\(
\str \sigma\indices{^{[\mu|}}
\bar \sigma\indices{^{ \tau \kappa}}
\bar \sigma\indices{^{|\nu]}}
=
\str 
\bar\sigma\indices{^{\nu\mu }}
\bar \sigma\indices{^{ \tau \kappa\;}}
=
- 
\str\bar\sigma\indices{^{\mu \nu}}
\bar \sigma\indices{^{ \tau \kappa} }
\). The term $\sim m^2$ has vanishing trace because it is proportional to   \( \tr  \sigma^{\nu\rho} = 0\).

Let us now evaluate the trace $V_\psi^2$, with
\begin{equation}
\begin{split}
V_\psi
& =  V_\psi^{\mu\mu}
= - 4 F_{\nu\rho} \sigma^{\nu\rho} 
	- \frac{1}{2} \bar \sigma^{\nu\rho} F_{\nu\rho} - 4 m^2,
\end{split}
\end{equation}
that contributes with
\begin{equation}
\begin{split}
\Tr V_\psi^2
& =
16 \tr F_{\nu\rho}  F_{\mu\kappa} \, \str\sigma^{\nu\rho}  \sigma^{\mu\kappa} 
	+ 32 m^4 d_\ad
\\
& = 
 64 \, C_2 \, F^a_{\mu\nu}  F^a_{\mu\nu} 
	+ 32 m^4 d_\ad;
\end{split}
\end{equation}
the mixed product vanishes again because \( \str \bar \sigma^{\nu\rho} = 0 \).

Now we consider  \( \Tr V_\psi^{\mu\nu} V_\psi^{\mu\nu} \) is much more complicated due to the various products of sigma matrices. Expanding the product we get
\begin{equation}\label{N=1-trVmunuVmunu}
\begin{split}
\Tr V_\psi^{\mu\nu} V_\psi^{\mu\nu} 
	= 
		\Tr
		\bigg[
		&	\frac{15}{4}
				F_{\nu\rho} F_{\mu\kappa} 
				\sigma^{\nu \rho} \sigma^{\mu \kappa}
			+ F_{\alpha\mu} F\indices{^{(\alpha|}_\rho}
				\sigma^\mu \bar \sigma^\beta  \sigma^\rho \bar \sigma^{|\beta)}
				+ 4 m^4
		\\
		&	\quad	
			+ \frac{1}{16} 
				F_{\nu\rho} F_{\mu\kappa} 
				\sigma^\alpha \bar \sigma^{\nu\rho}  \bar \sigma^\beta
				\sigma^{(\alpha|} \bar \sigma^{\mu\kappa} \bar \sigma^{|\beta)}
		\\
		&	\quad
			+ \frac{1}{2} 
				F_{\alpha\mu} F_{\nu\rho} 
				\sigma^\mu  \bar \sigma^{\beta}  \bar \sigma^\beta
				\sigma^{(\alpha|} \bar \sigma^{\mu\kappa} \bar \sigma^{|\beta)}
		\bigg]
\end{split}
\end{equation}
plus some other traceless terms (though this might not be evident). The first term in \eqref{N=1-trVmunuVmunu} is the sum of the square of the first term in \eqref{N=1-Vmunu} and of twice the mixed product between the first and the second one.
This expression can be manipulated with the techniques used in Section~2.3.4, and it is not really instructive to go through the details of the computation.  
Nonetheless, we mention that, after some algebra, one can find that the two terms in the second and third lines actually cancel against each other, so that the result is all due to to the contributions in the first line and reads
\begin{equation}
\begin{split}
\Tr V_\psi^{\mu\nu} V_\psi^{\mu\nu} 
	= 
		19\, C_2 F_{\mu\nu}^a F_{\mu\nu}^a
		+ 8 m^4 d_\ad.
\end{split}
\end{equation}
Applying \eqref{b4-coeff-3order-implicit}, we then get the result \eqref{b-4-spinor-n=1}.

\subsection{Auxiliary field}

The auxiliary field, up to the overall mentioned sign, contributes with the operator
\begin{equation}
\Delta_D = - \covD^2 + m^2,
\end{equation}
that is straightforward to evaluate, and its coefficient is
\begin{equation}
b_4(\Delta_D)
	= 
		\tr \left[
			\frac{1}{12}  F_{\mu\nu} F_{\mu\nu}
			+ \frac{1}{2} m^4
		\right]
	= 
		- \frac{1}{12} C_2 F^a_{\mu\nu} F^a_{\mu\nu}
		+ \frac{1}{2} m^4 d_\ad.
\end{equation}

\section[Higher-derivative \texorpdfstring{${ N=2}$}{N=1} \sym{} in \texorpdfstring{${ d=4}$}{d=4}]{Higher-derivative \texorpdfstring{$\boldsymbol{ N=2}$}{N=1} \sym{} in \texorpdfstring{$\boldsymbol{ d=4}$}{d=4}}

Here we consider the $N=2$ higher-derivative \sym{}, obtained, as anticipated, directly by dimensional reduction of the full theory \eqref{lagr-hd-sym-6d}.
The quantization, hence the linearisation about a background solution and the gauge fixing, can be carried out as in the case of $N=1$ supersmmetry, since there is no new contribution containing only the gauge field.

We therefore perform the shift \eqref{N=1-bkg-expansion} and we also expand the other spinor, the scalar and the auxiliary $F$ about a vanishing background solution.
The Lagrangian density  can be then cast in the form
\begin{equation}\label{lagr-quadr-N=2}
\Lagr^{N = 2}_{\HD\SYM, Q^2}
	=
\Lagr^{  }_A
+
\Lagr^{  }_\psi
+
\Lagr^{  }_\lambda
+
\Lagr^{  }_\phi
+
\Lagr^{  }_D
+
\Lagr^{  }_F
\end{equation}
where all the symbols have the usual meaning.

By $R$-symmetry the two spinors must enter symmetrically in the Lagrangian; since there is no interaction quadratic in the spinors, their contribution is exactly the same and there is no new mixing between them.

The general considerations for this model are quite similar to those of the $N=1$ case. The only comment that we will add here is that the auxiliary field $F$, now dynamical, has a ghost character, and of course shares the same problems of the other auxiliary field $D$. As we did before, we will apply \eqref{beta-function-generic} anyway, and therefore the computation of the beta function is equivalent to obtain the coefficient
\begin{equation}\label{b4-N=2}
\begin{split}
b_4^\tot
&	=
b_4(\Delta_A)
+
2 b_4(\Delta_\phi)
-
4 b_4(\Delta_{\psi/\lambda})
\\
& \quad
+
b_4(\Delta_D)
+
2 b_4(\Delta_F)
-
2 b_4(M_0)
-
b_4(H)\\
& =
\left.b_4^\tot\right|_{ N = 1 }
+
2 b_4(\Delta_\phi)
-
2 b_4(\Delta_\lambda)
+
2 b_4(\Delta_F)
\end{split}
\end{equation}
where the factors $2$ in front of the contributions of $F$ and $\phi$ are due to their complex nature.

The only heat kernel coefficients that we have to compute in order to `complete' the $N=1$ result  \eqref{b4-N=1} are those of the scalar field $\phi$ and of the auxiliary field $F$.
The Lagrangian density for them is quite simple, at least at the linearised level, and the operators are simply
\begin{align}
 \Delta_\phi  =  \covD^4  +  m^2 \covD^2  ,
\hspace{5em} \Delta_F  =  - \covD^2  +  m^2  ,
\end{align}
whose Seeley-deWitt $b_4$ coefficients are
\begin{align}
& b_4 (\Delta_\phi) = -\frac{1}{6} C_2 F_{\mu\nu}^a F_{\mu\nu}^a+\frac{1}{2} m^4 d_\ad,\\
& b_4 (\Delta_F) = -\frac{1}{12} C_2 F_{\mu\nu}^a F_{\mu\nu}^a + \frac{1}{2} m^4 d_\ad,
.
\end{align}

Putting together the different contributions, we get for the total coefficient \eqref{b4-N=2}
\begin{equation}
\left.b_4^\tot\right|_{ N = 2 }
	=
-\frac{41}{6} C_2 F_{\mu\nu}^a F_{\mu\nu}^a
\end{equation}
yielding upon renormalization to the running of $g$ described by the $\beta$ function \eqref{beta-function-generic}
\begin{equation}
\beta_{ N = 2 } = 
\frac{g^3}{32 \pi^2} \cdot \frac{41}{3} C_2.
\end{equation}
The $\beta$ function in this case is bigger than the unextended $N=1$ case, and it is again positive, corresponding to the presence of a UV Landau pole.
Also for the $N=2$ supersymmetric case, there is no contribution proportional to $m^4$; this is due to the vanishing of such coefficient in the $N=1$ case and for the scalar multiplet separately.

\section[Higher-derivative \texorpdfstring{${ N=4}$}{N=4} \sym{} in \texorpdfstring{${ d=4}$}{d=4}]{Higher-derivative \texorpdfstring{$\boldsymbol{ N=4}$}{N=4} \sym{} in \texorpdfstring{$\boldsymbol{ d=4}$}{d=4}}


In terms of $N=1$ supermultiplets, we have the vector multiplet and three scalar multiplets. All fields with the same spin must enter completely symmetrically in the component field Lagrangian, therefore we can gain some information relevant for the new fields to be introduced by looking at the $N=2$ model, at least considering the Lagrangian expanded up to quadratic fluctuations.

Our construction of the Lagrangian will be guided by symmetry principles and by the fact that truncating two chiral multiplets out we have to recover the Lagrangian density for the $N=2$ case \eqref{lagr-quadr-N=2}.

Since there is no interaction term between the spinors in the $N=2$ supersymmetric case, no new term can therefore arise even in $N=4$.
Similarly, for the complex fields, interactions among the scalar fields are of higher order, and interactions with the gauge field like $F_{\mu\nu}F_{\mu\nu}\phi_i\phi_j$ would imply a contribution with the structure $F_{\mu\nu}F_{\mu\nu}\phi^2$ in the $N=2$ Lagrangian density. We notice that such term is absent is not the case, and this is not accidental: It is not possible since it would require a term at least cubic in the field strength in the Lagrangian in $d=6$, as explained in Section~3.1.

Following this line of thought, in order to compute this contribution we have just to repeat the computation for the $N=2$ case considering four times the contribution of fermionic fields and three times the scalar and auxiliary fields. 

Modulo the problem connected to the ghost nature of the auxiliary fields, we can apply \eqref{b4-total} and obtain the Seeley-deWitt coefficient
\begin{equation}
\begin{split}
\left.b_4^\tot\right|_{ N = 4 }
&	=
b_4(\Delta_A)
+
6 b_4(\Delta_\phi)
-
4 b_4(\Delta_{\psi/\lambda})
\\
& \quad
+
b_4(\Delta_D)
+
6 b_4(\Delta_F)
-
2 b_4(M_0)
-
b_4(H)\\
& =
\left.b_4^\tot\right|_{ N = 2 }
+
4 b_4(\Delta_\phi)
-
4 b_4(\Delta_\lambda)
+
4 b_4(\Delta_F)
\end{split}
\end{equation}
and the result is
\begin{equation}
\left.b_4^\tot\right|_{ N = 4 }
=
 -17 C_2 F^a_{ \mu \nu }  F^a_{ \mu \nu } .
\end{equation}

The divergence can as usual be reabsorbed renormalising the coupling constant $g$, whose flow is regulated by the $\beta$  function
\begin{equation}
\beta(g) =  \frac{g_\mu^3}{16 \pi^2} 34 C_2 ,
\end{equation}
is even bigger than the $N=2$ case, so that the coupling $g$ grows much faster with the scale than in the previously considered cases.

Also in this case, for what we said in the $N=2$ case, there is no contribution from $m^4.$

\section{Concluding remarks}

In this Chapter we managed to formulate, at least to a linearised level, the $N=1$, $2$ and $4$ supersymmetric extension of the higher derivative \ym{} theory \eqref{lagr-hdym}. We evaluated the $\beta$ functions of the theories, obtaining
\begin{align}
\beta_{N=1} =  \frac{g^3}{16 \pi^2} \frac{7}{4} C_2,
\qquad\quad
\beta_{N=2} =  \frac{g^3}{16 \pi^2} \frac{41}{3} C_2 ,
\qquad\quad
\beta_{N=4} =  \frac{g^3}{16 \pi^2} 34 C_2 .
\end{align}

Comparing with the result for the non-supersymemtric case, that is
\begin{align}
\beta_{\HD\YM} = - \frac{g^3}{16 \pi^2} \frac{43}{6} C_2,
\end{align}
we see that already in the $N=1$ case the asymptotic freedom is broken, and the theories are not ultraviolet complete any more, but, at least according to this one loop analysis, the coupling increases and diverges in a so-called Landau pole. This is the same behaviour observed in QED (see \eg{} \cite{Ram}). The extended supersymmetry models give an even faster running for the coupling $g$, so that the Landau point is expected to be at lower energy. In the usual-derivative supersymmetric theories, as computed in Section~1.5, the $\beta$ functions never turn positive; moreover, none of the considered theories maintains the conformal symmetry at the quantum level, while this happens for $N=4$ \sym{}.

Another feature that we pointed out during the analysis, is that in all the supersymmetric invariant Lagrangians there is not contribution to the cosmological constant, that is all divergences are reabsorbed renormalising the coupling, while in the case of  \ym{} field only, such a contribution is present.

	\newpage
	\pagestyle{plain}

\chapter*{Conclusion}
\addcontentsline{toc}{chapter}{Conclusion}\markboth{Conclusion}{}

		\fancyhead{} 
			\fancyhead[LE]{\scshape \leftmark}
			\fancyhead[RO]{\scshape \rightmark}
			\fancyfoot[LE,RO]{\thepage}
			\fancyfoot[LO,CE]{ }
			\fancyfoot[CO,RE]{ }
		 		\renewcommand{\headrulewidth}{0.4pt}
			\renewcommand{\footrulewidth}{0.4pt}
		\pagestyle{fancy}
		\renewcommand{\sectionmark}[1]{\markright{\thesection.\ #1}}
		\renewcommand{\chaptermark}[1]{\markboth{\chaptername\ \thechapter.\ #1}{}}

In this work we studied the one-loop renormalization properties of a \ym{} theory whose quadratic term in the Lagrangian contains, besides the conventional kinetic term, also an extra quadratic operator weighted with a coefficient with negative mass dimension. We also considered another dimension-six interaction cubic in the field strength weighted with a dimensionless parameter, and then we extended the theory to make it supersymmetric.

We first gave a rapid review of the path integral formalism, presenting the method of background field quantization to compute one-loop effective actions. In order to evaluate of the divergence in the effective action, the heat kernel technique was also introduced.

We then analysed the basic aspects of the theory, in particular considering its renormalization properties. This theory turned out to be renormalizable, since the extra derivatives present in the Lagrangian density induce another factor of squared momentum in the propagator, improving the ultraviolet properties of the theory. 
We computed the one-loop effective action, thus deriving the necessary renormalization of the gauge coupling and the $\beta$ function. We showed that the $\beta$ function is a quadratic function of the parameter weighting the cubic contribution, and can acquire either sign, or even vanish. Our computation corrects a mistake of an earlier work in the literature, confirming other computations with a  diagrammatic approach.
We also considered the coupling with this kind of higher derivative matter fields, including their contribution in the $\beta$ function, confirming some literature results. We observed that such a model solves the hierarchy problem because the mass of the scalar does not get quadratic corrections on the cut-off.
The alternative formulation of higher derivative theories in terms of usual-derivative ones was also discussed, showing that one ghost field appears.

The supersymmetric generalization of the theory presents some technical difficulties, that can be partially overcame in formulating it in six spacetime dimensions and then performing a trivial dimensional reduction.  Six dimensional spacetime seems to be more a more natural framework in which to formulate this kind of higher derivative theories, as the coupling is dimensionless and the component field expression of the Lagrangian is much simpler than the four dimensional ones. This allowed us to get the $N=1$ and $N=2$ supersymmetric extension of the original theory. We also managed to gain enough information to reconstruct, at least at the linearised level, the $N=4$ supersymmetric model. The supersymmetric case is much different from those previously considered. To begin with, supersymmetry prohibits the presence of the contribution cubic in the field strength; then, the matter field contribution is such that the theory has a positive $\beta$ function, growing with the number of supersymmetries.

To conclude with, higher derivative theories are interesting from many points of view and they arise in many different contexts. Even though they suffer of some inconsistencies and ambiguities, and it is unlikely that the theories we studied actually are a fundamental theory, they may serve as a toy model for them and might as well describe the low energy behaviour of some fundamental theory. Surely this is enough to motivate further studies in this area, and in particularly much work is to be done in understanding the connection between supersymmetry and higher-derivative theories. For example, it would be interesting to study in greater detail how the supermultiplet structure relates with the higher derivative operators; also, the relation between supersymmetry breaking mechanisms and the presence of extra derivatives should be understood with more attention.

\appendix

\renewcommand\chaptername{Appendix}

	\fancyfoot[CE,CO]{\thepage}
	\fancyfoot[LO,RO]{ }
	\fancyfoot[LE,RE]{ }
\chapter{Notation}


For definiteness of the functional integral we consider the Wick-rotated space with metric therefore
\(
\delta^{\mu\nu}
= \1_4
=
\text{diag}(++++)
\)
.
However, as we mentioned in the introduction, the Wick rotation is ill-defines for the higher derivative theories we study in this work. Such problems are not of interest for this thesis and this operation might be thought as a formal operation done to the theory defined in Minkovsky spacetime with metric \(\eta^{\mu\nu}=\text{diag}(-+++)\) to slightly simplify the notation. The formal differences between the two treatments is an overall sign in the Lagrangian density and factors of $i$ in the path integral formulation.
Einstein's index convention applies otherwise stated.
Our notation for spinors and supersymmetry is basically that of \cite{WB}.

Considering gauge theories, the symmetry group will generically be indicated with $\GroupName{G}$ and assumed to be a compact Lie group. Its structure constants are
\begin{equation}
	[T^a,T^b] = f^{amb} T^m,
\end{equation}
where $T^a$ are a set of generators of any representation. 
The generators \(t^a\) of the fundamental representation are normalized so so that \begin{equation}\label{notation-tr-generatord-fundam}
\tr t^a t^b = - \delta^{ab}/2 ;
\end{equation}  those of the adjoint representation have components
\(
(T_\ad^m)^{ab} = f^{amb}
\).
The normalization of any representation $R$ are normalized such that
\begin{equation}
\tr T_R^a  T_R^b = - C_R \delta^{ab} 
\end{equation}
defining the Casimir invariants $C_R$.
In a given representation $R$ we will therefore have
\begin{equation}\label{notation-trFmunuFmunu}
\tr F_{\mu\nu} F_{\mu\nu}
	=
- C_R F_{\mu\nu}^a F_{\mu\nu}^a.
\end{equation}
The Casimir element for the adjoint representation $C_2$ can be expressed in terms of the structure constants as
\begin{equation}
f^{mna} f^{mnb} =  C_2 \delta^{ab}.
\end{equation}
We assume Einstein's convention to hold also for gauge indices.

The gauge field $A_\mu$ induces in the minimal coupling prescription the covariant derivative
\begin{equation}
\covD_\mu = \partial_\mu + A_\mu
\end{equation}
and the field strength tensor reads
\begin{equation}
F_{\mu\nu} = [\covD_\mu, \covD_\nu]
= \partial_\mu A_\nu - \partial_\nu A_\mu + [A_\mu , A_\nu ].
\end{equation}
The field strength tensor satisfy the Bianchi's identity
\begin{equation}
\covD_{[\mu} F_{\nu\rho]} = 0
\end{equation}
that can be used to prove the relation
\begin{equation}\label{CovDF-dentity}
2 \left( \covD_\mu F_{\mu\nu} \right)^2 
	=
\left( \covD_\mu F_{\nu\rho} \right)^2 
+ F_{\mu\nu} \left[ F_{ \nu \rho } , F_{ \rho \mu } \right].
\end{equation}

We will also employ natural units in which the speed of light $c$ and Planck's constant $\hbar$ are taken to 1. All quantities will then be measured in units of mass. The only exception is the first chapter, where Planck's constant is used as a formal parameter to study loop-expansions.

The trace over gauge indices is $\tr$; $\tilde \tr$ is that on spinor indices and $\Tr$ indicates the trace over all indices after the symbol.

\section[Spinors in \texorpdfstring{${d=4}$}{{{d=4}}} ]{Spinors in \texorpdfstring{$\boldsymbol{d=4}$}{{{d=4}}} }

We use here two-component Weyl spinors $\psi^\alpha$.
The conjugate spinor is \( \bar \psi^{\dot \alpha} := (\psi^\alpha)^* \) where \(*\) denotes complex conjugation.

The antisymmetric tensors are
\[ \varepsilon^{21} = \varepsilon_{12} = +1 ,
\hspace{5em}
\varepsilon^{\dot 2 \dot 1} = \varepsilon_{\dot 1 \dot 2} = +1
\]
which moreover they satisfy
\[
\varepsilon_{\alpha\beta} \varepsilon^{\beta\gamma} = \delta^\gamma_\alpha,
\hspace{5em}
\varepsilon_{\dot\alpha\dot\beta} \varepsilon^{\dot\beta\dot\gamma} = \delta^\gamma_\alpha
.
\]
Sigma matrices read
\begin{equation}\label{notation-sigma-matrices}
	\sigma^\mu_{\alpha \dot\beta}
		=
	( \mathbbm{1} , \sigma^i )_{\alpha \dot \beta},
\hspace{4em}
	\bar\sigma^{\mu\; \dot \alpha \beta}
		=
	( \mathbbm{1} ,  -\sigma^i )^{\dot \alpha \beta}
		=
	\varepsilon^{\dot \alpha \dot \beta}
	\varepsilon^{\beta \alpha}
	\sigma^\mu_{\alpha \dot \beta}
\end{equation}
Lorentz generators of the spinor representation are
\begin{equation}\label{notation-gener-lorentz-spinor}
\begin{split}
	{\sigma^{\mu \nu \; }}\indices{ _{ \alpha} ^{{\beta}} }
& =
	 {\sigma^{[\mu}\bar\sigma^{\nu]}}%
		\indices{ _\alpha ^\beta }
=
	\frac{1}{2} \left(
				\sigma^\mu \bar \sigma^\nu
				-
				\sigma^\nu \bar \sigma^\mu 
			\right)\indices{ _\alpha ^\beta },
\\
	{{\bar{\sigma}}}\indices{ ^{\mu \nu \; } ^{\dot \alpha} _{\dot{\beta}} }
& =
	 {\bar\sigma^{[\mu}\sigma^{\nu]}}
		\indices{ ^{\dot{\alpha}} _{\dot{\beta}}}
=
	\frac{1}{2} \left(
				\bar \sigma^\mu \sigma^\nu
				-
				\bar \sigma^\nu \sigma^\mu 
			\right)\indices{ ^{\dot{\alpha}} _{\dot{\beta}}}.
\end{split}
\end{equation}

The conventions for contracting spinor indices are
\begin{equation}
\psi \chi = \psi^\alpha \chi_\alpha,
\hspace{5em}
\bar \psi \bar \chi = \bar \psi_{\dot \alpha} \bar \chi^{\dot \alpha}.
\end{equation}

\subsection{Identities}
For a comprehensive treatment of the two-component spinor notation we found \cite{Dreiner:2008tw} very useful. In our notation, we recall that the following identities hold. 
\begin{align}
\label{identity-1}
\str \sigma^\mu \bar \sigma^\nu
& 	= \str \bar \sigma^\mu \sigma^\nu 
	=
		- 2 \eta^{\mu\nu}
\\
\label{identity-2}
(\sigma^{\mu\nu})\indices{_\alpha^\beta}
&	=
		\varepsilon_{\alpha\tau} \varepsilon^{\beta\kappa} 	
 			( \sigma^{\mu\nu} )\indices{_\kappa^\tau}
\\ 
\label{identity-3}
\sigma^{\mu}_{\alpha\dot\alpha} \bar\sigma^{\dot \beta \beta}_\mu 
&	=
	-	2 \delta_\alpha^\beta\delta^{\dot\beta}_{\dot\alpha}
\\
\label{identity-4}
\sigma^{\mu\; \alpha\dot\alpha} \sigma_\mu^{ \beta \dot \beta}
&	=
	-	2\varepsilon_{\alpha\beta}\varepsilon_{\dot\alpha\dot\beta}
\\
\label{identity-5}
\bar \sigma^{\mu\;\dot \alpha\alpha} \bar\sigma^{\dot \beta \beta}_\mu 
&	=
	-	2\varepsilon^{\alpha\beta}\varepsilon^{\dot\alpha\dot\beta}
\\
\label{identity-6}
[\sigma^{(\mu} \bar\sigma^{\nu)}]\indices{_\alpha^\beta}
&	=
	-	\eta^{\mu\nu} \delta^\beta_\alpha
\\
\label{identity-7}
[\bar \sigma^{(\mu} \sigma^{\nu)}]
	\indices{^{\dot \alpha}_{\dot \beta}}
&	=
	-	\eta^{\mu\nu} \delta^{\dot\alpha}_{\dot\beta}
\\
\label{identity-8}
\str \sigma^{\mu\nu} \sigma^{\tau\kappa}
&	= 
		-4 \left( \eta^{\mu\tau} \eta^{\nu\kappa}
			-	\eta^{\mu\kappa} \eta^{\tau\nu}
			- i \varepsilon^{\mu\nu\tau\kappa}
			\right)
\\
\label{identity-9}
\str \bar \sigma^{\mu\nu} \bar \sigma^{\tau\kappa}
&	= 
		-4 \left( \eta^{\mu\tau} \eta^{\nu\kappa}
			-	\eta^{\mu\kappa} \eta^{\tau\nu}
			+ i \varepsilon^{\mu\nu\tau\kappa}
			\right)
\end{align}
In particular that the last two identities imply
\begin{equation}\label{identity-spinor_trace-double-sigmamn}
\Tr{F_{\mu\nu} F_{\tau\kappa}}
		\sigma\indices{^{\mu \nu}} 
		\sigma\indices{^{\tau \kappa}} 
	=
\tr{F_{\mu\nu} F_{\tau\kappa}}
\str		\sigma\indices{^{\mu \nu}} 
		\sigma\indices{^{\tau \kappa}} 
	=
	- 4 \tr F_{\mu\nu} F^{\mu\nu}
\end{equation}
and
\begin{equation}\label{identity-spinor_trace-double-sigmamn-V}
\Tr{F_{\mu\nu} F_{\tau\kappa}}
		{\bar \sigma}\indices{^{\mu \nu}} 
		{\bar \sigma}\indices{^{\tau \kappa} }
		=
\tr{F_{\mu\nu} F_{\tau\kappa}}
\str	{\bar \sigma}\indices{^{\mu \nu} } 
		{\bar \sigma}\indices{^{\tau \kappa} }
	=
	- 4 \tr F_{\mu\nu} F^{\mu\nu}.
\end{equation}

\section[Supersymmetry in \texorpdfstring{${d=4}$}{{{d=4}}} ]{Supersymmetry in \texorpdfstring{$\boldsymbol{d=4}$}{{{d=4}}} }

The Poincar\'{e} algebra, consisting of generators of translations $P_\mu$ and Lorentz transformations $M_{\mu\nu}$  reads
\begin{align}
\label{susy1}
[M_{\mu\nu}, M_{\rho \sigma}] &= 
- i \eta_{\mu\rho} M_{\nu\sigma}
- i \eta_{\nu\sigma} M_{\mu\rho}
+ i \eta_{\mu\sigma} M_{\nu\rho}
+ i \eta_{\nu\rho} M_{\mu\sigma} 
\\
[M_{\mu\nu} , P_\rho]
&  =
-i \eta_{\rho\mu}  P_\nu + i \eta_{\rho\nu} P_\mu.
\end{align} 
The supersymmetric extension, in the cases of vanishing central charge, is
\begin{align}
\label{pedissequo0000}
[M_{\mu\nu}, Q^i_\alpha] 
& = i {\sigma}\indices{_{\mu\nu \, \alpha}^\beta} Q^i_\beta
\\
[M_{\mu\nu}, \bar{Q}_{i}^{\dot \alpha}] 
&  = i {\bar{\sigma}}%
\indices{_{\mu\nu \,} ^{\dot{\alpha}}_{\dot{\beta}}} \bar Q_{i}^{\dot{\beta}}
\\
\label{pedissequo}
\{Q^i_\alpha , \bar{Q}_{j \dot \beta} \} & = 2 \sigma^\mu_{\alpha \dot{\beta}}
P_\mu \delta^{i}_j,
\end{align}
where $Q^i_\alpha$ are the supercharges, with $i$ being the $R$-symmetry index and $\alpha$ the spinor index; $(Q^i_\alpha)^\dagger = \bar Q_{i\dot\alpha}$.

The covariant derivatives read
\begin{equation}
\covDs_{\alpha i} = \frac{\partial}{\partial \theta^{\alpha i}}
+ i \sigma^\mu_{\alpha\dot\alpha} \bar\theta^{\dot \alpha}_i \frac{\partial}{\partial x^\mu},
\hspace{3em}
\bar \covDs_{\dot \alpha}^i = - \frac{\partial}{\partial \bar{\theta}^{\dot \alpha}_i}
- i \theta^{\alpha i } \sigma^\mu_{\alpha \dot\alpha} \frac{\partial}{\partial x^\mu};
\end{equation}
they commute with the generators $Q$, $\bar Q$ and  satisfy the algebra
\begin{equation}
\{
\covDs_{\alpha i},
\bar \covDs_{\dot \alpha} ^ i\}
=
 - 2 \delta_i^j \sigma^\mu_{\alpha \dot \alpha} \, \partial_\mu.
\end{equation}

The $N=1$ relevant supermultiplets are the chiral and the gauge one. For the latter, the degrees of freedom can be collected into the gauge superfield
\begin{equation}\label{wz-V}
V(x,\theta,\bar\theta) = - \theta \sigma^\mu \bar \theta A_\mu(x) + i (\theta\theta\bar \theta\bar\lambda(x) - \bar \theta\bar \theta\theta\lambda(x))
+ \frac{1}{2} \theta \theta \bar \theta \bar \theta D(x)
\end{equation}
that can be used to define the superfield strength
\begin{equation}
W_\alpha = - \frac{1}{4} \bar \covDs \bar \covDs \left( e^{-V} \covDs_\alpha e^V \right).
\end{equation}
The action can then be written as
\begin{align}\label{N1sym-4d}
S^{N=1}_{\SYM}
& = \frac{1}{2}
\int \dd{^4x} \dd{^2\theta}\tr [ W W + \text{h.c.}]
=
\int \dd{^4x}
\Lagr^{N=1}_\SYM
\\
\Lagr^{N=1}_\SYM
&=
\tr
\left[
\frac{1}{2} F_{\mu\nu} F^{\mu\nu} + 2 i \bar \lambda \bar \sigma^\mu \covD_\mu  \lambda -  D^2
\right]
\end{align}
with $F_{\mu\nu} = [\covD_\mu, \covD_\nu]$, $\covD_\mu $ being in the adjoint representation.
We also recall that the Wick-rotated Lagrangian, with a suitable rescaling of the fields, can be written as
\begin{align}
\label{lagr-N1sym-4d}
\Lagr^{N=1}_\SYM
&=
-
\frac{1}{g^2}
\tr
\left[
\frac{1}{2} F_{\mu\nu} F^{\mu\nu} + 2 i \bar \lambda \bar \sigma^\mu \covD_\mu  \lambda -  D^2
\right]
\end{align}

In the Abelian case the superfield strength reads
\begin{equation}
W_\alpha = - \frac{1}{4} \bar \covDs \bar \covDs \covDs_\alpha V 
\end{equation}
and the Lagrangian for super-Maxwell theory is
\begin{align}
\label{lagr-n=1-smaxwell}
 \Lagr^{N=1}_{\text{sM}} &=
-\frac{1}{4} F_{\mu\nu} F^{\mu\nu} - i \bar \lambda \bar \sigma^\mu \partial_\mu  \lambda + \frac{1}{2} D^2
\end{align}
with $F_{\mu\nu} = \partial_\mu A_\nu - \partial_\nu A_\mu.$

The gauge theory with extended supersymmetry can be realised by adding to \eqref{N1sym-4d} one or three $N=1$  chiral multiplets in the adjoint representations of the gauge group, with the correct coefficients in order to have the required $R$-symmetry between spinors and scalars.

	\fancyfoot[CE,CO]{\thepage}
	\fancyfoot[LO,RO]{ }
	\fancyfoot[LE,RE]{ }

\chapter{Some technical computations}
\pagestyle{fancy}
	\fancyhead{} 
	\fancyhead[LE]{\scshape \leftmark}
	\fancyhead[RO]{\scshape \rightmark}
	\fancyfoot[CE,CO]{\thepage}
	\fancyfoot[LO,RO]{ }
	\fancyfoot[LE,RE]{ }
	\renewcommand{\headrulewidth}{0.4pt}
	\renewcommand{\footrulewidth}{0.4pt}

\section{Linearised Lagrangian  for the gauge field}

Here we derive the expansions \eqref{LW-DF2-expansion} and \eqref{LW-FFF-expansion}.
We are looking for an operator of the form \eqref{hk-fourth-order-generic}, so we will ignore total derivatives and $N$-like terms, that as discussed will not contribute to the divergence.

\subsection{Expansion of the terms in the Lagrangian}

Let us start with the first higher-derivative contribution
\(\tr (\covD_\mu F_{\mu \beta} )^2  \).
We start substituting the expansions  \eqref{ym-shift-covD} and \eqref{ym-shift-Fmunu} ad expand the square
\begin{align*}
\left(\covD_\mu F_{\mu \beta} \right)^2  
& \rightarrow 
	\left[
		\left(
			\covD_\mu + [A_\mu, \#]
		\right)
		\left(
			F_{\mu \beta} 
				+ 2 \covD_{ [ \mu } A_{ \beta ] }  
				+ [A_\mu, A_\beta  ]
		\right)
	\right]^2
	\\
& \quad \simeq
 	\left(
 		[A_\mu , F_{\mu\beta}]
 	\right)^2
 	+ 
 	\left(
 		2 \covD_\mu \covD_{ [ \mu } A_{ \beta ] }
 	\right)^2 
 	+ 4 [  F_{ \mu \beta} , A_\mu ] \covD_\alpha
 		 \covD_{ [ \alpha } A_{ \beta ] }
 	\\
& \qquad  \quad	
 	+ 2 
 	\left(
 		\covD_\mu F_{\mu \beta} 
	\right)
 	\covD_\alpha
 	\left(
 		[ A_\alpha , A_\beta ]
 	\right)
	+ 4
	\left( 
		\covD_\mu F_{\mu \beta}
	\right) 
	[ A_\alpha^c  , 
		\covD_{ [ \alpha } A_{ \beta ] } 
	].
\end{align*}
Consider the two terms in the last line. The first one can be rewritten symbolically as \( A \cdot ( \covD(\#) ) \cdot A + \text{total derivate} \), hence it would only contribute to \( U \) with a total derivative; similarly, the second one contributes to \( N \). We therefore drop these terms from now on.

Taking the trace we can then write
\begin{align}\label{LW1-expansion-intermediate}
\begin{split}
\left(\covD_\mu F^a_{\alpha\beta} \right)^2  
& \rightarrow 
 	\left(
 		f^{abc}  F^b_{\mu \beta} A_\mu^c
 	\right)^2
 	+ 4
 	\left(
 		\covD^2 A^a_{ \beta  } - \covD_\mu \covD_{ \beta  } A^a_{ \mu  } 
 	\right)^2 
	+ 4 f^{abc} A^c_\mu F^b_{ \mu \beta} \left( \covD_\alpha
 		 \covD_{ [ \alpha } A^a_{ \beta ] }
 	\right)\\
& \quad
\simeq
(F_{\mu\beta} A_\mu)^2
+ 4 \left( \covD^2 A_{ \beta  } - \covD_\mu \covD_{ \beta  } A_{ \mu  } \right)^2
+ 4 A_\mu \cdot F_{\mu\beta} ( \covD_\alpha \covD_{[\alpha} A_{\beta]} )
.
\end{split}
\end{align}
where in the second step we have written the fields in the adjoint representation.
The first and third terms are immediately rewritten as
\begin{align*}
\left(
	F_{\mu \beta} A_\mu
\right)^2
& =
	A_\alpha \cdot F_{\alpha \mu } F_{\mu \beta } A_\beta
\end{align*}
and
\begin{align*}
4  A_\mu \cdot F_{ \mu \beta} \covD_\alpha
 	 \covD_{ [ \alpha } A^a_{ \beta ] }
& = 
	2 A_\alpha \cdot F_{ \alpha \beta} \covD^2 A_{ \beta }
 	-
 	2 A_\alpha \cdot F_{ \alpha \mu } \covD_{ \beta } \covD_{ \mu } A_{ \beta },
 	\\
& = 
	2 A_\alpha \cdot F_{ \alpha \beta} \covD^2 A_{ \beta }
 	-
 	2 A_\alpha \cdot F_{ \alpha \nu } \covD_{ \nu } \covD_{ \mu  } A_{ \mu }	
 	-
 	2 A_\alpha \cdot F_{ \alpha \mu } F_{ \beta  \mu } A_{ \beta }	,
\end{align*}
where we used \(\covD_\beta \covD_\mu = \covD_\mu \covD_\beta + F_{\beta\mu}\).
Now we focus on the middle term, that needs some rearrangement. First we expand the square and we get, up to total derivatives,
\begin{align*}
\begin{split}
\left(	\covD^2 A_{ \beta  } - \covD_\mu \covD_{ \beta  } A_{ \mu  } 	\right)^2 
& =
	A_\beta \cdot \covD^4 A_\beta
	- 2 A_\beta \cdot \covD^2 \covD_\mu \covD_\beta A_\mu
	+ (\covD_\mu \covD_\beta A_\mu)\cdot (\covD_\nu \covD_\beta A_\nu)
\end{split}
\end{align*}
and then we employ the following two relations, that can be proved with this kind of techniques,
\begin{align}
\label{LW1-expansion-tchanical-1}
A_\beta \cdot \covD^2 \covD_\mu \covD_\beta A_\mu
		& =	 A_\alpha \cdot F_{ \beta \alpha } \covD^2 A_\beta
			- (\covD_\mu A_\mu) \cdot \covD^2 (\covD_\nu A_\nu ) 
			+ 2 A_\alpha \cdot F_{ \mu \alpha } \covD_\mu  \covD_\beta A_\beta, 
			\\
\nonumber			\\
(\covD_\mu \covD_\beta A_\mu)\cdot (\covD_\nu \covD_\beta A_\nu)	
\label{LW1-expansion-tchanical-2}			
		& = 	- (\covD_\mu A_\mu) \cdot \covD^2 (\covD_\nu A_\nu ) 
				+ 2 A_\alpha \cdot F_{\mu \alpha } \covD_\mu \covD_\beta A_\beta  
				+ A_\alpha \cdot F_{ \alpha \mu } F_{ \mu \beta } A_\beta;
\end{align}
substituting them back in \eqref{LW1-expansion-intermediate} (and remembering the extra factor $-2$ for \eqref{LW1-expansion-tchanical-1}), we finally get
\begin{equation}
\begin{split}
\left(\covD_\mu F_{\mu \beta}^a \right)^2  
 \rightarrow 
 	A_\alpha \cdot &\delta_{\alpha \beta} \covD^4 A_\beta
 	 + (\covD_\mu A_\mu) \cdot \covD^2 (\covD_\nu A_\nu ) \\
 	& + 4 A_\alpha \cdot F_{ \alpha \beta} \covD^2 A_\beta
 	+ 4 A_\alpha \cdot F_{ \alpha \mu} F_{\mu \beta } A_\beta.
\end{split}
\end{equation}
that is \eqref{LW-DF2-expansion}.

We now consider the second higher-derivative term in the Lagrangian density \eqref{lagr-hdym}, that is the term cubic in the gauge field strength.
Before writing the components explicitly, it is convenient to exploit the antisymmetry of $F_{\mu\nu}$  and the cyclicity of the trace to get
\begin{equation}
\tr  F_{\mu\nu} [ F_{\mu\lambda} , F_{\nu,\lambda} ]
 = 
2 \tr  F_{\mu\nu} F_{\nu\lambda} F_{\lambda \mu} .
\end{equation}
Computing expansion \eqref{ym-shift-Fmunu}, it is not difficult to take the contribution to the one-loop effective action. Indeed we have two kind of terms: one of the form $\sim F\ F\ [ A,  A] $ and the other with the structure $\sim \tr F\ \covD A\ \covD A$. The first one can be cast in the form
\begin{equation}
\begin{split}
\tr  F_{ \mu \nu } F_{ \nu \lambda } [ A_\lambda , A_\mu  ]
&	=
\frac{1}{2}\tr [ F_{ \mu \nu } ,  F_{ \nu \lambda } ] [ A_\lambda , A_\mu  ]
\\
&	=
\frac{1}{2}
A^m_\lambda
\left(
F^a_{ \mu \nu }  F^b_{ \nu \lambda }  f^{abc} f^{mnc}
\right)
A^n_\mu ;
\end{split}
\end{equation}
this contributes to $U$, but it is traceless (with respect to the gauge indices, \ie contracting $n$ and $m$). We can therefore ignore this contribution.

The second contribution reads
\begin{equation}\label{LW2-expansion-intermediate}
\begin{split}
\tr  F_{\mu\nu} [ F_{\mu\lambda} , F_{\nu,\lambda} ]
& \rightarrow
2  \tr 12  F_{\mu\nu} \covD_{[\nu} A_{\lambda]} \covD_{[\lambda} A_{\mu]}
\\
&  \quad =
\tr 12  F_{\mu\nu} 
\left[ 
	\covD_{[\nu} A_{\lambda]} , \covD_{[\lambda} A_{\mu]}
\right]
\end{split}
\end{equation}
where in the equality we used the antisymmetry of $F_{\mu\nu}.$

Computing the traces in \eqref{LW2-expansion-intermediate}, we then obtain
\begin{align}
f^{abc} F_{\mu\nu}^a F_{\mu\lambda}^b F_{\nu\lambda}^c
	& \rightarrow 
12 f^{abc} F^b_{\mu\nu} \covD_{[\mu} A^a_{\lambda]} \covD_{[\lambda} A^c_{\nu]} .
\end{align}
Expanding the antisymmetrizations of the second term, \eqref{LW2-expansion-intermediate} reads
\begin{align*}
\begin{split}
f^{abc} F_{\mu\nu}^a F_{\mu\lambda}^b F_{\nu\lambda}^c
 \rightarrow 
 6 & F^{ a b }_{\mu \beta}  ( \covD_{\mu} A_{\alpha} )^a ( \covD_{\alpha} A_{\beta} )^b
 \\
 & 
 - 3 F^{ a b }_{\mu\nu} \delta_{ \alpha \beta } ( \covD_{\mu} A_{\alpha} )^a ( \covD_{\nu} A_{\beta} )^b- 3 F^{ a b }_{\alpha \beta} ( \covD_{\lambda} A_{\alpha} )^a ( \covD_{\lambda} A_{\beta} )^b
\end{split}
\end{align*}
where we have also written the fields explicitly in the adjoint representation.

Up to total derivatives and contributions from irrelevant terms, the previous expression becomes
\begin{align}
f^{abc} F_{\mu\nu}^a F_{\mu\lambda}^b F_{\nu\lambda}^c
& 
\rightarrow
A_{\alpha}^a
	\left[
		3 F_{ \mu \nu } \delta_{ \alpha \beta }  
		+  3 F_{ \alpha \beta} \delta_{ \mu \nu } 
		- 6 F_{ \mu \beta }\delta_{ \alpha \nu } 
	\right]^{ a b }
	 \covD_{ \mu } \covD_{ \nu } A_{\beta}^b.
\end{align}
that is indeed in agreement with \eqref{LW-FFF-expansion}.

	\fancyfoot[CE,CO]{\thepage}
	\fancyfoot[LO,RO]{ }
	\fancyfoot[LE,RE]{ }


\end{document}